\documentclass[a4paper,12pt,final]{article}
\usepackage{etex}

\usepackage[sc]{mathpazo}

\usepackage[table]{xcolor}

\usepackage{threeparttable}
\usepackage{standalone}
\usepackage{booktabs, dcolumn}
\usepackage{booktabs}

\usepackage{afterpage}

\usepackage{rotating}
\usepackage{tikz}

\usepackage{latexsym}
\usepackage{amsfonts}
\usepackage{amssymb}
\usepackage{amsthm}
\usepackage{amsmath}
\usepackage[mathscr]{eucal}
\usepackage{multicol}
\usepackage{setspace,graphicx,amsmath,amsfonts,amssymb,amsthm}
\usepackage{pdfpages}
\usepackage{pdflscape}

\usepackage{anyfontsize}
\usepackage{t1enc}

\usepackage{caption}
\usepackage{subcaption}
\DeclareCaptionFont{captiofont}{\fontsize{9pt}{10.8pt}\selectfont}
\usepackage{verbatim}
\usepackage[plainpages=false,
breaklinks=true,
colorlinks=true,
urlcolor=gray,
citecolor=blue,
linkcolor=red,
bookmarks=true,
bookmarksopen=true,
bookmarksopenlevel=1,
pdfstartview=FitH,
pdfview=FitH]{hyperref}

\usepackage{color,hyperref}

\usepackage[round]{natbib}

\usepackage[lmargin=1in, rmargin=1in, tmargin=1.0in, bmargin=1.0in, paperwidth=220mm, paperheight=290mm]{geometry}

\usepackage{booktabs}
\usepackage{tabulary}
\usepackage{tabularx}
\usepackage{multirow}

\usepackage{enumerate}
\usepackage{sectsty}
\usepackage{lipsum}
\sectionfont{\centering}

\usepackage{tocloft}
\usepackage{titlesec}
\titleformat{\section}
{\large\bfseries}{\thesection}{0.5em}{}

\renewcommand\thesection{\arabic{section}}

\renewcommand\thesubsection{\thesection.\arabic{subsection}}
\titleformat{\subsection}
{\normalfont\itshape}{\thesubsection}{0.5em}{}

\renewcommand\thesubsubsection{\thesection.\arabic{subsection}.\arabic{subsubsection}}
\titleformat{\subsubsection}{\normalfont\itshape}{\thesubsubsection}{0.5em}{}

\usepackage{setspace}

\titlespacing\section{0pt}{10pt plus 4pt minus 2pt}{5pt plus 2pt minus 2pt}
\titlespacing\subsection{0pt}{10pt plus 4pt minus 2pt}{0pt plus 2pt minus 2pt}
\titlespacing\subsubsection{0pt}{10pt plus 4pt minus 2pt}{0pt plus 2pt minus 2pt}

\providecommand{\keywords}[1]{\textbf{Keywords:}  #1}
\providecommand{\JEL}[1]{\textbf{JEL:}  #1}

\makeatletter
\newcommand*{\myfnsymbolsingle}[1]{%
\ensuremath{%
\ifcase#1
\or 
*%
\or 
\dagger
\or 
\ddagger
\or 
\mathsection
\or 
\mathparagraph
\else 
\@ctrerr
\fi
}%
}
\makeatother

\usepackage{alphalph}
\newalphalph{\myfnsymbolmult}[mult]{\myfnsymbolsingle}{}

\renewcommand*{\thefootnote}{%
\myfnsymbolmult{\value{footnote}}%
}

\makeatletter
\def\@xfootnote[#1]{%
\protected@xdef\@thefnmark{#1}%
\@footnotemark\@footnotetext}
\makeatother

\usepackage{bbding} 		

\usepackage[symbol]{footmisc}

\usepackage{chngcntr}

\usepackage{scrwfile}
\TOCclone[\contentsname]{toc}{atoc}

\AfterTOCHead[toc]{%
}
\AfterTOCHead[atoc]{%
\edef\maintocdepth{\the\value{tocdepth}}%
\value{tocdepth}=-10000\relax%
}

\usepackage{blindtext}

\newcommand{\bPhi}{\ensuremath{\boldsymbol\Phi}}
\newcommand{\bPsi}{\ensuremath{\boldsymbol\Psi}}
\newcommand{\btheta}{\ensuremath{\boldsymbol\theta}}
\newcommand{\bepsilon}{\ensuremath{\boldsymbol\epsilon}}
\newcommand{\bbeta}{\ensuremath{\boldsymbol\eta}}
\newcommand{\bSigma}{\ensuremath{\boldsymbol\Sigma}}

\def\finsupport{We appreciate the valuable discussions of Saleem Bahaj (EFA discussant), Mete Kilic (WFA discussant), Manuela Pedio (Bristol discussant), Guofu Zhou (SGF discussant), and the helpful comments of Vikats Agarwal, Kevin Aretz, Mattia Bevilacqua, Pasquale Della Corte, Michael Ellington, Nicola Fusari, Janet Gao, Amit Goyal, Massimo Guidolin, Byeongju Jeong, Anastasios Kagkadis, Michal Kejak, Roman Kozhan, Daniele Massacci, Ingmar Nolte, Grzegorz Pawlina, Roberto Pinto, Bryan Routledge, Lucio Sarno, Mark Shackleton, Ivan Shaliastovich, Ctirad Slavik, Radu Tunaru, Michael Weber, and conference participants at the 2023 Western Finance Association (WFA) Meeting, the 2023 Conference of Swiss Society for Financial Market Research (SGF), the 2023 European Economic Association (EEA) Meeting, the 2022 European Finance Association (EFA) Meeting, the 2022 International Symposium on Forecasting (ISF), the 2022 Slovak Economic Association Meeting (SEAM), and 3rd Annual Bristol Financial Markets Conference and seminar participants at the Bank of England, Adam Smith Business School, Alliance Manchester Business School, CERGE-EI, and University of Liverpool Management School. A previous version of the paper circulated under the title ``Currency Network Risk''.
}

\begin{document}

\setcounter{footnote}{0}

\newpage
\hypersetup{colorlinks,linkcolor=black} 

\linespread{1}

\title{\Large \setcounter{footnote}{0}Volatility Shocks and Currency Returns\footnote{\finsupport}}
\vspace{20pt}

\author{\setcounter{footnote}{6}Mykola Babiak\footnote{Lancaster University Management School, Bailrigg, LA1 4YX, Lancaster, UK, E-mail: \texttt{m.babiak@lancaster.ac.uk}.}\\{\small\textit{Lancaster University Management School}}
\and
\and
\setcounter{footnote}{11}Jozef Barun\'{i}k\thanks{Institute of Economic Studies, Charles University, Opletalova 26, 110 00, Prague, CR and Institute of Information Theory and Automation, Academy of Sciences of the Czech Republic, Pod Vodarenskou Vezi 4, 18200, Prague, Czech Republic, E-mail: \texttt{barunik@utia.cas.cz}.} \\{\small\textit{Charles University}}}

\date{\hspace{2em}}

\maketitle

\linespread{1.0}

\begin{abstract}

This paper examines how shocks to currency volatilities predict exchange rates. Using option-implied volatilities, we construct a dynamic, directed network of volatility connections. Currencies that transmit more volatility shocks, which control for common correlation, earn lower excess returns. Buying the weakest and selling the strongest transmitters delivers high risk-adjusted performance, driven by spot exchange rate movements and not explained by standard factors. A general equilibrium model shows that volatility transmission related to idiosyncratic shocks proxies for priced country-specific risk. Assuming a monotonic amplification of domestic idiosyncratic risk, volatility transmission forecasts negatively future excess returns, consistent with the empirical evidence.

\vspace{10pt}

\keywords{Currency predictability, currency volatility, volatility shock, network, option-implied volatility, contagion, term structure}

\JEL{G12, G15, F31}
\end{abstract}







\newpage

\renewcommand{\thefootnote}{\arabic{footnote}}
\setcounter{footnote}{0}

\hypersetup{colorlinks,linkcolor=red} 

\section{Introduction}

\linespread{1.5}

Volatility has played a central role in economics and finance. In currency markets, the carry trade returns are strongly related to innovations in the global equity \citep*{lustig2011common} and foreign exchange (FX) \citep*{menkhoff2012carry} volatility. While the role of global volatility risk in foreign exchange markets is well-documented, there is little evidence on how individual currency volatilities relate to one another. This seems surprising, since a shock to the ex ante volatility of a particular currency might influence expectations about the future volatility of other currencies. These linkages consequently define weighted and directed structures that capture the transmission of volatility shocks among individual currencies, beyond fluctuations in global volatility.  Since the connections are likely asymmetric, investors cannot diversify away from country-specific shocks. Hence, differences in the strength of directional volatility connections become essential characteristics of exchange rates.

Guided by this insight, we aim to measure how expectations about future exchange rate fluctuations covary across currencies and how shocks to these expectations create a network and propagate through it. Using currency option prices, we first construct forward-looking measures of exchange rate volatilities for 20 countries from January 1996 to December 2021 as in \cite*{della2016volatility} and \cite*{della2020cross}.\footnote{Surprisingly, the literature on the information content of currency options is limited compared to the evidence on equity options. Notable exceptions making use of dollar options include \cite*{campa1995testing}, \cite*{della2016volatility}, \cite*{londono2017variance}, and \cite*{della2020cross}, whereas \cite*{jurek2014crash}, \cite*{mueller2017international}, and \cite*{della2022arbitrage} employ cross currency options.}$^{,}$\footnote{The highly liquid and large foreign exchange volatility market provides an excellent opportunity to synthesize such measures. As of June 2019, the daily average turnover was \$294 billion, and the notional amount outstanding was \$12.7 trillion \citep*{BIS1,BIS2}. A wide variety of strike and maturity options available in the market allows us to compute option-implied variances of exchange rates precisely.} We then estimate volatility linkages via time-varying parameter vector autoregression models \citep*{diebold2014,barunik2020dynamic}. The volatility network inferred from time-varying variance decompositions has several key attributes. First, the connections between the volatilities are weighted and directed. Since shocks influence exchange rate volatilities asymmetrically within the system, this network captures information beyond standard correlation-based measures. Second, international volatility comovement is driven by global shocks. In our analysis, we examine two types of volatility networks that either allow or control for common variation. Third, we can identify connections between shocks with various degrees of persistence. Thus, we shed light on the importance of linkages between transitory and persistent volatility shocks. 

\begin{figure}[t!]
\begin{center}
\includegraphics[width=1.0\textwidth]{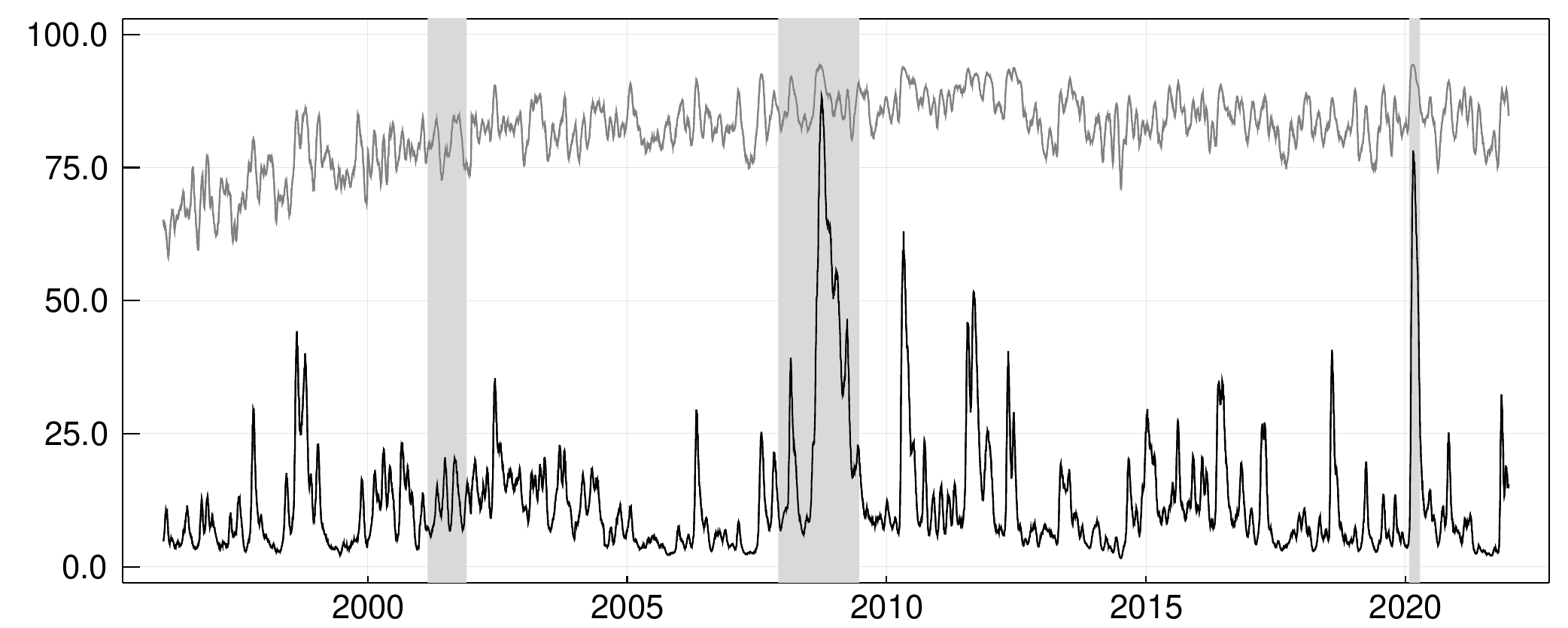}
\caption{\textbf{Volatility connectedness dynamics.}} 
\begin{minipage}{\textwidth} \footnotesize
The figure depicts the time variation in the volatility connectedness $\mathcal{C}(u)$ based on linkages that include (gray line) or exclude (black line) a common correlation component. The shaded areas denote the NBER recessions. The sample is from January 1996 to December 2021.
\end{minipage}
\label{fig:overall connectedness}
\vspace{-20pt}
\end{center}
\end{figure}

Figure \ref{fig:overall connectedness} depicts the aggregate connectedness, defined as the fraction of volatility shocks transmitted between individual currencies relative to all shocks, for the two volatility network types. When contemporaneous comovement is allowed, we observe high volatility transmission, reflecting strong unconditional correlations. Once we remove common correlation, the degree of volatility spillovers substantially declines and becomes counter-cyclical. It is worth emphasizing that, after controlling for strong correlation among currency volatilities, the volatility connectedness exhibits economically meaningful dynamics. The main contribution of this paper is to examine the implications of currency volatility linkages purged of strong commonality.

We document that such disaggregate volatility connections predict currency returns. Currencies that transmit more correlation-adjusted volatility shocks to others earn a lower average excess return. Buying the weakest and selling the strongest transmitters of volatility shocks, which controls for common correlation, delivers a high average return, a high Sharpe ratio, and strong cumulative performance. Interestingly, this strong performance is driven by both long and short portfolios of a zero-cost strategy and stems from predicting the spot exchange rates. Furthermore, sorting currencies by transmitted shocks yields a novel source of currency predictability relative to extant currency benchmarks.

We show additional implications of volatility transmission for predictability. First, we construct a long-short strategy for short- and long-term linkages. In both cases, the average return and the Sharpe ratio change only slightly and remain statistically significant. This suggests that the documented predictability is related to permanent volatility shocks. Second, we show substantial diversification benefits from combining standard benchmarks with our strategy, driven by a weak negative correlation with many currency factors. 

Next, we examine the frequency with which countries are identified as the strongest or weakest transmitters of volatility shocks. We find that volatility transmission is neither uniform nor static over time, exhibiting persistent yet evolving patterns. In another exercise, we relate the returns obtained from sorting on volatility transmission to global volatility, liquidity, and financial stress risks. The long-short portfolio shows a strong positive correlation with various measures of equity and foreign exchange volatility, while its comovement with liquidity is less pronounced. Thus, aggregate volatility and liquidity risk conditions cannot fully explain the information in correlation-adjusted volatility transmission. Our findings align well with recent evidence in \cite*{nucera2024currency}. Using statistical analysis, the authors examine a large currency cross-section and find a strong dollar factor, weak carry and momentum, and additional candidate factors related to volatility, uncertainty, and liquidity conditions, rather than macro variables. 

Inspired by the empirical analyses, we develop the theory of international contagion. In the model, cash-flow shocks in one country can affect cash flows in the domestic and foreign economies. Hence, currency risk premia and exchange rate volatility are linked through cross-country exposure to jumps. The theoretical model predicts that volatility transmission associated with idiosyncratic jumps proxies for the priced risk of country-specific shocks, generating cross-sectional return predictability unrelated to global risk. Under a monotonic domestic amplification of idiosyncratic risk, stronger volatility transmitters of country-specific shocks earn lower average excess returns, consistent with the empirical evidence. Consistent with theoretical predictions, we empirically demonstrate that the excess returns of the volatility transmission strategy are negatively associated with the likelihood of idiosyncratic jumps. Thus, they can be regarded as the compensation for country-specific risk. 

This paper contributes to the literature on currency return predictability.\footnote{The literature proposes strategies, among others, based on interest rates \citep{lustig2007cross,lustig2011common,menkhoff2012carry}, momentum \citep{menkhoff2012currency,asness2013value,dahlquist2020economic}, business cycles \citep*{colacito2020business}, and global imbalances \citep*{corte2016currency}.} The two volatility-related strategies use global foreign exchange volatility \citep*{menkhoff2012carry} and currency volatility risk premium \citep{della2016volatility,londono2017variance}. \cite*{gabaix2015international} and \cite*{colacito2018currency} build the models explaining these strategies. We contribute to this literature by empirically and theoretically showing that correlation-adjusted shocks to exchange rate volatility predict the cross-section of currency excess returns. This predictability cannot be explained by existing currency predictors. In a related paper, \cite{della2020cross} document a global risk factor --- the implied volatility slope --- in the cross-section of implied volatility returns. We instead examine shocks to implied volatilities and their predictive power for spot currency returns. 

Furthermore, \cite{mueller2017international} propose a strategy based on the currency sensitivity to the cross-sectional dispersion of conditional correlations between spot exchange rates. They find interesting results regarding compensation for exposure to high- or low-dispersion states. In contrast, we focus on dependencies in currency volatilities. Additionally, our volatility linkages are directional, unlike symmetric correlation-based proxies.

Our paper is related to the model of \cite*{richmond2019trade}, which explains the carry trade premium via a country's position in the global trade network. Note that the predictability induced by volatility transmission in our paper stems from changes in spot exchange rates, not interest rate differentials, making it distinct from trade links. More recently, \cite*{hou2024trade} and \cite*{bahaj2024beyond} extend the theory presented in \cite{gabaix2015international} and explore network structures in trade imbalances and financial intermediation. Unlike their networks based on international trade, we study the network of shocks to exchange rate implied volatilities, which are clearly very different from macroeconomic fundamentals in terms of the information content. 

Finally, our paper is related to the literature on downside risk in currency markets. \cite*{jurek2014crash}, \cite*{burnside2011peso}, \cite*{farhi2015crash} and \cite*{chernov2018crash} investigate currency crash risk, while \cite*{fan2021equity} document an option-based equity tail factor in currency returns. Although we use currency options data, our paper examines dependencies among currency volatilities and does not specifically focus on downside risk. Nevertheless, our results show that the strategy based on the transmission of correlation-adjusted volatility shocks provides excellent diversification benefits, particularly during periods of high volatility and low liquidity.

\section{Shocks to currency volatilities}

This section describes the construction of option-based currency volatilities, outlines the procedure for estimating shocks to individual volatilities, and introduces the volatility network measures used in the core analysis.\footnote{Appendix \ref{app:estimate} provides a detailed methodology for estimating a dynamic volatility network.}

\subsection{Currency implied volatilities}
\label{sec: currency option-implied volatilities}

We begin by synthesizing the risk-neutral expectation of exchange rate variances from quoted currency options by applying a model-free approach of \cite*{BJN2000} and \cite*{BKM2003}. We use prices of European call and put options expiring at time
$t+\tau$
to compute the risk-neutral expectation of the return variance of an exchange rate
$k$
versus the USD between
$t$
and
$t+\tau:$
\begin{equation}
E_t^\mathbb{Q}\left[(RV_{t,\tau}^k)^2\right] = \frac{2}{B^k(t,t+\tau)}\left\{\int\limits_{F^k(t,t+\tau)}^{\infty} \frac{C^k(t,t+\tau,K)}{K^2} dK  + \int\limits_{0}^{F^k(t,t+\tau)} \frac{P^k(t,t+\tau,K)}{K^2} d K\right\},
\label{eq:ivdecomp}
\end{equation}
where
$RV_{t,\tau}^k$
is the realized volatility of the underlying asset, 
$E_t^\mathbb{Q}\left[\cdot\right]$
denotes the expectation operator under the risk-neutral probability measure
$\mathbb{Q},$
$C^k(t,t+\tau,K)$ 
and 
$P^k(t,t+\tau,K)$ 
denote the prices of call and put contracts at time
$t$
with a strike price 
$K$
and maturity
$\tau,$ 
$B^k(t,t+\tau)$
is the price of a country's bond at time
$t$
with maturity
$\tau,$
$F^k(t,t+\tau)$
is the forward exchange rate of the currency
$k$
at time 
$t$ 
with maturity 
$\tau.$
We discretize the integral in Equation (\ref{eq:ivdecomp}) by adopting call and put option prices interpolated around the $\tau$ maturity, and by considering a range of strike prices for the currency $k.$ 

Two comments are noteworthy. First, large swings in volatilities will produce even larger fluctuations in variances due to convexity. This may artificially imply a stronger transition of shocks. Following the definition of the volatility risk premium strategy in \cite{della2016volatility} and the forward volatility contract for exchange rates studied by \cite{della2020cross}, we examine the linkages between implied volatilities, $\sqrt{E_t^\mathbb{Q}\left[(RV_{t,\tau}^k)^2\right]}.$\footnote{In robustness checks, we demonstrate that our results remain unchanged if we use implied variances. We use volatility measures to facilitate comparison with the existing literature.} Second, the important characteristic of the network of shocks to volatilities is the forward-looking nature of risk-neutral volatilities synthesized from derivatives. It is intuitively evident that their forward-looking nature should be distinct from backward-looking realized volatilities. In robustness checks, we demonstrate that, indeed, the predictive power of network structures estimated from realized volatilities is substantially weaker. Thus, we emphasize the importance of shocks to forward-looking volatilities, which can be synthesized from options data.

\subsection{A dynamic volatility connectedness}
\label{sec: dynamic volatility network}

Having constructed forward-looking currency volatilities, we aim to identify the network of shocks that propagate across them. The knowledge of how volatility shocks to a currency 
$j$ 
transmit to a currency 
$k$  
defines a directed link. This disaggregate connection between two currency pairs characterizes one of them as a recipient and the other as a transmitter of volatility shocks. Aggregating information across all pairs yields a system-wide measure of the connectedness among currency volatilities. 

A dynamic network of volatility shocks can be well characterized using variance decompositions from a time-varying parameter vector autoregression (TVP-VAR) approximation model \citep*{diebold2014}. The variance decompositions provide information about how much of the future variance of a variable 
$j$ 
is due to shocks in a variable 
$k$. 
In this paper, dynamic networks are distinct from classical networks in the financial econometrics literature. In a typical network, the adjacency matrix is symmetric and contains zero and one entries, indicating whether nodes are linked. In our work, however, the links are directed, meaning that the $j$ to $k$ link is not necessarily the same as the $k$ to $j$ link; hence, the adjacency matrix is not symmetric. This is the key to our analysis, as the difference between recipients and transmitters stems directly from network asymmetries. Furthermore, the variance decompositions in our work incorporate weighted links, indicating the strength of the connections.

\subsubsection{Volatility network definition}

We construct a dynamic volatility network from a TVP-VAR model estimated using currency implied volatilities, following the methodology of \cite*{barunik2020dynamic}. We consider a locally stationary TVP-VAR of a lag order $p$ describing the dynamics as:
\begin{equation}\label{eq:VAR1}
\mathbf{CIV}_{t,T}=\bPhi_{1}(t/T)\mathbf{CIV}_{t-1,T}+\ldots+\bPhi_{p}(t/T)\mathbf{CIV}_{t-p,T} + \bepsilon_{t,T},
\end{equation}
where $\mathbf{CIV}_{t,T}=\left(\mathrm{CIV}_{t,T}^{(1)},\ldots,\mathrm{CIV}_{t,T}^{(N)}\right)^{\top}$ is a double indexed $N$-variate time series of currency volatilities, 
$\bepsilon_{t,T}=\bSigma^{-1/2}(t/T)\bbeta_{t,T}, \bbeta_{t,T}\sim NID(0,\boldsymbol{I}_M)$ are normally distributed shocks, $\bPhi(t/T)=(\bPhi_{1}(t/T),\ldots,\bPhi_{p}(t/T))^{\top}$
are the time-varying autoregressive coefficients. Note that $t$ refers to a discrete-time index $1\le t \le T$ and $T$ is an additional index indicating the sharpness of the local approximation of the time series by a stationary process. Rescaling time such that the continuous parameter $u \approx t/T$ is a local approximation of the weakly stationary time-series \citep*{dahlhaus1996kullback}, we approximate $\mathbf{CIV}_{t,T}$ in a neighborhood of $u_0=t_0/T$ by a stationary process:
\begin{equation}\label{eq:VAR2}
\widetilde{\mathbf{CIV}}_t(u_0)=\bPhi_1(u_0)\widetilde{\mathbf{CIV}}_{t-1}(u_0)+\ldots+\bPhi_p(u_0)\widetilde{\mathbf{CIV}}_{t-p}(u_0) + \bepsilon_t.
\end{equation}

The TVP-VAR process defined by Equation (\ref{eq:VAR1}) has a time varying Vector Moving Average VMA($\infty$) representation \citep*{dahlhaus2009empirical}:
\begin{equation}
\mathbf{CIV}_{t,T} = \sum_{h=-\infty}^{\infty} \bPsi_{t,T}(h)\bepsilon_{t-h}
\end{equation}
where a parameter vector $\bPsi_{t,T}(h) \approx\bPsi(t/T,h)$ is a time-varying impulse response function characterized by a bounded stochastic process.\footnote{Since $\bPsi_{t,T}(h)$ contains an infinite number of lags, we approximate the moving average coefficients at $h=1,\ldots,H$ horizons.} Information contained in $\bPsi_{t,T}(h)$ permits the measurement of the contribution of shocks with various degrees of persistence. Since a shock to a variable in the model need not occur in isolation, an identification scheme is crucial for identifying the network. We adapt the extension of the generalized identification scheme of \cite*{pesaran1998generalized} to a locally stationary process as proposed by \cite*{barunik2020dynamic}. Furthermore, we transform local impulse responses in the system into local impulse transfer functions via the Fourier transform. This enables us to measure the horizon-specific dynamics of the network using the heterogeneous persistence of shocks in the system. A dynamic representation of the variance decomposition of shocks from a currency $j$'s volatility to a currency $k$'s volatility then establishes a dynamic horizon-specific adjacency matrix, which is central to connectedness measures.

Specifically, the element of a dynamic horizon-specific adjacency matrix, which captures how shocks from the volatility of a currency $j$ propagate to the volatility of a currency $k$ at a given point of time $u=t_0/T$ and a given horizon $d_i \in \mathcal{H},$ is defined as:
\begin{equation} \label{eq:dynamicadjmatrix}
\Big[ \btheta(u,d_i) \Big]_{j,k} = \frac{\widehat{\sigma}_{kk}^{-1} \displaystyle \sum_{\omega \in d_i} \left( 
\bigg[ \widehat\bPsi(u,\omega) \widehat \bSigma(u) \bigg]_{j,k} \right)^2 }{ \displaystyle \sum_{\omega \in \mathcal{H}} \Bigg[ \widehat\bPsi(u,\omega) \widehat \bSigma(u) \widehat\bPsi^{\top}(u,\omega)  \Bigg]_{j,j} },
\end{equation}
where $\widehat\bPsi(u,\omega) = \sum_{h=0}^{H-1} \sum_h \widehat \bPsi(u,h) e^{-i\omega h}$ is an impulse transfer function estimated from Fourier frequencies $\omega$ of impulse responses that cover a specific horizon $d_i,$ which can be defined for arbitrarily chosen bands of frequencies.\footnote{Note that $i=\sqrt{-1}$.} It is important to note that $\Big[ \btheta(u,d) \Big]_{j,k}$ is a natural disaggregation of traditional variance decompositions to a time-varying and $h$-horizon adjacency matrix. This is because the portion of the local error variance of the $j$-th variable at horizon $h$ due to shocks in the $k$-th variable is scaled by the total variance of the $j$-th variable. Next, we normalize the elements in the $j$-th row by the corresponding row sum and multiply by the implied volatility of a currency  $j$: 
\begin{equation}
\label{eq:dynamicadjmatrix normalized}
\Big[ \widetilde \btheta(u,d) \Big]_{j,k} = \frac{\Big[ \btheta(u,d) \Big]_{j,k}}{ \sum\limits_{k=1}^N\Big[  \btheta(u,d) \Big]_{j,k}} \cdot \mathrm{CIV}^{(j)}(u).
\end{equation}
We can interpret each element
 $\Big[ \widetilde \btheta(u,d) \Big]_{j,k}$
as the amount (expressed in volatility units) of the volatility of a currency $j$ due to shocks to the volatility of a currency $k$. By construction, the sum of elements in each row equal the corresponding currency's implied volatility.   

\subsubsection{The role of heterogeneous persistence and contemporaneous correlations}
\label{sec: the role of heterogeneous persistence and contemporaneous correlations}

The adjacency matrix defined by Equations (\ref{eq:dynamicadjmatrix})-(\ref{eq:dynamicadjmatrix normalized}) possess the two attributes -- a chosen horizon
 $d_i$ 
and a covariance matrix 
 $\widehat{\bSigma}$ 
-- that are important for our empirical analysis. First, economists commonly measure the impact of transitory and persistent shocks on asset prices. The horizon-specific adjacency matrix allows us to estimate the volatility connections stemming from shocks with various degrees of persistence by choosing arbitrarily chosen bands for  $d_i.$ In the empirical estimation, we define the short horizon $(S)$ as a 1-day to 1-month interval and the long horizon $(L)$ as longer than 1 month. Thus, these two frequencies specify transitory and more persistent shocks.

Second, financial markets exhibit strong comovement, particularly during adverse states, when asset price volatility tends to increase synchronously. In particular, we observe strong synchronicity in option-implied volatilities for exchange rates. The TVP-VAR model would capture these strong correlations between individual currency volatilities through the covariance matrix
$\widehat{\bSigma}.$
As a result, the strength of volatility connections identified by the adjacency matrix might be significantly amplified by contemporaneous effects. We disentangle the impact of contemporaneous correlations on volatility spillovers by considering two types of time-varying covariance matrices. We first allow the contemporaneous effects among volatility shocks and then remove the contemporaneous correlations by diagonalizing the covariance matrix. 

\subsubsection{Numerical implementation}

We estimate the approximating model in Equation  (\ref{eq:VAR2}) using Quasi-Bayesian Local-Likelihood (QBLL) methods \citep*{petrova2019quasi} to obtain the time-varying coefficient estimates
$\widehat{\bPhi}_{1}(u),...,\widehat{\bPhi}_{p}(u)$
and the time-varying covariance matrix
$\widehat{\bSigma}(u)$
at a given point of time $u=t_{0}/T$. Specifically, we use a kernel weighting function that assigns larger weights to observations near the period whose coefficient and covariance matrices are of interest. Using conjugate priors, the (quasi) posterior distribution of the model's parameters is available analytically. This alleviates the need for a Markov Chain Monte Carlo (MCMC) simulation algorithm and enables parallel computing. We provide a detailed discussion of the estimation algorithm in Appendix \ref{app:estimate}. We also publish the computationally efficient packages $\mathtt{DynamicNets.jl}$ in $\mathtt{JULIA}$ and  $\mathtt{DynamicNets}$ in $\mathtt{MATLAB}$ that can be used to replicate volatility network measures.\footnote{The packages are available at \href{https://github.com/barunik/DynamicNets.jl}{https://github.com/barunik/DynamicNets.jl} and \href{https://github.com/mte00/DynamicNets}{https://github.com/mte00/DynamicNets}.} 


\subsection{Volatility connectedness measures}
\label{sec: connectedness measures}

The elements of the adjacency matrix displayed in Table \ref{table: adjacency matrix} completely specify the horizon-specific connections in each period. We begin by defining the aggregate volatility connectedness as the ratio of the off-diagonal elements to the sum of all values:
\begin{equation}
\label{eq: local}
\mathcal{C}(u,d) = \displaystyle \sum_{\substack{j,k=1\\ k\ne j}}^N \Big[\widetilde \btheta(u,d)\Big]_{j,k} \Big/ \sum_{\substack{j,k=1}}^N \Big[\widetilde \btheta(u,d)\Big]_{j,k} \quad d \in \mathcal{H} = \left\{S,L\right\},
\end{equation}
where the short-term and long-term horizons are determined by chosen bands of frequencies in Section \ref{sec: the role of heterogeneous persistence and contemporaneous correlations}. The sum of the measures across the two frequency bands 
 $\mathcal{C}(u) = \mathcal{C}(u,S) + \mathcal{C}(u,L)$ 
defines the connectedness of shocks with any degree of persistence. A higher value of the aggregate volatility connectedness measure indicates greater shock transmission among individual currency volatilities. 

\begin{table}[t!]
\centering
\caption{Dynamic adjacency matrix}
\begin{minipage}{\textwidth} 
This table presents a time-varying adjacency matrix. The element $\Big[ \widetilde \btheta(u,d) \Big]_{j,k}$ of such a matrix captures a portion of the local error variance of the volatility of a currency $j$ due to shocks in the volatility of a currency $k$ at a given point of time $u$ and a given horizon $d \in \mathcal{H}.$  The from-directional connectedness (to-directional) measure of a currency $j$ is the sum of elements in the row (column) $j,$ excluding the one on the main diagonal.
\end{minipage}
\vspace{\medskipamount}
\footnotesize

\begin{tabular}{lccccc}
\toprule
Currency & 1     & 2     & $\cdots$ & N     & $\mathcal{F}_{j}(u,d)$ \\
\midrule
1 & $\Big[\widetilde \btheta(u,d)\Big]_{1,1}$ & $\Big[\widetilde \btheta(u,d)\Big]_{1,2}$ & $\cdots$ & $\Big[\widetilde \btheta(u,d)\Big]_{1,N}$ & $\sum\limits_{k\neq 1} \Big[\widetilde \btheta(u,d)\Big]_{1,k}$ \\
2 & $\Big[\widetilde \btheta(u,d)\Big]_{2,1}$ & $\Big[\widetilde \btheta(u,d)\Big]_{2,2}$ & $\cdots$ & $\Big[\widetilde \btheta(u,d)\Big]_{2,N}$ & $\sum\limits_{k\neq 2} \Big[\widetilde \btheta(u,d)\Big]_{2,k}$ \\
$\vdots$ & $\vdots$ & $\vdots$ & $\ddots$ & $\vdots$ & $\vdots$ \\
$N$ & $\Big[\widetilde \btheta(u,d)\Big]_{N,1}$ & $\Big[\widetilde \btheta(u,d)\Big]_{N,2}$ & $\cdots$ & $\Big[\widetilde \btheta(u,d)\Big]_{N,N}$ & $\sum\limits_{k\neq N} \Big[\widetilde \btheta(u,d)\Big]_{N,k}$ \\
\midrule
$\mathcal{T}_{j}(u,d)$ & $\sum\limits_{k\neq 1} \Big[\widetilde \btheta(u,d)\Big]_{k,1}$ & $\sum\limits_{k\neq 2} \Big[\widetilde \btheta(u,d)\Big]_{k,2}$ & $\cdots$ & $\sum\limits_{k\neq N} \Big[\widetilde \btheta(u,d)\Big]_{k,N}$ &   \\
\bottomrule
\end{tabular}
\label{table: adjacency matrix}
\end{table}

We also define disaggregate volatility connectedness quantities that measure the degree to which an individual currency transmits (receives) volatility shocks. The to-directional volatility connectedness quantifies the contribution of each currency's $j$ volatility to the volatilities of other currencies:
\begin{equation}
\label{eq: to-directional risk}
\mathcal{T}_{j}(u,d) = \displaystyle \sum_{\substack{k=1\\ k\ne j}}^N \Big[\widetilde \btheta(u,d)\Big]_{k,j}  \quad d \in \mathcal{H} = \left\{S,L\right\}.
\end{equation}
The from-directional volatility connectedness captures how much of each currency's volatility is due to shocks of other currencies' volatilities in the cross-section:
\begin{equation}
\label{eq: from-directional risk}
\mathcal{F}_{j}(u,d) = \displaystyle \sum_{\substack{k=1\\ k\ne j}}^N \Big[\widetilde \btheta(u,d)\Big]_{j,k} \quad d \in \mathcal{H} = \left\{S,L\right\}.
\end{equation}
Adding these measures across the horizons yields the impact of shocks of any persistence:
\begin{equation}
\label{eq: from-directional and to-directional risk}
\mathcal{F}_{j}(u) = \mathcal{F}_{j}(u,S)+\mathcal{F}_{j}(u,L)
\quad\wedge\quad
\mathcal{T}_{j}(u) = \mathcal{T}_{j}(u,S)+\mathcal{T}_{j}(u,L).
\end{equation}

The volatility connectedness measures capture multiple novel features. First, unlike prior work on correlation risk, the volatility connections in our paper can be asymmetric, allowing the separation of transmitted and received shocks. Second, we can remove contemporaneous effects and identify shocks that are not driven by the common (global) factor, which has been shown to drive carry trade returns. Third, our econometric methodology allows us to disentangle the effect of horizon-specific volatility connections. A large body of literature examines the role of shocks with varying persistence. For example, long-term fluctuations in expected growth and the volatility of cash flows \citep* {BanYar2004} have played a central role in understanding returns on equity, bonds, and currencies.

\section{Data and currency portfolios}

\subsection{Currency options data}
We begin our empirical investigation by collecting daily OTC option-implied volatilities for exchange rates against the USD from Bloomberg. Following \cite{della2016volatility} and \cite{della2020cross}, we consider a sample of the following 20 developed and emerging market countries: Australia, Brazil, Canada, the Czech Republic, Denmark, Euro Area, Hungary, Japan, Mexico, New Zealand, Norway, Poland, Singapore, South Africa, South Korea, Sweden, Switzerland, Taiwan, Turkey, and the United Kingdom. The data cover the sample period from January 1996 to December 2021. The cross-section begins with 10 currencies and gradually increases over time, with data on all exchange rates available from August 2004 to December 2021.\footnote{We greatly appreciate the help of Roman Kozhan with the currency implied volatilities for the sample before 2013. We then extend the sample until 2021 using currency options from Bloomberg.}

We synthesize spot implied variances using a model-free approach of \cite*{BJN2000}, which requires currency option prices for a range of strike prices. Quotes for OTC currency options are expressed in terms of \cite*{garman1983foreign} implied volatilities for selected combinations of plain-vanilla options (at-the-money and 10- and 25-delta put and call options).  We recover strike prices from deltas and option prices from implied volatilities using Bloomberg interest rates, as well as spot and forward exchange rates from Barclays and Reuters via Datastream. Using this recovery procedure, we obtain plain-vanilla European calls and puts on exchange rates against the USD for maturities of 1 month, 3 months, 6 months, 12 months, and 24 months. 

Since our portfolio sorts are conducted monthly, it is natural to assume that traders prefer to use 1-month implied volatilities for exchange rates rather than longer-maturity data. We therefore use 1-month volatilities in our empirical analysis. Furthermore, we estimate the volatility network using daily volatilities to increase the number of observations and better capture the dynamic nature of shock propagation. 

\subsection{Exchange rate data}
We retrieve daily bid, mid, and ask spot and forward exchange rates versus the USD from Barclays and Reuters via Datastream. We obtain daily nominal interest rates for domestic (the US in our case) and foreign countries from Bloomberg. The core empirical analysis is conducted at a monthly frequency; therefore, we sample end-of-month observations for all time series. We match exchange and interest rate data with currency options data for a cross-section of 20 countries, covering the period from January 1996 to December 2021.

\subsection{Currency excess returns}
We denote the spot and forward exchange rates of foreign currency
 $k$ 
at time 
 $t$
as
 $S_{k,t}$
and
 $F_{k,t}.$
Exchange rates are expressed as USD per unit of a foreign currency. Thus, an increase in  
 $S_{k,t}$
indicates an appreciation of the foreign currency. The one-period ahead excess return to a US investor for holding foreign currency 
 $k$ 
at time 
 $t$
is
\begin{equation}
\label{eq: exchange rates}
rx_{k,t+1} = \Delta s_{k,t+1} + i_{k,t} - i_t \approx s_{k,t+1} - f_{k,t},
\end{equation} 
in which
 $i_{k,t}$
and 
 $i_t$
represent the risk-free rates of the foreign country 
 $k$
and the US,
 $\Delta s_{k,t+1}$
is the log change in the spot exchange rate, 
 $f_{k,t}$
and
 $s_{k,t+1}$
denote the log spot and forward rates. Under covered interest rate parity (CIP), the interest rate differential 
 $i_{k,t} - i_t$
is equal to a forward discount
 $s_{k,t} - f_{k,t}.$
Thus, the approximation in Equation (\ref{eq: exchange rates}) states that the currency excess return equals the difference between the future spot rate and the current forward rate. The early literature documented that CIP held even for very short horizons \citep*{akram2008arbitrage}, while recent evidence has shown CIP deviations in the post-global financial crisis period \citep*{du2018deviations,andersen2019funding}. We demonstrate that the profitability of the strategies studied in our paper stems primarily from predictability in spot exchange rates. Therefore, our key results are independent of the validity of the CIP condition.

\section{Volatility shocks and currency returns}
\label{section: core portfolio analysis}
\subsection{Volatility connectedness}

Figure \ref{fig:overall connectedness} illustrates the volatility connectedness measure
$\mathcal{C}(u)$
based on shocks that include or exclude a contemporaneous effect. The high correlations among risk-neutral currency volatilities induce strong and persistent volatility spillovers. As shown in the figure, shocks propagating among individual currency volatilities account for 80\% of the total volatility. However, it does not necessarily reflect a strong impact of individual currencies' implied volatilities on one another. Instead, it may be an artifact of a significant exposure to a common shock. 

After removing correlated effects, volatility connectedness declines and exhibits countercyclical dynamics, with prominent spikes during periods of distress. For instance, we observe surges in connectedness during the Asian financial crisis in October 1997, the Russian financial crisis in 1998, the Global Financial Crisis in 2008-2009, the European sovereign debt crisis between 2010 and 2012, and the onset of the COVID-19 outbreak in February 2020. This evidence strongly supports potentially relevant information in volatility linkages, controlling for common variation. The novel contribution of this paper is to examine the asset pricing implications of these connections among currency volatilities.  

\subsection{Out-of-sample construction of volatility connectedness portfolios}

It is worth noting that the volatility connectedness illustrated in Figure \ref{fig:overall connectedness} is based on the full-sample estimation. Specifically, we divide the entire sample into subsamples with balanced panels of implied volatilities, estimate the TVP-VAR model for each subperiod, and combine the subsample estimates. Since we employ the Quasi-Bayesian Local-Likelihood approach to estimate time-varying parameters using only observations surrounding the period of interest, the risk of a look-ahead bias is minimal. Nevertheless, we completely refute this concern by obtaining out-of-sample estimates and using them to construct volatility-connectedness-sorted portfolios. 

Our empirical methodology requires balanced data. For this reason, the out-of-sample portfolio analysis is based on data from August 2004 to December 20021, a period during which all 20 currencies are available. We employ recursive estimation with a rolling window of $5 \times 26 = 130$ daily observations (half a year of trading days), using data from August 2004 to January 2005 for the initial estimation. We retain the last daily estimate from the TVP-VAR model within a rolling window; hence, no future information can enter the parameter estimates for the last observation. We repeat this procedure for each day in the out-of-sample period from February 2005 to December 2021. We use out-of-sample estimates from the TVP-VAR model to construct daily volatility connectedness measures. We then average them over the last two weeks of each month to obtain monthly time series and use these series to construct volatility-connectedness-sorted portfolios.

Each month, we sort currencies into quintiles using to- or from-directional volatility connectedness measures described in Section \ref{sec: connectedness measures}. The first quintile portfolio 
 $\mathcal{P}_1$
comprises 20\% of currencies with the highest values of a particular characteristic, whereas the fifth quintile portfolio 
$\mathcal{P}_5$ 
contains 20\% of currencies with the lowest values. Each 
$\mathcal{P}_i$ 
is an equally weighted portfolio of the corresponding currencies. We next form a long-short strategy that buys 
 $\mathcal{P}_5$
and sells
 $\mathcal{P}_1.$
We consider the portfolios for the to-directional and from-directional connectedness measures estimated from shocks of any persistence (denoted by
 $\mathcal{T}$
and
 $\mathcal{F}$)
as well as short-term and long-term persistence (labeled as
 $\mathcal{T}(d)$
and
 $\mathcal{F}(d)$
in which
$d \in \{S, L\}$).

\subsection{Alternative currency strategies}

We compare the performance of volatility-connectedness-sorted portfolios with that of alternative strategies. The first group comprises common currency factors in the existing literature. Following \cite{lustig2011common}, we consider the dollar factor (DOL), defined as the average excess return across all available currencies, and the carry trade strategy (CAR), defined as buying (selling) currencies with the highest (lowest) forward discounts. Next, we implement a currency value strategy (VAL) that buys (sells) undervalued currencies, defined as those with the lowest (highest) past nominal exchange-rate returns. We also implement short- and long-term momentum strategies (MOM (ST) and MOM (LT)) based on past returns over 1- and 12-month horizons. 

The second group considers volatility-related factors as well as those derived from currency options data. It is natural to conjecture that volatility spillovers are strongly correlated with currency volatilities. Hence, we consider a realized (implied) volatility strategy, labeled VOL (IVOL), that buys and sells currencies with the highest and lowest one-month realized (implied) volatility. Next, we define the one-month volatility risk premium (VRP) as the difference between one-month realized and implied volatility. The volatility risk premium strategy goes long (short) currencies with the highest (lowest) VRPs, which is equivalent to buying (selling) low-insurance-cost (high-insurance-cost) currencies \citep{della2016volatility}.\footnote{Two comments are noteworthy. First, \cite*{Li_Sarno_Zinna_2025} acknowledge in the Appendix that the original VRP strategy does not work anymore, and hence, they implement an opposite sorting scheme from \cite{della2016volatility}. We instead follow the original study's sorting. Second, we implement the one-month VRP strategy to ensure consistency with volatility-connectedness-sorted portfolios, even though \cite{della2016volatility} focus on the one-year VRP.} We define a skewness risk premium (SRP) of a currency as the difference between the positive and negative risk-neutral semi-variances. The SRP strategy then buys (sells) currencies with the highest (lowest) 
SRPs \citep{Li_Sarno_Zinna_2025}. 

Furthermore, we define risk reversal (RR) as the three-month currency option implied volatility difference between out-of-the-money (10 delta) call and out-of-the-money (10 delta) put options \citep{della2022exchange}. A currency with a high (low) RR can be considered low (high) risk, in that it is more likely to experience strong appreciation (depreciation). The RR strategy then buys (sells) currencies with 
the lowest (highest) RRs. Finally, we compute the ratio of net foreign assets (NFA) to gross domestic product (GDP) for a foreign country and construct an NFA strategy by buying (selling) debtor (creditor) currencies, i.e., countries with the lowest (highest) NFA-to-GDP ratios. To ensure consistency across portfolio sorts, we construct the aforementioned strategies using the sample of 20 currencies. Specifically, we construct DOL, CAR, MOM (ST), MOM (LT), VOL, IVOL, and VRP using own datasets, whereas we obtain VAL, SRP, RR, and NFA using the replication package of \cite{Li_Sarno_Zinna_2025}.\footnote{We are thankful to \cite{Li_Sarno_Zinna_2025} for sharing the currency characteristics and codes used in their study, which enables us to reproduce several factors for our sample.} 

We also consider a third group of factors related to global trade. The trade-related strategies may not appear to be natural counterparts for volatility-connectedness-sorted portfolios due to a disconnect between options and trade data (e.g., differences in data frequency, financial and macroeconomic characteristics, among others). Nevertheless, we ensure that the predictability arising from implied volatility spillovers is not attributable to countries' trade balances and flows. Recently, \cite{hou2024trade} propose a novel predictor of currency risk premia called a centrality based characteristic (CBC), based on the centrality of the trade imbalance network and the variance–covariance matrix of currency returns. Next, we consider several related strategies using the total trade network centrality (TTNC) \citep{richmond2019trade}, the trade imbalance (TImb) \citep{gabaix2015international}, global imbalances (GImb) \citep{corte2016currency}, and the variance–covariance weighted trade imbalance (V-weighted TImb). Finally, we consider an intermediary asset pricing factor from \cite{he2017intermediary} (HKM). We retrieve the data from the replication package and refer to their study for details on how to construct these factors.

\begin{table}[t!]
\centering
\caption{To-directional volatility connectedness portfolios.} 
\begin{minipage}{\textwidth} 
\footnotesize
This table presents descriptive statistics for quintile $(\mathcal{P}_i: i = 1,\dots,5)$ and long-short $(\mathcal{P}_{5-1})$ portfolios sorted by to-directional volatility connectedness measures based on volatility shocks that exclude contemporaneous correlations. Panels A, B, and C report the results for shocks of any, short-, and long-term persistence. $\mathcal{P}_1 (\mathcal{P}_5)$ comprises currencies with the highest (lowest) levels of volatility transmission. $\mathcal{P}_{5-1}$ buys $\mathcal{P}_5$ and sells $\mathcal{P}_1.$ Mean, standard deviation, and Sharpe ratio are annualized, but the \cite{newey1987simple} t-statistic of mean, skewness, kurtosis, and the first-order autocorrelation are based on monthly returns. We also report the annualized mean of the exchange rate ($\text{fx} = \Delta s^k $) and interest rate ($\text{ir} = i^k - i$) components of excess returns and the average to- and from-directional volatility connectedness measures of portfolios. The sample spans March 2005 to December 2021, which corresponds to the out-of-sample estimation period.
\end{minipage}
\vspace{\medskipamount}
\footnotesize

\begin{tabular}{llD{.}{.}{4.2}D{.}{.}{4.2}D{.}{.}{4.2}D{.}{.}{4.2}D{.}{.}{4.2}D{.}{.}{4.2}}
\toprule
&       & \multicolumn{1}{r}{$\mathcal{P}_1$} & \multicolumn{1}{r}{$\mathcal{P}_2$} & \multicolumn{1}{r}{$\mathcal{P}_3$} & \multicolumn{1}{r}{$\mathcal{P}_4$} & \multicolumn{1}{r}{$\mathcal{P}_5$} & \multicolumn{1}{r}{$\mathcal{P}_{5-1}$} \\
\midrule
\multicolumn{8}{l}{\textbf{Panel A: $\mathcal{T}$-sorted portfolios}} \\
\midrule
mean (\%) &       & -2.70 & 0.26  & -0.50 & -0.46 & 1.76  & 4.46 \\
t-stat &       & -1.16 & 0.13  & -0.27 & -0.21 & 0.81  & 2.74 \\
\midrule
\multicolumn{1}{r}{fx (\%)} &       & -4.59 & -1.05 & -1.73 & -1.92 & -0.80 & 3.79 \\
\multicolumn{1}{r}{ir (\%)} &       & 1.89  & 1.32  & 1.23  & 1.46  & 2.56  & 0.67 \\
\multicolumn{1}{r}{$\mathcal{T}$} &       & 0.34  & 0.22  & 0.18  & 0.14  & 0.10  & -0.24 \\
\multicolumn{1}{r}{$\mathcal{F}$} &       & 0.30  & 0.28  & 0.27  & 0.27  & 0.26  & -0.04 \\
\midrule
Sharpe &       & -0.26 & 0.03  & -0.05 & -0.05 & 0.20  & 0.61 \\
std (\%) &       & 10.44 & 8.92  & 9.47  & 9.42  & 8.66  & 7.27 \\
skew  &       & -0.85 & -0.46 & -0.33 & -0.45 & -0.38 & 0.41 \\
kurt  &       & 5.86  & 4.58  & 5.77  & 3.82  & 4.52  & 3.46 \\
ac1   &       & 0.02  & -0.03 & 0.03  & 0.00  & 0.10  & -0.05 \\
\midrule
\multicolumn{8}{l}{\textbf{Panel B: $\mathcal{T}(S)$-sorted portfolios}} \\
\midrule
mean (\%) &       & -2.46 & -0.22 & -0.34 & -0.49 & 1.88  & 4.34 \\
t-stat &       & -1.06 & -0.11 & -0.18 & -0.23 & 0.86  & 2.70 \\
\midrule
\multicolumn{1}{r}{fx (\%)} &       & -4.34 & -1.53 & -1.58 & -1.97 & -0.68 & 3.67 \\
\multicolumn{1}{r}{ir (\%)} &       & 1.88  & 1.30  & 1.24  & 1.48  & 2.56  & 0.68 \\
\multicolumn{1}{r}{$\mathcal{T}(S)$} &       & 0.27  & 0.18  & 0.14  & 0.12  & 0.08  & -0.19 \\
\multicolumn{1}{r}{$\mathcal{F}(S)$} &       & 0.24  & 0.22  & 0.22  & 0.22  & 0.22  & -0.02 \\
\midrule
Sharpe &       & -0.24 & -0.02 & -0.04 & -0.05 & 0.22  & 0.60 \\
std (\%) &       & 10.30 & 9.05  & 9.56  & 9.43  & 8.58  & 7.30 \\
skew  &       & -0.81 & -0.48 & -0.59 & -0.45 & -0.38 & 0.41 \\
kurt  &       & 5.85  & 4.55  & 5.50  & 3.87  & 4.55  & 3.43 \\
ac1   &       & 0.02  & -0.06 & 0.05  & 0.01  & 0.09  & -0.04 \\
\midrule
\multicolumn{8}{l}{\textbf{Panel C: $\mathcal{T}(L)$-sorted portfolios}} \\
\midrule
mean (\%) &       & -2.30 & -0.60 & 0.37  & -0.61 & 1.52  & 3.82 \\
t-stat &       & -1.01 & -0.29 & 0.20  & -0.28 & 0.71  & 2.33 \\
\midrule
\multicolumn{1}{r}{fx (\%)} &       & -4.22 & -1.91 & -0.87 & -2.09 & -1.02 & 3.20 \\
\multicolumn{1}{r}{ir (\%)} &       & 1.92  & 1.30  & 1.23  & 1.47  & 2.54  & 0.62 \\
\multicolumn{1}{r}{$\mathcal{T}(L)$} &       & 0.04  & 0.02  & 0.02  & 0.02  & 0.01  & -0.03 \\
\multicolumn{1}{r}{$\mathcal{F}(L)$} &       & 0.04  & 0.04  & 0.04  & 0.04  & 0.03  & -0.01 \\
\midrule
Sharpe &       & -0.23 & -0.07 & 0.04  & -0.07 & 0.18  & 0.53 \\
std (\%) &       & 10.15 & 9.19  & 9.47  & 9.39  & 8.64  & 7.15 \\
skew  &       & -0.85 & -0.46 & -0.30 & -0.49 & -0.34 & 0.47 \\
kurt  &       & 6.16  & 4.11  & 5.57  & 3.90  & 4.55  & 3.57 \\
ac1   &       & 0.02  & -0.02 & 0.01  & 0.00  & 0.10  & -0.04 \\
\bottomrule
\end{tabular}%
\label{tab: to-directional volatility connectedness portfolios}%
\end{table}%

\subsection{To-directional and from-directional volatility connectedness portfolios}
\label{sec: to- and from-directional connectedness}

Panel A of Table \ref{tab: to-directional volatility connectedness portfolios} reports summary statistics for the quintile
 $(\mathcal{P}_i: i = 1,\dots,5)$
and long-short $(\mathcal{P}_{5-1})$ portfolios sorted by to-directional volatility connectedness based on volatility shocks that exclude common correlations. The cross-sectional currency premia decrease with the amount of transmitted volatility. The average annual return difference between the weakest and strongest transmitters is 4.46\%, which is statistically different from zero at the 1\% confidence level. Notably, both long and short lags contribute to this significant return spread. Furthermore, the ``fx (\%)'' and ``ir (\%)'' rows indicate that the performance is primarily driven by predicting the spot exchange rates. For instance, the spread in the exchange rate component of the 
 $\mathcal{T}$
portfolio is 
 $3.79\%$ 
per annum, whereas the interest rate differential is  only
 $0.67\%.$ 
This observation indicates that our strategy likely differs from the carry trade, and hence,  the to-directional volatility connectedness provides incremental information to innovations in the foreign exchange volatility. The long-short portfolio's excess returns also exhibit the lowest volatility relative to those of the quintile portfolios. Thus, its annualized Sharpe ratio of 0.61 arises from a high average return and a moderate variability. The positive skewness indicates that the strategy does not suffer big losses, unlike many common currency factors.  

Panels B and C of Table \ref{tab: to-directional volatility connectedness portfolios} report the portfolio sorts for horizon-specific volatility connectedness. The key findings remain qualitatively unchanged for transitory and persistent volatility shocks. Quantitatively, the predictability arising from short-term to-directional volatility connectedness is slightly stronger, though the effects remain economically and statistically significant in both cases. For instance, the average returns and Sharpe ratios of the spread portfolio formed on short-term (long-term) volatility connections are 4.34\% (3.82\%) and 0.60 (0.53) per annum. This near-flat term structure echoes the finding of \cite{della2020cross}, demonstrating that the volatility carry premium does not vary with horizon. It also complements the literature on the price of variance risk in equity markets \citep*{dew2017price,johnson2017risk}. 

We now turn to portfolios sorted by from-directional volatility connectedness. To save space, we report the results in Appendix \ref{section: additional exercises appendix} and discuss the key findings here. Table \ref{tab: from-directional volatility connectedness portfolios} shows that the cross-sectional currency premia increase with the amount of received volatility, whereas the opposite pattern is observed when sorting on transmitted volatility. The ``fx (\%)'' row of Table \ref{tab: from-directional volatility connectedness portfolios} demonstrate that the exchange rate component tends to decline with the from-directional volatility connectedness, similarly to the results reported in Table \ref{tab: to-directional volatility connectedness portfolios}. However, the ``ir (\%)'' row of Table \ref{tab: from-directional volatility connectedness portfolios} shows that the interest rate differential strongly increases as we move from the weakest to the strongest recipients of volatility shocks from others. This strong carry trade component is the reason for the upward trend. Nevertheless, the average return spread between extreme quintile portfolios (only 2.02\% per annum) is economically small and statistically insignificant. 

These findings demonstrate that directional volatility connectedness is strongly associated with spot rate changes: currencies that receive (transmit) more volatility tend to depreciate more strongly. However, only to-directional volatility connectedness contains economically and statistically meaningful predictive power for cross-sectional currency risk premia. Specifically, stronger volatility transmitters earn significantly lower average excess returns, primarily driven by exchange rate predictability. In contrast, stronger volatility recipients tend to have higher interest rate differentials, reversing the pattern in currency premia and offsetting predictability in exchange rates. Thus, the empirical results highlight differences in the effects of volatility connectedness: the to-directional transmission likely provides a novel source of currency predictability, whereas the from-directional transmission may pertain to the carry trade and, hence, global volatility. We test these hypotheses in the next sections.

\subsection{Benchmark strategies and diversification gains}
\label{section: benchmark strategies and diversification gains}

We study the relationship between volatility-connectedness-sorted portfolios and existing currency factors. Table \ref{tab:benchmarks summary statistics} reports summary statistics of alternative benchmarks. The three trade-related strategies -- CBC, GImb, and HKM -- yield the highest Sharpe ratios of slightly below 0.40 per annum. The annualized average returns are modest across all factors, with the highest values slightly above 2\%, which is comparable to (significantly below) the average returns of a from-directional (to-directional) volatility connectedness strategy. Interestingly, only the trade imbalance network factor generates an economically and statistically significant average return with a high Sharpe ratio. Many well-known strategies --- MOM, VRP, and TImb --- have been unprofitable in the recent period. Consistent with prior literature, most factors exhibit negative skewness and excess kurtosis, indicating crash risk. 

\begin{table}[t!]
\centering
\caption{Alternative currency factors.}
\begin{minipage}{\textwidth} 
This table presents descriptive statistics for common currency factors. The top panel includes the dollar (DOL), carry trade (CAR), value (VAL), short- and long-term momentum (MOM (ST) and MOM (LT)). The middle panel reports realized and implied volatility (VOL and IVOL), volatility and skewness risk premia (VRP and SRP), risk-reversal (RR), and net-foreign asset (NFA) strategies. The bottom panel demonstrates trade imbalance network (CBC), total trade network centrality (TTNC), trade imbalance (TImb), global imbalance (GImb), and variance–covariance weighted trade imbalance (V-weighted TImb), as well as the intermediary capital risk (HKM) factors. Mean, standard deviation, and Sharpe ratio are annualized, while the \cite{newey1987simple} t-statistic of mean, skewness, kurtosis, and the first-order autocorrelation are based on monthly returns. The sample spans March 2005 to December 2021, which corresponds to the out-of-sample estimation period. 
\end{minipage}
\vspace{\medskipamount}
\footnotesize

\begin{tabular}{llD{.}{.}{4.2}D{.}{.}{4.2}D{.}{.}{4.2}D{.}{.}{4.2}D{.}{.}{4.2}D{.}{.}{4.2}D{.}{.}{4.2}}
\toprule
\multicolumn{1}{r}{ } &       & \multicolumn{1}{r}{mean (\%)} & \multicolumn{1}{r}{t-stat} & \multicolumn{1}{r}{Sharpe} & \multicolumn{1}{r}{std (\%)} & \multicolumn{1}{r}{skew} & \multicolumn{1}{r}{kurt} & \multicolumn{1}{r}{ac1} \\
\midrule
DOL   &       & -0.33 & -0.17 & -0.04 & 8.49  & -0.54 & 4.65  & 0.03 \\
CAR   &       & 1.72  & 0.68  & 0.16  & 11.08 & -0.90  & 5.15  & 0.09 \\
VAL   &       & 1.80   & 0.91  & 0.22  & 8.21  & -0.72 & 4.88  & 0.02 \\
MOM (ST) &       & -0.58 & -0.35 & -0.07 & 7.73  & 0.82  & 7.24  & -0.04 \\
MOM (LT) &       & -1.27 & -0.69 & -0.15 & 8.70   & -0.19 & 3.66  & -0.09 \\
\midrule
VOL   &       & 2.16  & 1.21  & 0.24  & 8.83  & -0.94 & 6.22  & -0.01 \\
IVOL  &       & 1.08  & 0.59  & 0.11  & 9.40  & -0.89 & 5.60  & -0.01 \\
VRP   &       & -0.29 & -0.11 & -0.03 & 8.66  & -0.13 & 7.25  & 0.10 \\
SRP   &       & 1.02  & 0.51  & 0.10  & 9.79  & -0.87 & 6.12  & 0.13 \\
RR    &       & 2.33  & 1.19  & 0.24  & 9.78  & -0.56 & 4.82  & 0.02 \\
NFA   &       & 1.25  & 0.76  & 0.16  & 7.68  & -0.97 & 6.34  & 0.04 \\
\midrule
CBC   &       & 2.24  & 1.98  & 0.38  & 5.95  & -0.87 & 5.76  & -0.05 \\
TTNC  &       & 0.46  & 0.41  & 0.10  & 4.65  & -0.63 & 5.14  & 0.12 \\
TImg  &       & -0.14 & -0.12 & -0.03 & 4.79  & -0.43 & 4.54  & 0.02 \\
GImb  &       & 1.77  & 1.77  & 0.39  & 4.59  & 0.74  & 6.57  & 0.02 \\
V-weighted TImb &       & 1.50   & 0.76  & 0.16  & 9.53  & 0.86  & 6.30  & -0.03 \\
HKM   &       & 1.48  & 1.43  & 0.36  & 4.13  & 0.14  & 3.50  & 0.13 \\
\bottomrule
\end{tabular}%
\label{tab:benchmarks summary statistics}%
\end{table}%

Next, we examine how well standard factors explain the returns associated with directional volatility connectedness. Table \ref{tab: to-directional network portfolios and benchmark strategies} performs contemporaneous regressions of the excess returns of the to-directional volatility connectedness long-short portfolio on alternative benchmarks, estimated one factor at a time. Across all specifications, the strategy delivers economically and statistically significant abnormal returns, with estimated alphas ranging from 3.99\% (t-stat = 2.49) to 4.99\% (t-stat = 3.07) per annum. Thus, none of the common currency risk premia subsumes the profitability of the long-short portfolio formed on to-directional volatility connectedness.

The strategy exhibits negative and statistically significant exposure to realized and implied volatility, risk reversal, and net foreign asset factors. These negative betas show that the strategy attenuates currency volatility and credit risks rather than amplifying them. The long-short portfolio also demonstrates a weaker negative association with carry, variance and skewness risk premia, trade imbalance network, and total trade network centrality factors, indicating diversification benefits with a wider set of currency factors. The strategy is also positively related to the short-term momentum, global imbalance, and variance–covariance weighted trade imbalance factors. This weak association with benchmark factors results in low 
$R^2$
statistics from 0.05\% to 9.06\%. 

\begin{table}[t!]
\centering
\caption{To-directional volatility connectedness and alternative currency strategies.}
\begin{minipage}{\textwidth} 
This table presents the results of time-series regressions of the excess returns of the $\mathcal{T}$ strategy on alternative benchmarks, estimated one factor at a time. The values reported in the ``$\alpha$ (\%)'' column are expressed in percentage per annum. The t-statistics are based on \cite{newey1987simple} standard errors. The sample spans March 2005 to December 2021, which corresponds to the out-of-sample estimation period.
\end{minipage}
\vspace{\medskipamount}
\footnotesize

\begin{tabular}{llD{.}{.}{4.2}D{.}{.}{4.2}D{.}{.}{4.2}D{.}{.}{4.2}D{.}{.}{4.2}}
\toprule
\multicolumn{1}{r}{ } &       & \multicolumn{1}{r}{$\alpha$ (\%)} & \multicolumn{1}{r}{t-stat} & \multicolumn{1}{r}{$\beta$} & \multicolumn{1}{r}{t-stat} & \multicolumn{1}{r}{$R^2$ (\%)} \\
\midrule
DOL   &       & 4.39  & 2.69  & -0.21 & -2.76 & 5.94 \\
CAR   &       & 4.65  & 2.82  & -0.11 & -1.91 & 2.94 \\
VAL   &       & 4.53  & 2.76  & -0.04 & -0.37 & 0.18 \\
MOM (ST) &       & 4.51  & 2.77  & 0.09  & 1.28  & 1.00 \\
MOM (LT) &       & 4.43  & 2.69  & -0.02 & -0.22 & 0.05 \\
\midrule
VOL   &       & 4.99  & 3.07  & -0.25 & -3.61 & 9.06 \\
IVOL  &       & 4.69  & 2.83  & -0.21 & -3.45 & 7.60 \\
VRP   &       & 4.41  & 2.96  & -0.15 & -1.66 & 3.32 \\
SRP   &       & 4.59  & 2.81  & -0.13 & -1.93 & 3.23 \\
RR    &       & 4.87  & 3.09  & -0.18 & -2.50 & 5.69 \\
NFA   &       & 4.77  & 2.87  & -0.25 & -3.06 & 7.20 \\
\midrule
CBC   &       & 4.80  & 2.89  & -0.15 & -1.44 & 1.53 \\
TTNC  &       & 4.54  & 2.81  & -0.18 & -1.56 & 1.28 \\
TImg  &       & 4.44  & 2.78  & -0.14 & -0.93 & 0.84 \\
GImb  &       & 3.99  & 2.49  & 0.26  & 1.91  & 2.80 \\
V-weighted TImb &       & 4.20  & 2.72  & 0.17  & 2.51  & 4.92 \\
HKM   &       & 4.23  & 2.63  & 0.15  & 0.99  & 0.77 \\
\bottomrule
\end{tabular}%
\label{tab: to-directional network portfolios and benchmark strategies}%
\end{table}%

\begin{figure}[t!]
\begin{center}
\includegraphics[width=0.65\textwidth]{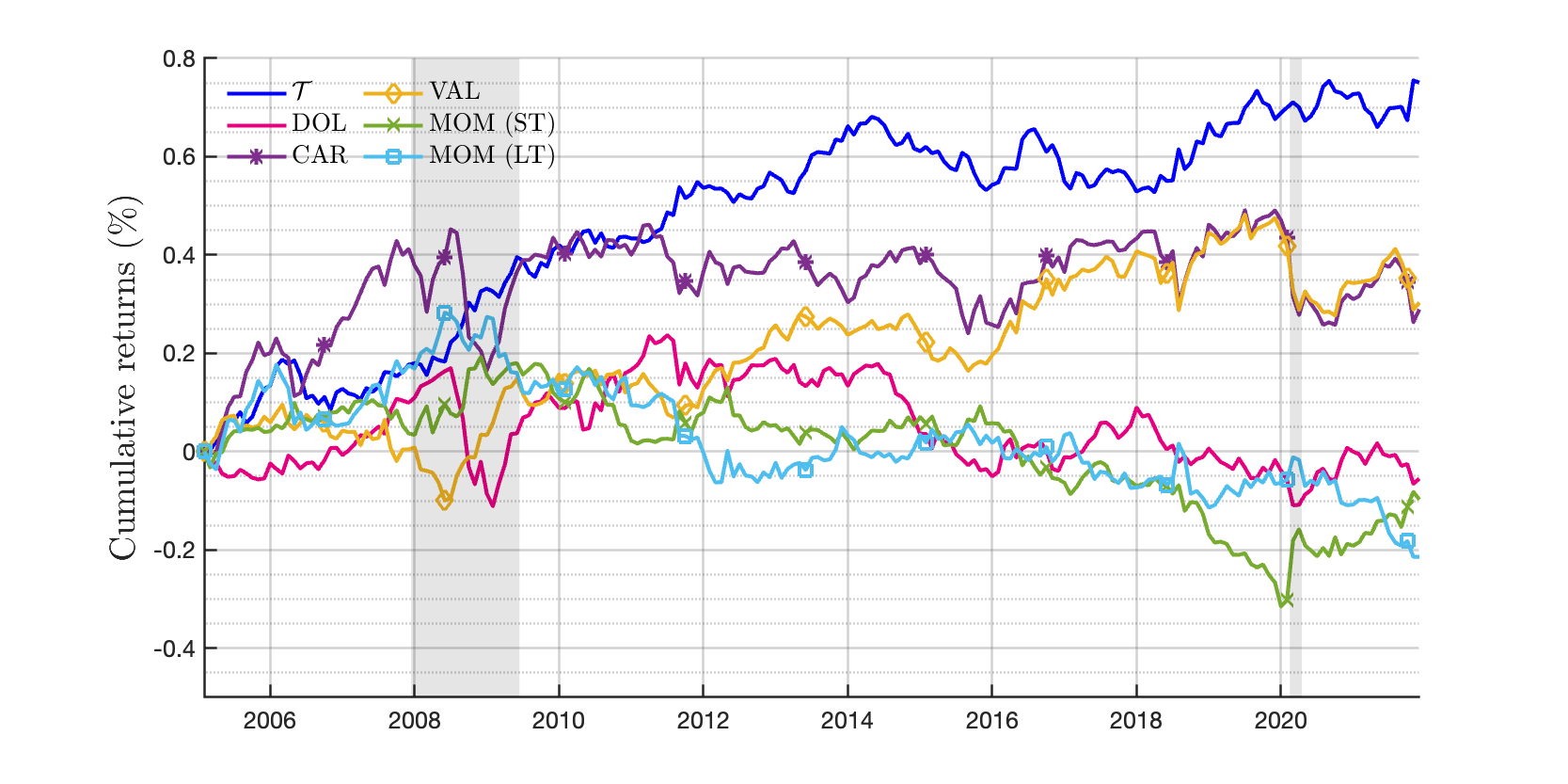}
\hspace{-20pt}
\includegraphics[width=0.65\textwidth]{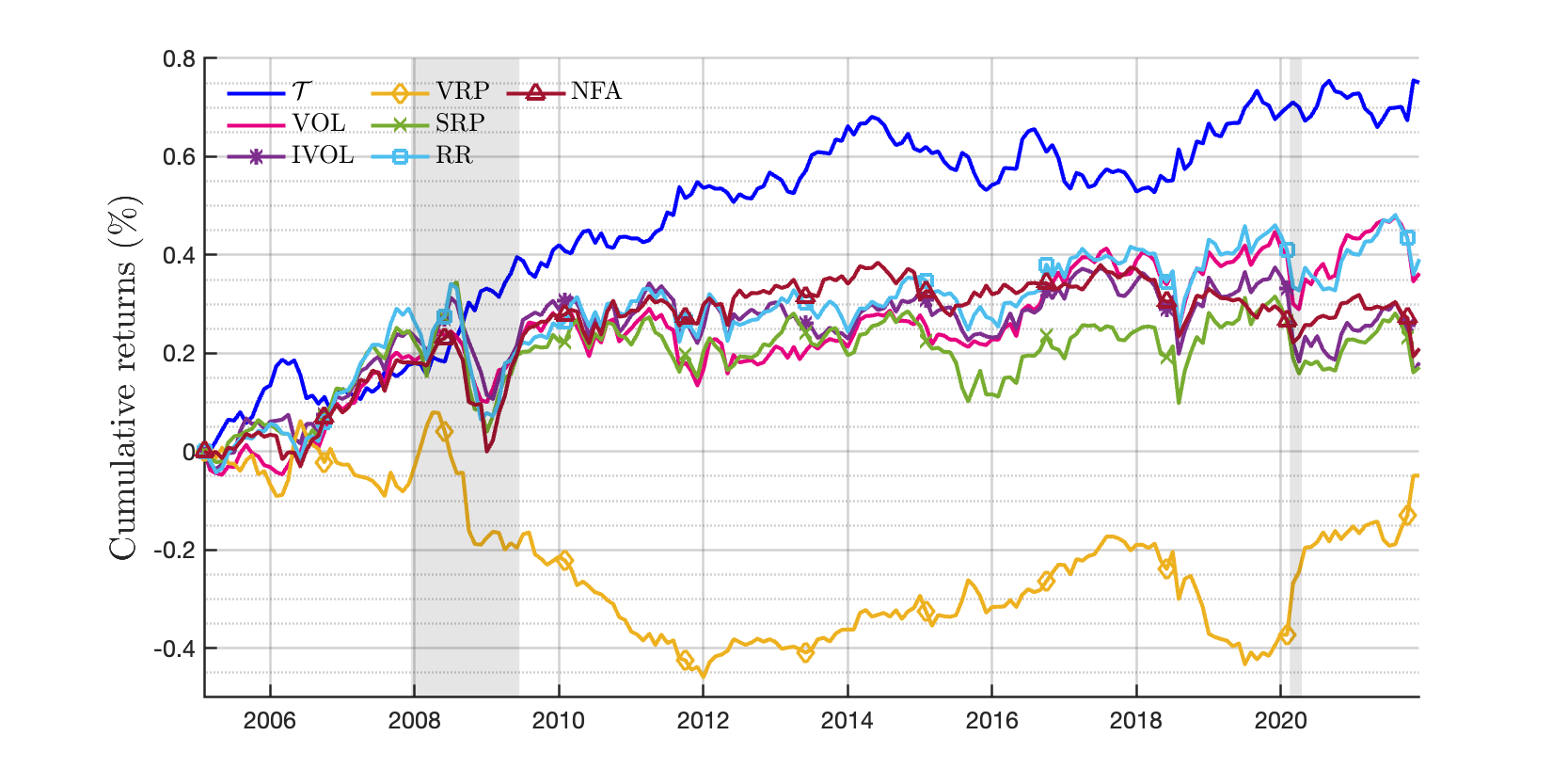}
\hspace{-20pt}
\includegraphics[width=0.65\textwidth]{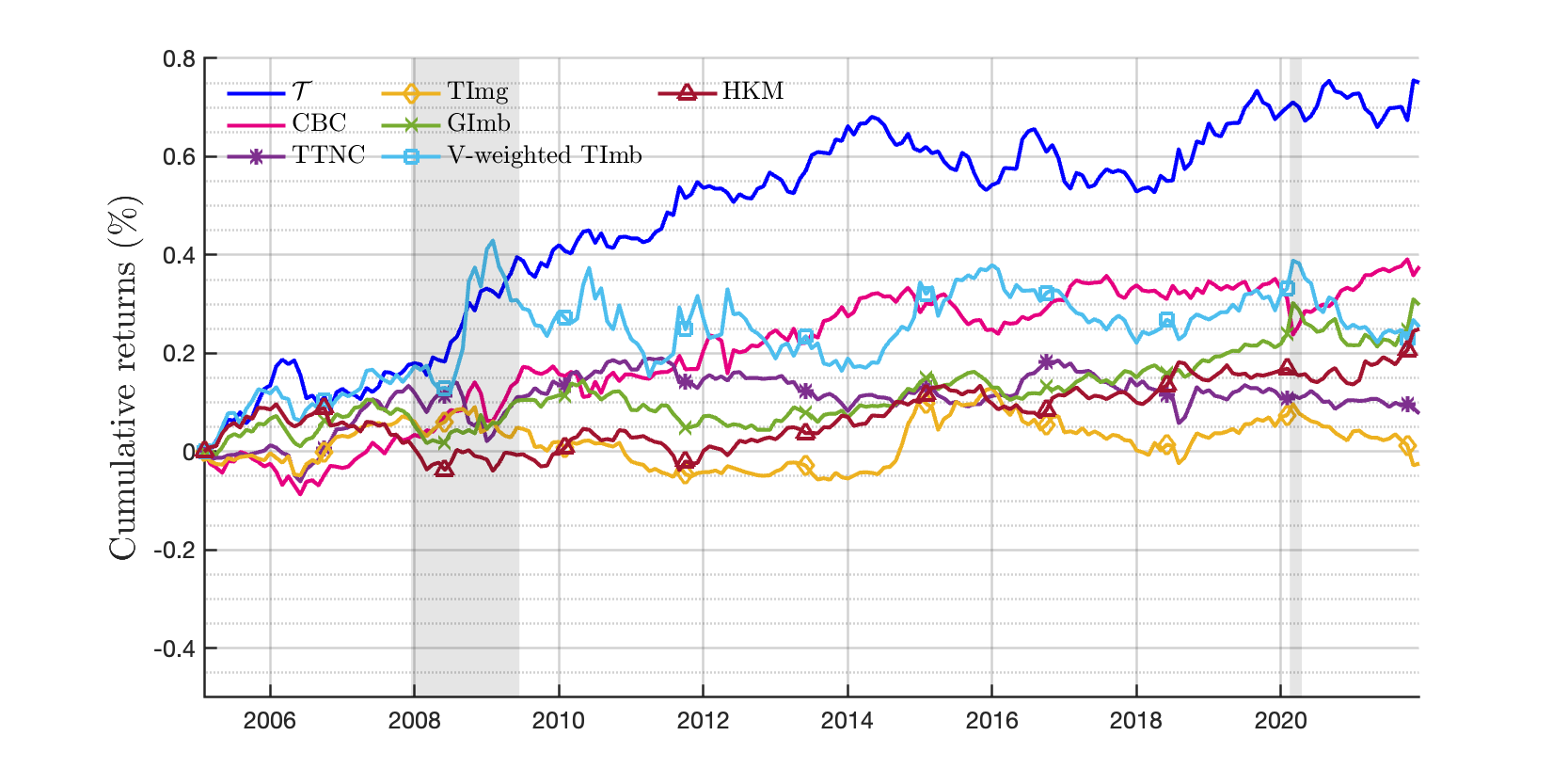}
\caption{\textbf{Cumulative returns of currency factors.}} 
\begin{minipage}{\textwidth}
The figure depicts the cumulative returns of currency factors. The top panel compares the performance of the $\mathcal{T}$ strategy with the dollar (DOL), carry trade (CAR), value (VAL), short- and long-term momentum (MOM (ST) and MOM (LT)). The middle panel displays realized and implied volatility (VOL and IVOL), volatility and skewness risk premia (VRP and SRP), risk-reversal (RR), and net foreign asset (NFA) strategies. The bottom panel illustrates trade imbalance network (CBC), total trade network centrality (TTNC), trade imbalance (TImb), global imbalance (GImb), and variance–covariance weighted trade imbalance (V-weighted TImb), as well as the intermediary capital risk (HKM) factors. The shaded areas denote the NBER recessions. The sample spans March 2005 to December 2021, which corresponds to the out-of-sample estimation period.
\end{minipage}
\label{fig: cumulative returns}
\vspace{-20pt}
\end{center}
\end{figure}

Overall, the results demonstrate that to-directional volatility connectedness contains novel cross-sectional information for currency risk premia. The strategy exploiting this information generates highly significant performance, both economically and statistically, that cannot be understood through the lens of the benchmarks.\footnote{We also replace currency benchmarks with the five equity factors of \cite{fama1993common} and momentum (constructed for the U.S., developed or emerging economies) or the hedge fund factors of \cite{fung2004hedge}. As shown in Appendix \ref{section: additional exercises appendix}, these alternative factors cannot explain the to-directional volatility connectedness strategy. Furthermore, the strategy is meaningfully (negatively) related only to stock market returns in the U.S., developed, and emerging market specifications.} We augment this evidence by displaying the cumulative returns of various currency portfolios in Figure \ref{fig: cumulative returns}. Consistent with previous findings, the to-directional volatility connectedness strategy outperforms others across the entire sample.

\begin{table}[t!]
\centering
\caption{Diversification gains.}
\begin{minipage}{\textwidth} 
This table illustrates the impact of adding the $\mathcal{T}$ strategy to common currency factors. Specifically, we construct a naive 50\%-50\% portfolio of the $\mathcal{T}$ strategy and another factor. Mean, standard deviation, and Sharpe ratio are annualized, but the \cite{newey1987simple} t-statistic of mean, skewness, kurtosis, and the first-order autocorrelation are based on monthly returns. The last column shows the percentage increase in the Sharpe ratio of a naive portfolio relative to the original currency strategy. The sample spans March 2005 to December 2021, which corresponds to the out-of-sample estimation period.
\end{minipage}
\vspace{\medskipamount}
\footnotesize

\begin{tabular}{llD{.}{.}{4.2}D{.}{.}{4.2}D{.}{.}{4.2}D{.}{.}{4.2}D{.}{.}{4.2}D{.}{.}{4.2}D{.}{.}{4.2}D{.}{.}{4.2}}\toprule
\multicolumn{1}{r}{ } &       & \multicolumn{1}{r}{mean (\%)} & \multicolumn{1}{r}{t-stat} & \multicolumn{1}{r}{Sharpe} & \multicolumn{1}{r}{std (\%)} & \multicolumn{1}{r}{skew} & \multicolumn{1}{r}{kurt} & \multicolumn{1}{r}{ac(1)} & \multicolumn{1}{r}{$\%\Delta$ Sharpe} \\
\midrule
DOL   &       & 2.07  & 1.71  & 0.42  & 4.87  & 0.09  & 3.35  & 0.08  & \multicolumn{1}{c}{\quad\quad n.a.} \\
CAR   &       & 3.09  & 2.04  & 0.51  & 6.09  & 0.25  & 4.28  & 0.06  & 227.18 \\
VAL   &       & 3.13  & 2.20  & 0.58  & 5.37  & 0.30  & 3.65  & 0.06  & 165.70 \\
MOM (ST) &       & 1.94  & 1.60  & 0.35  & 5.56  & 0.70  & 5.43  & -0.03 & \multicolumn{1}{c}{\quad\quad n.a.} \\
MOM (LT) &       & 1.59  & 1.21  & 0.28  & 5.61  & 0.13  & 4.20  & -0.02 & \multicolumn{1}{c}{\quad\quad n.a.} \\
\midrule
VOL   &       & 3.31  & 3.02  & 0.69  & 4.80  & 0.20  & 4.44  & 0.02  & 181.94 \\
IVOL  &       & 2.77  & 2.31  & 0.54  & 5.09  & 0.14  & 4.02  & 0.04  & 375.10 \\
VRP   &       & 2.08  & 1.79  & 0.41  & 5.12  & 1.09  & 7.47  & 0.05  & \multicolumn{1}{c}{\quad\quad n.a.} \\
SRP   &       & 2.74  & 2.16  & 0.49  & 5.55  & 0.43  & 4.38  & 0.06  & 372.87 \\
RR    &       & 3.39  & 2.94  & 0.63  & 5.36  & 0.44  & 4.47  & 0.02  & 166.32 \\
NFA   &       & 2.85  & 2.53  & 0.63  & 4.53  & -0.03 & 3.78  & 0.07  & 288.42 \\
\midrule
CBC   &       & 3.35  & 3.17  & 0.76  & 4.40  & 0.11  & 3.61  & -0.01 & 102.07 \\
TTNC  &       & 2.46  & 2.63  & 0.60  & 4.09  & 0.49  & 3.92  & -0.01 & 509.43 \\
TImg  &       & 2.16  & 2.42  & 0.52  & 4.17  & 0.01  & 3.34  & -0.02 & \multicolumn{1}{c}{\quad\quad n.a.} \\
GImb  &       & 3.11  & 3.15  & 0.68  & 4.61  & 0.84  & 5.97  & -0.03 & 74.97 \\
V-weighted TImb &       & 2.98  & 2.06  & 0.45  & 6.60  & 0.75  & 5.39  & -0.05 & 185.96 \\
HKM   &       & 2.97  & 3.11  & 0.69  & 4.34  & 0.65  & 4.76  & -0.04 & 90.88 \\
\bottomrule
\end{tabular}%
\label{tab:benchmark strategies diversification gains}%
\end{table}%

We now examine the diversification benefits of the strategy by combining it with traditional currency factors, one at a time, in equal weights. Table \ref{tab:benchmark strategies diversification gains} shows that the to-directional volatility connectedness strategy substantially improves the risk-adjusted performance of alternative benchmarks. The annualized Sharpe ratios of these naive portfolios range from 0.28 for short-term momentum to 0.76 for the trade imbalance network strategy, exceeding 0.60 in several other cases (VOL, RR, NFA, CBC, GImb, and HKM). In relative terms, these values are in stark contrast to the negative Sharpe ratios for some factors. In other instances, we observe particularly pronounced improvements in the Sharpe ratios of IVOL (375.10\%), SRP (372.87\%), and TTNC (509.43\%). Additionally, including the to-directional volatility connectedness strategy leads to lower volatility and positive skewness of the combined returns relative to standalone factors. In general, the return predictability generated by directional volatility transmission yields substantial portfolio gains and improves the efficiency of common currency factors.

We next examine the relationship between the from-directional volatility connectedness and alternative currency factors. Table \ref{tab: from-directional portfolios and benchmark strategies} in the Appendix \ref{section: additional exercises appendix} demonstrates that the strategy exhibits a strong exposure to several benchmarks. Most prominently, the loadings are close to unity, and the $R^2$ values are 78.44\% and 91.40\% for VOL and IVOL, respectively. Thus, the return predictability captured by directional volatility reception is completely subsumed by currency volatility risk. Previous literature has shown that carry trade returns are strongly associated with innovations in foreign exchange volatility. Consequently, we observe an economically and statistically significant loading of 0.65 (t-stat = 8.32) and a high $R^2$ of 60.60\% for CAR. This is consistent with a strong interest rate component in the excess returns based on from-directional volatility connectedness shown in Table \ref{tab: from-directional volatility connectedness portfolios}.

\subsection{Additional exercises}
\label{section: additional exercises}

We discuss additional exercises in this section and delegate the detailed results to the Appendix \ref{section: additional exercises appendix}. We apply the TVP-VAR model to two-month implied volatilities. Tables \ref{tab: summary stats and alphas for two-month implied volatilities} and \ref{tab: diversification benefits for two-month implied volatilities} demonstrate that the key findings remain qualitatively and quantitatively unchanged. Next, we replicate our analysis using realized volatilities based on daily spot rates in the previous month. This exercise evaluates the incremental contribution of using ex-ante expectations of currency volatility synthesized from currency options. Volatility connectedness estimated from realized volatilities contains some predictive power for currency returns. However, this predictability is completely subsumed by the carry trade or volatility factors.  Thus, the forward-looking information of options data becomes instrumental for our results. Making effective use of information from currency options markets helps us discover novel predictive information in investor expectations about future currency fluctuations. We hope that the findings of this paper will advance the usage of options data in the foreign exchange literature.

\subsection{Asset Pricing}
This section presents the time-series and cross-sectional asset-pricing tests conducted on the excess returns of to-directional volatility connectedness portfolios. We begin by applying principal component analysis to understand the factor structure of the novel cross-section. We then test the ability of linear factor models, using existing benchmarks, to explain the cross-sectional variation in quintile portfolio returns. Finally, we report the results of the time-series regressions.  

\subsubsection{Principal component analysis}
\label{section: pca}

We investigate whether average excess returns of to-directional volatility connectedness quintiles are associated with a small set of statistical factors. Following \cite{lustig2011common}, we conduct a principal component (PC) decomposition of test portfolios. Furthermore, we examine the correlations between principal components and various currency factors.

\begin{table}[t!]
\centering
\caption{Principal components}
\begin{minipage}{\textwidth}
This table presents the loadings of the principal components
$(\text{PCi} : \text{i} = 1, \dots, 5)$
for quintile portfolios 
$(\mathcal{P}_i: i = 1,\dots,5)$ 
sorted by the to-directional volatility connectedness measure based on volatility shocks that exclude contemporaneous correlations. $\mathcal{P}_1 (\mathcal{P}_5)$ comprises currencies with the highest (lowest) levels of volatility transmission. The table also reports correlations of principal components with the to-directional volatility connectedness strategy and alternative currency factors. The sample spans March 2005 to December 2021, which corresponds to the out-of-sample estimation period.
\end{minipage}
\vspace{\medskipamount}
\footnotesize

\begin{tabular}{lD{.}{.}{2.2}D{.}{.}{2.2}D{.}{.}{2.2}D{.}{.}{2.2}D{.}{.}{2.2}llD{.}{.}{2.2}D{.}{.}{2.2}D{.}{.}{2.2}D{.}{.}{2.2}D{.}{.}{2.2}}
\toprule
& \multicolumn{1}{r}{PC1}   & \multicolumn{1}{r}{PC2}   & \multicolumn{1}{r}{PC3}   & \multicolumn{1}{r}{PC4}   & \multicolumn{1}{r}{PC5}   &       &       & \multicolumn{1}{r}{PC1}   & \multicolumn{1}{r}{PC2}   & \multicolumn{1}{r}{PC3}   & \multicolumn{1}{r}{PC4}   & \multicolumn{1}{r}{PC5} \\
\cmidrule{2-6}\cmidrule{9-13}          & \multicolumn{5}{c}{PC loadings}       &       &       & \multicolumn{5}{c}{Correlations} \\
\cmidrule{2-6}\cmidrule{9-13}    
$\mathcal{P}_1$    & 0.50  & 0.70  & -0.47 & 0.00  & -0.20 &       & $\mathcal{T}$     & -0.26 & -0.85 & 0.18  & 0.43  & 0.00 \\
$\mathcal{P}_2$    & 0.42  & 0.30  & 0.75  & 0.23  & 0.34  &       & DOL   & 1.00  & -0.01 & 0.01  & 0.01  & 0.00 \\
$\mathcal{P}_3$    & 0.46  & -0.29 & 0.30  & -0.53 & -0.58 &       & CAR   & 0.50  & 0.06  & -0.23 & 0.12  & 0.00 \\
$\mathcal{P}_4$    & 0.46  & -0.34 & -0.32 & -0.34 & 0.68  &       & VAL   & 0.17  & 0.05  & -0.03 & 0.11  & -0.02 \\
$\mathcal{P}_5$    & 0.40  & -0.47 & -0.17 & 0.74  & -0.21 &       & MOM (ST) & -0.22 & -0.05 & 0.03  & 0.00  & 0.07 \\
CV    & 82.10 & 88.44 & 92.86 & 96.84 & 100.00 &       & MOM (LT) & -0.14 & -0.03 & -0.04 & -0.18 & 0.11 \\
\cmidrule{2-6}\cmidrule{9-13}          & \multicolumn{5}{c}{Correlations}       &       &       & \multicolumn{5}{c}{Correlations} \\
\cmidrule{2-6}\cmidrule{9-13}    
\cmidrule{2-6}    
VOL   & 0.62  & 0.19  & -0.07 & 0.09  & 0.03  &       & CBC   & 0.48  & 0.05  & 0.02  & 0.09  & 0.06 \\
IVOL  & 0.64  & 0.15  & -0.14 & 0.09  & 0.04  &       & TTNC  & 0.46  & -0.01 & -0.19 & 0.08  & 0.05 \\
VRP   & 0.11  & 0.15  & 0.01  & -0.06 & 0.02  &       & TImg  & -0.33 & 0.14  & -0.26 & -0.02 & -0.02 \\
SRP   & 0.56  & 0.04  & -0.23 & 0.10  & -0.02 &       & GImb  & -0.14 & -0.16 & -0.06 & 0.00  & 0.08 \\
RR    & 0.53  & 0.15  & -0.13 & 0.11  & 0.05  &       & V-weighted TImb & -0.83 & 0.00  & -0.10 & 0.05  & 0.04 \\
NFA   & 0.73  & 0.10  & -0.18 & 0.10  & -0.01 &       & HKM   & -0.18 & -0.07 & -0.05 & -0.02 & 0.03 \\
\bottomrule
\end{tabular}%
\label{tab:pca}%
\end{table}%

Table \ref{tab:pca} reports the results. The PC loadings indicate a strong factor structure. The first principal component (PC1) accounts for most of the time variation in quintile returns and has similar loadings across the five portfolios. The second principal component (PC2) displays a pronounced monotonic pattern in loadings as we move from 
$\mathcal{P}_1$
to
$\mathcal{P}_5.$ 
The first two principal components explain approximately 82\% (88\%) of the common variation in portfolios. PC1 is perfectly correlated with the dollar factor. PC2 exhibits the strongest correlation with the to-directional volatility connectedness strategy. The results suggest that the returns formed on directional volatility transmission can be summarized by a small number of principal components. We can approximate the first using the average returns across currency portfolios and interpret it as a ``level'' factor and the second using the spread between 
$\mathcal{P}_5$
and
$\mathcal{P}_1$
portfolios and interpret it as a ``slope'' factor.

\subsubsection{Time-series and cross-sectional regressions}

We now implement a two-step Fama–MacBeth (FMB) estimation procedure \citep{fama1973risk} using the novel cross-section of quintile portfolios formed on directional volatility transmission.\footnote{For robustness checks, we also perform cross-sectional tests based on the generalized method of moments estimation. These results are similar to the cross-sectional regressions.} Motivated by principal component analysis, we consider a two-factor specification with DOL as the first factor and the $\mathcal{T}$ portfolio as a second factor.

\begin{table}[t!]
\centering
\caption{Pricing to-directional volatility connectedness portfolios}
\begin{minipage}{\textwidth} 
This table presents the results of the two-stage procedure of \cite{fama1973risk}. The test portfolios are quintile portfolios
$(\mathcal{P}_i: i = 1,\dots,5)$ 
sorted by the to-directional volatility connectedness measure based on volatility shocks that exclude contemporaneous correlations. $\mathcal{P}_1 (\mathcal{P}_5)$ comprises currencies with the highest (lowest) levels of volatility transmission. In the first stage, we estimate time-series regressions of test portfolios on the DOL and $\mathcal{T}$ factors. In the second stage, we estimate the cross-sectional regression of average returns on the slope coefficients. We exclude the intercept in the second stage. Panels A and B present the outputs of the first and second stage regressions. The values reported in the “$\alpha$ (\%)” column and the “$\lambda$ (\%)” row are expressed in percentage per annum. The t-statistics are based on \cite{newey1987simple} standard errors. Goodness-of-fit statistics include the
$R^2,$ RMSE, 
$\chi^2$ test statistic and associated p-value. The sample spans March 2005 to December 2021, which corresponds to the out-of-sample estimation period.
\end{minipage}
\vspace{\medskipamount}
\footnotesize

\begin{tabular}{llD{.}{.}{4.2}D{.}{.}{4.2}D{.}{.}{4.2}D{.}{.}{4.2}}
\toprule
\multicolumn{6}{l}{Panel A: Time-series regressions} \\
\midrule
&       & \multicolumn{1}{r}{$\alpha$ (\%)} & \multicolumn{1}{r}{$\beta_{\text{DOL}}$} & \multicolumn{1}{r}{$\beta_{\mathcal{T}}$} & \multicolumn{1}{r}{$R^2 (\%)$} \\
\midrule
$\mathcal{P}_1$    &       & 0.02  & 1.00     & -0.54 & 94.94 \\
t-stat &       & 0.04  & 51.06 & -19.76 &  \\
$\mathcal{P}_2$    &       & 0.87  & 0.93  & -0.07 & 80.25 \\
t-stat &       & 1.21  & 17.46 & -1.66 &  \\
$\mathcal{P}_3$    &       & -0.47 & 1.04  & 0.07  & 84.76 \\
t-stat &       & -0.53 & 26.61 & 1.54  &  \\
$\mathcal{P}_4$    &       & -0.45 & 1.04  & 0.07  & 84.71 \\
t-stat &       & -0.50  & 30.49 & 1.36  &  \\
$\mathcal{P}_5$    &       & 0.02  & 1.00     & 0.46  & 92.64 \\
t-stat &       & 0.04  & 51.06 & 17.02 &  \\
\midrule
\multicolumn{6}{l}{Panel B: Cross-sectional regressions} \\
\midrule
&       &       & \multicolumn{1}{r}{DOL} & \multicolumn{1}{r}{$\mathcal{T}$} & \multicolumn{1}{r}{$\chi^2$} \\
\cmidrule{4-6}
$\lambda$ (\%) &       &       & -0.35 & 4.21  & 0.97 \\
t-stat &       &       & -0.16 & 2.35  & 0.81 \\
RMSE  &       &       & 0.48  &       &  \\
$R^2 (\%)$    &       &       & 88.90  &       &  \\
\bottomrule
\end{tabular}%
\label{tab:asset pricing test fama-macbeth}%
\end{table}%

Panel A in Table \ref{tab:asset pricing test fama-macbeth} reports the time-series regressions of the first step of the FMB procedure. The 
$\beta_{\text{DOL}}$
coefficients are close to one, whereas the $\beta_{\mathcal{T}}$
coefficients display pronounced monotonicity when we move from 
$\mathcal{P}_1$
to 
$\mathcal{P}_5,$ increasing from $-0.54$ (a t-stat of $-19.76$) to $0.46$ (a t-stat of $17.02$). The two factors capture a lot of variation of quintile portfolios, ranging from $80.28\%$ for 
$\mathcal{P}_2$
to $94.94\%$ for
$\mathcal{P}_1.$
The estimated alphas of quintile portfolios become economically and statistically insignificant when controlling for the DOL and 
$\mathcal{T}$
factors. 

Panel B in Table \ref{tab:asset pricing test fama-macbeth} reports the cross-sectional results of the second step of the FMB procedure. The $\mathcal{T}$ factor has a positive and statistically significant price of 4.21\% per annum (a t-stat of 2.35). Since the spread portfolio is actually tradable, we can apply the Euler equation to the factor excess returns and derive that its price of risk must be equal to the average excess return. We verify that this no-arbitrage condition indeed holds: the annual average return of 4.46\% in Table \ref{tab: to-directional volatility connectedness portfolios} is close to the estimated price. Regarding the DOL factor, its price of risk is insignificant, but the estimated
$\lambda_{\text{dol}}$
matches the factor's average excess return of -0.33\% per annum. Although the dollar factor does not help explain average excess returns, it serves as a constant that captures common mispricing in the cross-sectional regression. The two-factor specification accounts for 88.90\% of the cross-sectional variation in average returns across quintiles and cannot be rejected at any conventional confidence level.

Overall, the time-series results are consistent with cross-sectional tests. The dollar and to-directional volatility connectedness factors fully explain the cross-section formed on directional volatility transmission.

\section{Understanding volatility-connectedness-sorted returns}

This section presents additional checks to illustrate the mechanism driving our results. 

\subsection{Allocation analysis} 
\label{subsubsection: allocation analysis}

As a first step, Table \ref{table: allocation analysis} presents the allocation analysis of quintiles formed on to-directional volatility connectedness. Poland and the United Kingdom belong to the lowest-transmission portfolio ($\mathcal{P}_1$) 36\% and 25\% of the time, whereas Denmark and the Euro Area spend 8\% and 10\%. Brazil, Taiwan, and Turkey are most frequently classified as the strongest volatility transmitters, occurring 30\%, 30\%, and 35\% of periods, respectively, in the highest-transmission portfolio ($\mathcal{P}_5$). Several countries --- Norway, Poland, and Sweden -- appear to be the strongest contributors to the volatility of other currencies in fewer than 10\% of months. Several other countries --- the Czech Republic, Singapore, and South Africa --- tend to have balanced allocations of approximately 20\% across quintiles. Although we observe some differences in currency frequencies, they do not appear extreme. Consequently, to-directional volatility connectedness exhibits dynamically evolving patterns. This further underscores its relevance beyond common currency predictors (e.g., interest rates, trade links), which are typically static. 

\begin{table}[t!]
\centering
\caption{Allocation analysis.}
\begin{minipage}{\textwidth} 
This table presents the allocation frequencies of currencies across quintile $(\mathcal{P}_i: i = 1,\dots,5)$ portfolios sorted by the to-directional volatility connectedness measure based on volatility shocks that exclude contemporaneous correlations. $\mathcal{P}_1 (\mathcal{P}_5)$ comprises currencies with the highest (lowest) levels of volatility transmission. The columns report the fraction of months each currency belongs to the portfolios considered. The sample spans March 2005 to December 2021, which corresponds to the out-of-sample estimation period.
\end{minipage}
\vspace{\medskipamount}
\footnotesize

\begin{tabular}{llD{.}{.}{2.2}D{.}{.}{2.2}D{.}{.}{2.2}D{.}{.}{2.2}D{.}{.}{2.2}D{.}{.}{4.0}}
\toprule
&& \multicolumn{1}{r}{$\mathcal{P}_1$} & \multicolumn{1}{r}{$\mathcal{P}_2$} & \multicolumn{1}{r}{$\mathcal{P}_3$} & \multicolumn{1}{r}{$\mathcal{P}_4$} & \multicolumn{1}{r}{$\mathcal{P}_5$}  & \multicolumn{1}{r}{Obs.} \\
\midrule
Australia &       & 0.14  & 0.24  & 0.20  & 0.27  & 0.15  & 202 \\
Brazil &       & 0.24  & 0.15  & 0.14  & 0.17  & 0.30  & 202 \\
Canada &       & 0.21  & 0.20  & 0.19  & 0.27  & 0.13  & 202 \\
Czech Republic &       & 0.20  & 0.19  & 0.23  & 0.17  & 0.21  & 202 \\
Denmark &       & 0.08  & 0.21  & 0.27  & 0.25  & 0.19  & 202 \\
Euro Area &       & 0.10  & 0.23  & 0.28  & 0.23  & 0.15  & 202 \\
Hungary &       & 0.19  & 0.24  & 0.14  & 0.23  & 0.20  & 202 \\
Japan &       & 0.14  & 0.11  & 0.21  & 0.21  & 0.32  & 202 \\
Mexico &       & 0.21  & 0.15  & 0.17  & 0.18  & 0.29  & 202 \\
New Zealand &       & 0.17  & 0.20  & 0.27  & 0.23  & 0.13  & 202 \\
Norway &       & 0.23  & 0.31  & 0.24  & 0.15  & 0.06  & 202 \\
Poland &       & 0.36  & 0.18  & 0.23  & 0.13  & 0.09  & 202 \\
Singapore &       & 0.21  & 0.23  & 0.14  & 0.23  & 0.19  & 202 \\
South Africa &       & 0.22  & 0.21  & 0.20  & 0.20  & 0.18  & 202 \\
South Korea &       & 0.21  & 0.18  & 0.12  & 0.14  & 0.35  & 202 \\
Sweden &       & 0.20  & 0.30  & 0.26  & 0.19  & 0.05  & 202 \\
Switzerland &       & 0.15  & 0.18  & 0.23  & 0.28  & 0.15  & 202 \\
Taiwan &       & 0.26  & 0.15  & 0.14  & 0.15  & 0.30  & 202 \\
Turkey &       & 0.21  & 0.12  & 0.13  & 0.19  & 0.35  & 202 \\
United Kingdom &       & 0.25  & 0.22  & 0.20  & 0.13  & 0.20  & 202 \\
\bottomrule
\end{tabular}%
\label{table: allocation analysis}
\end{table}%

For comparison, we look at the allocation frequencies of currencies based on from-directional volatility connectedness. Table \ref{table: allocation analysis from-directional volatility} in the Appendix demonstrates pronounced asymmetries in the classification of different currencies. Several high-interest-rate countries --- Brazil, Hungary, South Africa, and Turkey --- spend a substantial fraction of the sample in the highest volatility exposure quintile ($\mathcal{P}_5$). Additionally, they are rarely among the weakest recipients of volatility shocks. In contrast, Singapore and Taiwan  --- the low-interest-rate countries --- exhibit the lowest levels of received volatility throughout the sample. The results show that the dynamics of from-directional volatility connectedness are highly persistent and strongly differentiated across some countries. These patterns support the interpretation that countries' exposure to external volatility shocks is associated with underlying macro-financial characteristics rather than transitory fluctuations.

\subsection{Relating to-directional volatility connectedness to liquidity and volatility}
\label{subsubsection: liquidity and volatity risk}

As a second step, we examine the contributions of two drivers of currency risk premia --- global volatility and liquidity --- in explaining the currency excess returns obtained from portfolios sorted by to-directional volatility connectedness. 

We use the VIX index, the risk-neutral expectation of stock market volatility, as a proxy for US investors' risk aversion to volatility. We employ the equivalent implied volatility indices for foreign exchange markets of the G7 countries (VXY-G7) and emerging markets (VXY-EM), developed by JP Morgan. We also compute global FX volatility (VOL) as defined in \cite{menkhoff2012carry}.\footnote{The global VOL measure in a month 
$t$ is formally defined as:
\begin{equation*}
VOL_t = \frac{1}{T_t}\sum\limits_{\tau\in T_t}\left[\sum\limits_{k\in K_{\tau}}\left(\frac{|r_{\tau}^k|}{K_{\tau}}\right)\right],
\end{equation*}
in which 
$|r_{\tau}^k| = |\Delta s_{\tau}|$ 
is the absolute daily log return for a currency 
$k$
on a day 
$\tau,$
$K_{\tau}$
is the number of available currencies on a day
$\tau,$
and 
$T_t$
is the number of days in a month 
$t.$
}
We consider several liquidity measures. We use the TED spread, defined as the difference between three-month euro interbank deposits (LIBOR) and three-month Treasury bills, as an illiquidity proxy in the funding market for carry trades. We employ the liquidity-based measure of financial commercial paper outstanding (FCPO). We consider the St. Louis Federal Reserve Bank Financial Stress Index (FSI) to capture the degree of financial stress in the US economy. The FSI is based on various interest rates, yield spreads, and other indicators. We construct a global proxy for FX market illiquidity using the bid-ask spread (BAS) data of individual currencies.\footnote{A global BAS measure in a month $t$ is formally defined as: 
\begin{equation*}
BAS_t = \frac{1}{T_t}\sum\limits_{\tau\in T_t}\left[\sum\limits_{k\in K_{\tau}}\left(\frac{BAS_{\tau}^k}{K_{\tau}}\right)\right],
\end{equation*}
in which 
$BAS_{\tau}^k$
is the daily percentage BAS of a currency 
$k$
on a day
$\tau,$
$K_{\tau}$
is the number of available currencies on a day 
$\tau,$ 
and 
$T_t$
is the number of days in a month 
$t.$} 

\begin{table}[t!]
\centering
\caption{Volatility, liquidity, and financial stress.}
\begin{minipage}{\textwidth} 
This table presents correlations between the $\mathcal{T}$ strategy and proxies for volatility, liquidity, and financial stress. 
Panel A reports the correlations with the VIX index (VIX), implied volatility indices for foreign exchange markets of G7 countries (VXY-G7) and emerging markets (VXY-EM), global 
FX volatility (VOL) and a tradable currency factor associated with VOL ($f_{\text{VOL}}$). Panel B reports the correlations with the TED spread (TED), the St. Louis Federal Reserve Bank Financial Stress Index (FSI), financial commercial paper outstanding (FCPO), global FX liquidity (BAS), and a tradable currency factor associated with BAS ($f_{\text{BAS}}$). The sample spans March 2005 to December 2021, which corresponds to the out-of-sample estimation period.
\end{minipage}
\vspace{\medskipamount}
\footnotesize

\begin{tabular}{lcD{.}{.}{4.2}D{.}{.}{4.2}D{.}{.}{4.2}D{.}{.}{4.2}D{.}{.}{4.2}}
\toprule
\multicolumn{7}{l}{\textbf{Panel A: Volatility}} \\
\midrule
&       & \multicolumn{1}{r}{VIX} & \multicolumn{1}{r}{VYX-G7} & \multicolumn{1}{r}{VXY-EM} & \multicolumn{1}{r}{VOL} & \multicolumn{1}{r}{$f_{\text{VOL}}$} \\
\cmidrule{3-7}    
$\mathcal{T}$ &       & 0.20  & 0.24  & 0.28  & 0.19  & -0.30 \\
t-stat &       & 2.89  & 3.23  & 3.86  & 2.72  & -4.45 \\
p-value &       & 0.00  & 0.00  & 0.00  & 0.01  & 0.00 \\
\midrule
\multicolumn{7}{l}{\textbf{Panel B: Liquidity and financial stress }} \\
\midrule
&       & \multicolumn{1}{r}{TED} & \multicolumn{1}{r}{FSI} & \multicolumn{1}{r}{FCPO} & \multicolumn{1}{r}{BAS} & \multicolumn{1}{r}{$f_{\text{BAS}}$} \\
\cmidrule{3-7}   
$\mathcal{T}$ &       & 0.10  & 0.17  & -0.15 & 0.11  & -0.17 \\
t-stat &       & 1.37  & 2.37  & -2.13 & 1.56  & -2.40 \\
p-value &       & 0.17  & 0.02  & 0.03  & 0.12  & 0.02 \\
\bottomrule
\end{tabular}%
\label{tab:global liquidity and volatility risk}%
\end{table}%

Panels A and B in Table \ref{tab:global liquidity and volatility risk} report correlations between the excess returns of the $\mathcal{T}$ strategy and changes in volatility and liquidity measures.\footnote{We express BAS in terms of first differences and compute log differences for all other variables.} The correlations with volatility proxies are uniformly positive and 
statistically significantly different from zero. This positive association shows that the excess returns of the $\mathcal{T}$ strategy are higher during periods of higher volatility. The small difference in correlations with VXY-G7 and VXY-EM highlights that profitability is enhanced equally during periods of higher uncertainty in the foreign exchange markets of the G7 and emerging economies. In turn, the strategy exhibits a weaker relationship with liquidity and stress conditions, though it also earns higher average returns in liquidity-constrained periods (note that the FCPO measure is procyclical, unlike other countercyclical liquidity measures). 

Although the strongest correlations are far from perfect (e.g., 0.28 for VXY-EM), the $\mathcal{T}$ returns could still be explained by the fact that global volatility (liquidity) loads with lags across currencies. For this reason, we construct a global volatility (liquidity) factor formed on time-varying loadings on a global volatility (liquidity) shock, following \cite{della2016volatility}.\footnote{For a global volatility (liquidity) factor, each month we estimate the regression of excess returns of individual currencies on a constant and global volatility (liquidity) innovations using a 36-month rolling window. Having estimated the slope coefficients, we sort the currencies into quintile portfolios based on these betas. Then, we create a zero-cost strategy by buying (selling) the lowest (highest) beta quintile of currencies.} As shown in the columns labeled $``f_{\text{VOL}}$'' and $``f_{\text{BAS}}$'', the long-short portfolio exhibits negative correlations of $-0.30$ and $-0.17$ with the tradable volatility and liquidity factors, which are still far from 1 in absolute terms. In unreported results, we also document economically large and statistically significant alphas when controlling for volatility and liquidity factors. 

In sum, the empirical evidence indicates that the returns from the $\mathcal{T}$ strategy cannot be understood as compensation for global risk. When global volatility and liquidity risks are high, we find that the expected return from the strategy increases. Intuitively, as we remove the common correlation component in volatility connections, which are likely driven by a global shock, volatility transmission should capture more local (country- or region-specific) than global information.

\subsection{Currency risk premia and volatility shocks in general equilibrium}

This section develops a general equilibrium model demonstrating that volatility transmission through correlation-adjusted linkages proxies for the priced risk of country-specific jumps. In the model, the strongest transmitters hedge against priced domestic idiosyncratic jumps, appreciating when such shocks occur and therefore earning lower average risk premia. Meanwhile, the weakest transmitters depreciate in response to foreign idiosyncratic jumps and must compensate investors with higher expected returns. The strategy of buying the latter and selling the former yields positive returns on average but delivers lower returns as the likelihood of idiosyncratic jumps increases. We demonstrate empirical support for this theoretical mechanism.

\subsubsection{Endowments}

We present an $n$-country economy to analyze the relationship between cross-sectional currency risk premia and volatility shocks in general equilibrium. The central feature of the model is the presence of jumps in country-level cash flows. Specifically, aggregate consumption in a given country is affected by domestic and foreign jumps. Formally, we assume a Lucas endowment world economy with $n$ countries, indexed by $i$, where the first country is the United States. Denoting log aggregate consumption in a country $i$ by $y_{i,t} = \ln\left(Y_{i,t}\right),$  we define the dynamics for endowments as:
\begin{equation}
\label{eq: dy it main}
\mathrm{~d} y_{i, t} = \mu \mathrm{~d} t+\sum_{j=1}^n K_{i,j} \mathrm{~d} N_{j, t},
\end{equation}
where 
$\mu$ 
is a constant drift, $N_{j,t}$ denotes a jump in a country $j$ whose impact on cash flows in a country $i$ is constant $K_{i,j}.$ For simplicity, we assume that all jump intensities are constant $\overline{\ell}.$ The coefficients $K_{i,j}$ are assumed to be negative, $K_{i,j}<0;$ therefore, jumps are associated with the adverse impact. 

Before advancing further, it is essential to discuss the endowment processes given by Equations (\ref{eq: dy it main}). First, jumps are a common, but certainly not the only, modeling choice for capturing shock propagation. We select a parsimonious model with a jump component solely for clear mechanism inspection, thereby isolating the impact of jumps from other factors.\footnote{In the Appendix, we develop a dynamic framework with mutually excited jump processes that allow us to extend jump propagation across countries and over time. Using the affine solution approach of \cite*{ES2008} and the log-linearization procedure, we derive a closed-form solution for the pricing kernel and exchange rates. However, this makes the derivations and exposition unnecessarily complex. Instead, we can demonstrate the key mechanism under contemporaneous jumps and constant intensities.} Since jumps are the only source of uncertainty, they simultaneously generate cross-sectional currency risk premia and exchange rate volatility. Therefore, the latter are completely governed by the matrix of cross-country exposure coefficients $[K_{i,j}]_{i,j=1}^n.$

\subsubsection{A representative agent}
\label{section: equilibrium in a dynamic model}

A representative agent with recursive preferences populates each country. The pricing kernel satisfies the equation: 
\begin{equation}
\mathrm{~d} \ln M_{i,t}=-\delta \theta \mathrm{~d} t-(1-\theta) \mathrm{~d} \ln R_{i,t}-\frac{\theta}{\psi} \mathrm{~d} y_{i,t},
\end{equation}
where $R_{i,t}$ is the return on the claim to aggregate consumption, $\delta$ is the subjective discount factor, $\gamma > 1$ is the coefficient of relative risk aversion, $\psi > 1$ is the elasticity of intertemporal substitution (EIS), and $\theta = \frac{1-\gamma}{1-\frac{1}{\psi}} < 0.$ Given the restrictions on parameter values, the representative agent has a preference for early resolution of uncertainty, that is, $\gamma > \frac{1}{\psi}.$

\subsubsection{Stochastic discount factor and exchange rates}
In what follows, we focus on pricing kernels, exchange rates, and their connection with jumps in the world economy. Using the general framework in Appendix \ref{appendix: model derivations} (in particular the simplified model in Appendix \ref{appendix: static model}), the following proposition holds. 

\textbf{Proposition 1.} 
\begin{itemize}
\item[1.]
\textit{The stochastic discount factor (SDF) in a country 
$i$
satisfies:
\begin{equation}
\label{eq: pricing kernel}
\frac{\mathrm{~d} M_{i,t}}{M_{i,t}} = -r_{i,t} \mathrm{~d} t - \sum_{k=1}^n\mathrm{MPJR}_{i,k} \left(\mathrm{~d} N_{k,t}-\overline{\ell} \mathrm{~d} t\right)
\end{equation}
where 
$r_{i,t}$
is the risk-free rate, 
$\mathrm{MPJR}_{i,k}$
is the market price of the country $k$'s jump in the country $i$ given by:
\begin{equation}
\label{eq: rho main text}
\mathrm{MPJR}_{i,k} = 1 - \exp\left[-\gamma K_{i,k}\right].
\end{equation}
\item[2.]  $\mathrm{MPJR}_{i,k} <0.$
\item[3.]  $|K_{j,k}|>|K_{i,k}| ~ \Rightarrow ~ |\mathrm{MPJR}_{j,k}| > |\mathrm{MPJR}_{i,k}|.$
}
\item[] Proof: See Appendix \ref{appendix: model derivations}.
\end{itemize}
The first property states that jumps in all countries drive the dynamics of the SDF for the country $i$. Intuitively, jumps affect cash flows across all countries, and hence, the pricing kernel of a particular country compensates for all jumps. The compensation of the country $i$'s pricing kernel for jumps originating in a country $k$ $(\mathrm{MPJR}_{i,k})$ depends on the jump magnitude $K_{i,k}.$ 

For a country $i,$ we define the real exchange rate as the ratio of SDF in this country $(M_{i,t})$ and the US $(M_{1,t}):$
\begin{equation*}
Q_{i,t} = \frac{M_{i,t}}{M_{1,t}}.    
\end{equation*}
Given this definition, the following proposition holds.

\textbf{Proposition 2.} 
\begin{itemize}
\item[1.]
\textit{The real exchange rate has dynamics:
\begin{align*}
\frac{\mathrm{~d} Q_{i,t}}{Q_{i,t}} & = \mu_{i,t} \mathrm{~d} t + \sum_{k=1}^n \theta_{i,k} \left(\mathrm{~d} N_{k,t} - \overline{\ell} \mathrm{~d} t\right), \\
\mu_{i,t} &  = - (r_{i,t} - r_{1,t}) + \sum_{k=1}^n \mathrm{MPJR}_{1,k}  \theta_{i,k} \overline{\ell}, \\
\theta_{i,k} &  =  -\Big[1 - \exp\left[- \gamma (K_{i,k} - K_{1,k})\right]\Big],
\end{align*}
where $r_{i,t}$ and $r_{1,t}$ are the risk-free rates in a country $i$ and US.
\item[2.]  $|K_{i,k}|>|K_{1,k}| ~ \Rightarrow ~ \theta_{i,k} > 0 ~  \wedge ~ |K_{1,k}|>|K_{i,k}| ~ \Rightarrow ~ \theta_{i,k} < 0.$
\item[3.] $|K_{j,k}|>|K_{i,k}|>|K_{1,k}| ~ \vee ~ |K_{1,k}|>|K_{i,k}|>|K_{j,k}| ~ \Rightarrow ~ |\theta_{j,k}| > |\theta_{i,k}|.$
}
\item[] Proof: See Appendix \ref{appendix: model derivations}.
\end{itemize}

Note that 
$\mu_{i,t}$
is equal to the expected instantaneous depreciation rate of the exchange rate $i.$ Consequently, the expected excess return (the currency risk premium) is given by:
\begin{equation}
\label{eq: rx}
rx_{i,t} = \mathbb{E}_t\left(\frac{\mathrm{~d} Q_{i,t}}{Q_{i,t}}\Big/\mathrm{~d} t\right) + r_{i,t} - r_{1,t} = \sum_{k=1}^n\mathrm{MPJR}_{1,k} \theta_{i,k} \overline{\ell}.
\end{equation}    
Furthermore, the expected instantaneous variance takes the form:
\begin{equation}
\label{eq: var fx}
\sigma^2_{i,t} = \mathbb{E}_t\left(\left[\frac{\mathrm{~d} Q_{i,t}}{Q_{i,t}}\right]^2\Big/\mathrm{~d} t\right) = \sum_{k=1}^n \theta_{i,k}^2 \overline{\ell}.
\end{equation}
In the empirical analysis, we identify shocks to exchange rate volatilities and examine their predictive power for excess returns. Hence, these identities provide a direct bridge between volatility connectedness and currency risk premia in our model.\footnote{Note that the empirical analysis uses the risk-neutral volatilities on exchange rates. It is possible to rewrite Equation (\ref{eq: var fx}) in terms of a risk-neutral measure. Doing so will not change the intuition but will introduce additional notations. We keep the discussion in terms of physical volatilities to avoid more cumbersome expressions.} 

Equation (\ref{eq: rx}) predicts that the expected excess return is driven by the degree to which investors in the US dislike jumps in the world economy (the market prices of jump risks, $\mathrm{MPJR}_{1,k}$) and the exposure of a particular currency to jumps (exchange rate loadings, $\theta_{i,k}$) (as well as the likelihood of jumps occurring, but it is the same across countries). Equation (\ref{eq: var fx}) shows that exchange rate volatility is fully determined by the squared exposure coefficients $\theta_{i,k}^2,$ which measure the differential effect of a country $k$'s jump on a country $i$ relative to the US. In our setting, ``volatility shocks'' represent the economic importance of jump risk, that is, the magnitude of the loadings,  
 $|\theta_{i,k}|.$
Although the cross-sectional differences in volatility shocks and expected returns depend on $[K_{i,j}]_{i,j=1}^n,$ the relationship between them is not purely mechanical. The reason is that volatility shocks increase with the square of exposure, whereas expected returns increase linearly and depend on the sign of exposure.

\subsubsection{Connecting the theory with the empirical analysis}
\label{section: example}

We now present three examples to illustrate the theoretical mechanism and connect it to the empirical results. 

\textbf{Example 1.} We first assume that jumps in the US are global (common to all countries), whereas other jumps are country-specific (idiosyncratic). The matrix of jump sizes is defined as:
\begin{equation*}
\left(\begin{array}{lllll}
K_{1,1} & 0 & 0 & \ldots & 0 \\
K_{2,1} & K_{2,2} & 0 & \ldots & 0 \\
K_{3,1} & 0 & K_{3,3} & \ldots & 0 \\
\vdots & \vdots & \vdots & \ddots & \vdots \\
K_{n,1} & 0 & 0 & \ldots & K_{n,n} \\
\end{array}\right)_{n \times n}.
\end{equation*}
The expected excess return and volatility of a currency $i$ are given by:
\begin{align*}
rx_{i,t} & = \mathrm{MPJR}_{1,1} \theta_{i,1} \overline{\ell} = - \left(1 - e^{- \gamma K_{1,1}}\right) \left(1 - e^{- \gamma (K_{i,1} - K_{1,1})}\right) \overline{\ell}, \quad i = 2, ..., n, \\ 
\sigma^2_{i,t} & = \theta_{i,1}^2 \overline{\ell} + \theta_{i,i}^2 \overline{\ell} = \left(1 - e^{- \gamma (K_{i,1} - K_{1,1})}\right)^2 \overline{\ell} + \left(1 - e^{- \gamma K_{i,i}}\right)^2 \overline{\ell}, \quad i = 2, ..., n.
\end{align*}
Since the US SDF prices only the global shock, country-specific jumps do not command a premium. Hence, currency risk premia depend exclusively on differential exposure to a global shock, while currency volatility comprises systematic and idiosyncratic components. 

In the empirical analyses, we identify volatility shocks that control for the common component across currency volatilities. Mapping empirical directional volatility connections to model-based quantities, the natural counterparts of from- and to-directional volatility connectedness measures in the model are:
\begin{equation*}
\mathcal{F}_i = 0, \quad \mathcal{T}_i = 0, \quad i = 2, ..., n.
\end{equation*}
Indeed, country-specific jumps do not spill over to other currencies and do not affect the exchange rate volatility of foreign economies. Once we remove the global common component in currency volatilities, there are no residual linkages. Thus, the model makes no meaningful predictions about the predictive power of directional volatility linkages, controlling for common variation, for cross-sectional currency premia. Any differences in average excess returns arise from exposure to the common global shock rather than from the volatility network structures driven by country-specific jumps.

\textbf{Example 2.}  We now assume that jumps in a country $n$ affect cash flows in the US, while the remaining structure remains unchanged. A modified matrix of jump sizes is then:
\begin{equation*}
\left(\begin{array}{lllll}
K_{1,1} & 0 & 0 & \ldots & K_{1,n} \\
K_{2,1} & K_{2,2} & 0 & \ldots & 0 \\
K_{3,1} & 0 & K_{3,3} & \ldots & 0 \\
\vdots & \vdots & \vdots & \ddots & \vdots \\
K_{n,1} & 0 & 0 & \ldots & K_{n,n} \\
\end{array}\right)_{n \times n}.
\end{equation*}
The expected excess return and volatility of exchange rates are given by:
\begin{align}
\label{eq: rx 2}
rx_{i,t} & = \mathrm{MPJR}_{1,1} \theta_{i,1} \overline{\ell} + \mathrm{MPJR}_{1,n} \theta_{i,n} \overline{\ell}, \quad i = 2, ..., n, \\ 
\nonumber
\sigma^2_{i,t} & = \theta_{i,1}^2 \overline{\ell} + \theta_{i,i}^2 \overline{\ell} + \theta_{i,n}^2 \overline{\ell}, \quad i = 2, ..., n-1, \\
\nonumber
\sigma^2_{n,t} & = \theta_{n,1}^2 \overline{\ell} + \theta_{n,n}^2 \overline{\ell},
\end{align}
in which
\begin{align}
\label{eq: theta i n}
\theta_{i,n} & = -\Big[1 - \exp\left[- \gamma (0 - K_{1,n})\right]\Big] < 0, \quad i = 2, ..., n - 1, \\
\label{eq: theta n n}
\theta_{n,n} & = - \Big[1 - \exp\left[- \gamma (K_{n,n} - K_{1,n})\right]\Big] > \theta_{i,n}.
\end{align}
The model-based from- and to-directional volatility connectedness measures are:
\begin{align*}
\mathcal{F}_i & = \theta_{i,n}^2 \overline{\ell}, \quad \mathcal{T}_i = 0, \quad i = 2, ..., n-1, \\
\mathcal{F}_n & = 0, \quad \mathcal{T}_n = (n-2)\theta_{i,n}^2 \overline{\ell}.
\end{align*}
Country 
$n$ is the sole transmitter of volatility shocks, while others act purely as receivers.

The average excess return in Equation (\ref{eq: rx 2}) consists of the two components: compensation for jumps in the US (global) and the country $n$ (idiosyncratic). Taking into account Equations (\ref{eq: theta i n})-(\ref{eq: theta n n}), the cross-sectional variation associated with the latter satisfies:
\begin{equation}
\mathrm{MPJR}_{1,n} \theta_{n,n} \overline{\ell} < \mathrm{MPJR}_{1,n} \theta_{i,n} \overline{\ell}, \quad i = 2, ..., n-1.
\end{equation}
Thus, sorting currencies on to-directional volatility transmission isolates the idiosyncratic shock component of $rx_{i,t}.$ However, currency excess returns continue to depend on differential exposure to the global shock. As shown in Table \ref{tab: to-directional volatility connectedness portfolios}, the interest rate differentials exhibit a small variation across to-directional volatility connectedness portfolios. Since carry trade returns are closely associated with innovations in global foreign exchange volatility, it is reasonable to assume that the average compensation for common jumps in the model does not differ substantially across $\mathcal{T}$-sorted portfolios. Consequently, the cross-sectional currency premia primarily reflect differences in the component associated with country-specific jumps, and therefore inherit the ordering induced by $\mathcal{T}$ sorting.

\textbf{Example 3.}  We now extend the previous setting by allowing jumps in countries $n-1$ and $n$ to affect cash flows in the US. Consequently, the expected excess return and volatility of exchange rates incorporate additional terms:
\begin{align}
\label{eq: rx 2 ex 3}
rx_{i,t} & = \mathrm{MPJR}_{1,1} \theta_{i,1} \overline{\ell} + \mathrm{MPJR}_{1,n-1} \theta_{i,n-1} \overline{\ell} + \mathrm{MPJR}_{1,n} \theta_{i,n} \overline{\ell}, \quad i = 2, ..., n, \\ 
\nonumber
\sigma^2_{i,t} & = \theta_{i,1}^2 \overline{\ell} + \theta_{i,i}^2 \overline{\ell} + \theta_{i,n-1}^2 \overline{\ell} + \theta_{i,n}^2 \overline{\ell}, \quad i = 2, ..., n-2, \\
\nonumber
\sigma^2_{n-1,t} & = \theta_{n-1,1}^2 \overline{\ell} + \theta_{n-1,n-1}^2 \overline{\ell} + \theta_{n-1,n}^2 \overline{\ell}, \\
\nonumber
\sigma^2_{n,t} & = \theta_{n,1}^2 \overline{\ell} + \theta_{n,n-1}^2 \overline{\ell} + \theta_{n,n}^2 \overline{\ell}.\footnotemark
\end{align}
\footnotetext{The remaining parameters are given by:
\begin{align}
\nonumber
\mathrm{MPJR}_{1,n-1} & = 1 - \exp\left[-\gamma K_{1,n-1}\right], \quad \mathrm{MPJR}_{1,n} = \ - \exp\left[-\gamma K_{1,n}\right] \\
\nonumber
\theta_{i,n-1} & = -\Big[1 - \exp\left[- \gamma (0 - K_{1,n-1})\right]\Big] < 0, \quad \theta_{i,n} = -\Big[1 - \exp\left[- \gamma (0 - K_{1,n})\right]\Big] < 0, \quad i = 2, ..., n - 2, \\
\nonumber
\theta_{n-1,n-1} & = - \Big[1 - \exp\left[- \gamma (K_{n-1,n-1} - K_{1,n-1})\right]\Big], \quad \theta_{n-1,n} = - \Big[1 - \exp\left[- \gamma (0 - K_{1,n})\right]\Big] < 0, \\
\nonumber
\theta_{n,n-1} & = - \Big[1 - \exp\left[- \gamma (0 - K_{1,n-1})\right]\Big] < 0, \quad \theta_{n,n} = - \Big[1 - \exp\left[- \gamma (K_{n,n} - K_{1,n})\right]\Big].
\end{align}
}
The model-based from- and to-directional volatility connectedness measures are:
\begin{align*}
\mathcal{F}_i & = \theta_{i,n-1}^2 \overline{\ell} + \theta_{i,n}^2 \overline{\ell}, \quad \mathcal{T}_i = 0, \quad i = 2, ..., n-1, \\
\mathcal{F}_{n-1} & = \theta_{n-1,n}^2 \overline{\ell}, \quad \mathcal{T}_{n-1} = (n-3)\theta_{n-2,n-1}^2 \overline{\ell} + \theta_{n,n-1}^2 \overline{\ell} \overset{\footnotemark[\value{footnote}]}{=} (n-2)\theta_{n,n-1}^2 \overline{\ell}, \\
\mathcal{F}_n & = \theta_{n,n-1}^2 \overline{\ell}, \quad \mathcal{T}_n =  (n-3)\theta_{n-2,n}^2\overline{\ell}  + \theta_{n-1,n}^2 \overline{\ell} \overset{\footnotemark[\value{footnote}]}{=} (n-2)\theta_{n-1,n}^2\overline{\ell}. 
\end{align*}
Any country $i = 2, ..., n-2$ appears to be the weakest volatility transmitter and earns the highest risk premium associated with jumps in countries $n-1$ and $n.$ Assuming $|K_{1,n-1}|<|K_{1,n}|,$ we can conclude $|\theta_{n,n-1}|<|\theta_{n,n}|;$ hence, $\mathcal{T}_{n-1}<\mathcal{T}_n$ and $|\mathrm{MPJR}_{1,n-1}| < |\mathrm{MPJR}_{1,n}|.$ Thus, the market price of country-specific jump risk (in absolute terms) is positively associated with volatility transmission. Equation (\ref{eq: rx 2 ex 3}), however, demonstrates that the cross-sectional currency risk premia, which are related to idiosyncratic jumps, depend on the differential effect of jumps on each country relative to the US ($|K_{n-1,n-1} - K_{1,n-1}|$ and $|K_{n,n} - K_{1,n}|$). In general, establishing the relationship between volatility transmission and average excess returns for the last two countries requires additional restrictions on jump sizes. We can draw some conclusions under two economically plausible assumptions.
\begin{itemize}
\item[A1:] \textbf{Domestic amplification:} country-specific shocks have a larger impact on the domestic country than on the US ($|K_{i,i}|>|K_{i,1}|$), implying $\theta_{i,i} > 0$. 
\item[A2:] \textbf{Monotonic idiosyncratic volatility scaling:} countries with larger own jump prices (larger $|K_{1,i}|$) exhibit larger idiosyncratic volatility shocks (larger $|K_{i,i} - K_{1,i}|$).
\end{itemize}
If $|K_{1,n-1}|<|K_{1,n}|,$ the two assumptions imply $\mathrm{MPJR}_{1,n-1} \theta_{n-1,n-1} \overline{\ell} > \mathrm{MPJR}_{1,n} \theta_{n,n} \overline{\ell}$ due to $\mathrm{MPJR}_{1,n} < \mathrm{MPJR}_{1,n} < 0$ and $0 < \theta_{n-1,n-1} < \theta_{n,n}.$ Thus,  we again reach the conclusion that volatility transmission across individual currencies negatively predicts the cross-sectional currency premia attributable to idiosyncratic jumps. 

\subsubsection{Discussion} In general, we can extend the previous example by allowing all country-specific shocks to affect cash flows in the US as represented by the matrix of jump sizes:
\begin{equation*}
\left(\begin{array}{lllll}
K_{1,1} & K_{1,2} & K_{1,3} & \ldots & K_{1,n} \\
K_{2,1} & K_{2,2} & 0 & \ldots & 0 \\
K_{3,1} & 0 & K_{3,3} & \ldots & 0 \\
\vdots & \vdots & \vdots & \ddots & \vdots \\
K_{n,1} & 0 & 0 & \ldots & K_{n,n} \\
\end{array}\right)_{n \times n}.
\end{equation*}
Under assumptions (A1)-(A2), the theory continues to predict that the average excess returns associated with idiosyncratic jumps align monotonically negatively with to-directional volatility transmission. The striking consistency between the empirical and theoretical predictions establishes the adequacy of these assumptions.

Another prediction of the model is that the strongest transmitters act as a hedge against own (domestic) priced idiosyncratic jumps, appreciating when such jumps occur and, on average, generating a lower risk premium. In contrast, the weakest transmitters depreciate when priced idiosyncratic jumps in other (foreign) countries happen, and hence they must compensate with the higher average returns. This explains the positive risk premium of the strategy that buys the weakest and sells the strongest transmitters of correlation-adjusted volatility shocks. 

To validate this theoretical prediction, we construct the empirical likelihood of idiosyncratic jumps in exchange rates and connect it to the returns of the $\mathcal{T}$ strategy. We implement a four-step procedure to identify extreme exchange-rate movements that cannot be attributed to broad common shocks. First, we remove common return components in daily excess returns using a factor model. For each currency, we estimate a time-series regression of currency excess returns on the dollar and carry trade factors. We employ daily returns over the full sample in the estimation and define the idiosyncratic component as the residual, which can be interpreted as the portion of the return orthogonal to common movements. Second, we standardize residual returns by the realized one-month volatility of daily residuals. This scaling allows us to classify unusually large moves relative to the local volatility rather than elevated unconditional variance. Third, we define an idiosyncratic jump indicator for each currency that equals 1 if the magnitude of the standardized residual exceeds a threshold. Note that the unconditional volatility of normalized residuals is approximately 1. We set the threshold to 1.5 to balance small and tail-like jumps.\footnote{ The results remain qualitatively and quantitatively similar for thresholds of 1 or 2} Fourth, we use the idiosyncratic jump indicators across different currencies to identify systematic and idiosyncratic jump days. Intuitively, if the minority of currencies experience jumps, this day is unlikely to reflect market-wide shocks. The day is labeled as non-systematic (systematic) if at most (least) $\tau$ idiosyncratic jump indicators are equal to one. 

We construct the monthly time-series $J_t^{\text{idio},\tau}$ and $J_t^{\text{syst},\tau}$ as the average number of idiosyncratic and systematic jump days in a month $t.$ We can interpret these measures as the probabilities of experiencing idiosyncratic and systematic shocks. Table \ref{tab: idiosyncratic jump days} shows correlations between the $\mathcal{T}$ strategy and several measures of $J_t^{\text{idio},\tau}$ and $J_t^{\text{syst},\tau}$ constructed for various choices of $\tau.$ The results demonstrate the negative correlation between the to-directional volatility connectedness strategy and the idiosyncratic jump probability, consistent with the theory. This negative association is stronger when fewer currencies are allowed to experience extreme jumps on the non-systematic jump day, whereas it becomes statistically insignificant when 10 currencies are considered. Intuitively, the larger number of exchange rates experiencing extreme returns reflects market-wide shocks. Since the $\mathcal{T}$ strategy performs well in times of elevated global risks, the negative association with $J_t^{\text{idio},\tau}$ becomes weaker as expected. In relation to this, the probability of market-wide shocks common to many currencies is positively correlated with the returns of the $\mathcal{T}$ strategy, consistent with stronger performance in periods of higher global risk.

\begin{table}[t!]
\centering
\caption{Idiosyncratic and systematic jumps.}
\begin{minipage}{\textwidth} 
This table presents correlations between the $\mathcal{T}$ strategy and the probabilities of idiosyncratic ($J_t^{\text{idio},\tau}$) and systematic ($J_t^{\text{syst},\tau}$) jump months. The sample spans March 2005 to December 2021, which corresponds to the out-of-sample estimation period.
\end{minipage}
\vspace{\medskipamount}
\footnotesize

\begin{tabular}{lcD{.}{.}{4.2}D{.}{.}{4.2}D{.}{.}{4.2}D{.}{.}{4.2}D{.}{.}{4.2}}
\toprule
&       & \multicolumn{1}{r}{$J_t^{\text{idio},3}$} & \multicolumn{1}{r}{$J_t^{\text{idio},5}$} & \multicolumn{1}{r}{$J_t^{\text{idio},7}$} & \multicolumn{1}{r}{$J_t^{\text{idio},10}$} & \multicolumn{1}{r}{$J_t^{\text{syst},10}$} \\
\cmidrule{3-7}    
$\mathcal{T}$ &       & -0.18 & -0.17 & -0.16 & -0.06 & 0.21 \\
t-stat &       & -2.54 & -2.48 & -2.33 & -0.91 & 3.02 \\
p-value &       & 0.01  & 0.01  & 0.02  & 0.37 & 0.00 \\
\bottomrule
\end{tabular}%
\label{tab: idiosyncratic jump days}
\end{table}%

In summary, exchange rate volatility is driven by global and country-specific shocks. The former accounts for a substantial fraction of total volatility and is related to carry trade returns. We complement this evidence by demonstrating that the latter is not merely a noisy residual but contains valuable information about currency excess returns. In particular, we empirically and theoretically show that, after controlling for a common component, currencies with stronger (weaker) transmission of volatility shocks tend to earn lower (higher) average excess returns. In the model, this predictability arises from the heterogeneity in the sizes of idiosyncratic jumps in a country's cash flows and their relative impact on the US. 

\subsubsection{Extensions and future research} It remains possible that the volatility network structures may capture additional dimensions of volatility linkages and their relationships with market prices of risk, which could violate our assumptions. Motivated by the theory, we describe an empirical procedure for forecasting cross-sectional currency risk premia within the general volatility network. Assuming the fully unconstrained matrix of jump sizes, the expected excess returns and variances are:
\begin{align}
\label{eq: rx 3}
rx_{i,t} & = \sum_{k=1}^n\Bigg[\frac{\mathrm{MPJR}_{1,k}}{\theta_{i,k}}\Bigg]\theta_{i,k}^2 \overline{\ell} = \sum_{k=1}^n \overline{\mathrm{MPJR}}_{i,k} \theta_{i,k}^2 \overline{\ell}, \\
\label{eq: var fx 3}
\sigma^2_{i,t} & = \sum_{k=1}^n \theta_{i,k}^2 \overline{\ell}.
\end{align}
In the empirical analysis, we identify the elements $\theta_{i,k}^2 \overline{\ell}$. However, the key challenge is to connect them to the cross-sectional currency premia, which depend on unknown exposure signs and jump prices in the US. 

We therefore suggest a calibration strategy similar to that in \cite{hou2024trade}. Using the estimated volatility shocks in Equation (\ref{eq: var fx 3}), one could first identify the parameters $\overline{\mathrm{MPJR}_{i,k}}$ to obtain the theoretically motivated currency premia in Equation (\ref{eq: rx 3}) and then search all combinations that result in the highest correlation with the realized currency risk premia in the data. The key benefit of this procedure is that it remains suitable even in the presence of bilateral, correlated shocks across countries. The main drawback, however, is the complexity of calibrating a matrix of adjusted prices for jump risk. While implementing this refined strategy is possible, we leave this investigation to future research.

Furthermore, decomposing exposures by sign, the following identities from the model
\begin{align*}
rx_{i,t} & = \sum_{\theta_{i,k}<0} \mathrm{MPJR}_{1,k} \theta_{i,k} \overline{\ell} + \sum_{\theta_{i,k}>0} \mathrm{MPJR}_{1,k} \theta_{i,k} \overline{\ell}, \\
\sigma^2_{i,t} & = \sum_{\theta_{i,k}<0}  \theta_{i,k}^2 \overline{\ell} + \sum_{\theta_{i,k}>0}  \theta_{i,k}^2 \overline{\ell}.
\end{align*}
suggest a natural link to semi-variance connectedness. Specifically, we propose estimating the connections among currency semi-variances associated with currency appreciation or depreciation. Then, the currencies with the strongest and weakest exposure to upside and downside volatility shocks are expected to earn the highest average excess returns. One can capture this predictability using the $\mathcal{F}^{+}  - \mathcal{F}^{-}$ measure for portfolio sorts. We conjecture that this sorting based on linkages, including a strong common correlation, should be closely related to the information content of semi-variance asymmetry. Although the latter has been shown to correlate strongly with carry trade risk premia \citep{Li_Sarno_Zinna_2025}, removing contemporaneous effects may uncover a novel source of currency predictability arising from asymmetry in the received semi-variance shocks.\footnote{Unfortunately, we are unable to construct the exchange rate implied semi-variances for the earlier period of our sample because we possess only the implied variances and not the currency options at the beginning of our time period. Thus, we must leave this investigation for future research due to data limitations.}

\section{Conclusion}

Global foreign exchange volatility is strongly associated with currency excess returns sorted by interest rate differentials. However, little is known about the relationship between individual exchange rate volatilities and their risk premia. This paper demonstrates, both empirically and theoretically, that cross-sectional currency risk premia are negatively related to the degree to which a currency transmits correlation-adjusted volatility shocks. Buying the weakest and selling the strongest transmitters of volatility shocks generates positive, statistically significant average returns and a high Sharpe ratio. The strategy's excess returns cannot be explained by the standard currency benchmarks. The predictive power of volatility transmission to others is equally strong for transitory and persistent components of volatility shocks. The novel investment strategy offers excellent diversification benefits, as it exhibits a weak negative correlation with most common currency factors. 

Although our findings complement the existing evidence on the global volatility risk factor in foreign exchange markets, our understanding of currency volatility spillovers requires further work. Theoretically, our model predicts the additional source of predictability stemming from shocks to currency semi-variances. In particular, the strongest and weakest receivers of shocks to negative and positive variances, respectively, should earn the highest currency premiums. Empirically, one can test this prediction. More broadly, our methodology can be directly applied to other asset classes. Finally, it is equally important to study the implications of volatility connectedness in financial data for asset prices and the real economy. We leave these interesting avenues for future research. 



\vspace{20pt}

\begingroup
\linespread{1}
\setlength{\bibsep}{0pt}
\setlength{\bibhang}{1.0em}
\bibliographystyle{chicago}
\bibliography{mybib}
\endgroup

\appendix


\newpage
\setcounter{section}{0}
\setcounter{equation}{0}
\setcounter{figure}{0}
\setcounter{table}{0}

\def\thesection{\Alph{section}}
\def\thesubsection{\thesection.\arabic{subsection}}
\def\thesubsubsection{\thesubsection.\arabic{subsubsection}}
\renewcommand{\theequation}{\Alph{section}.\arabic{equation}}
\renewcommand{\thetable}{A\arabic{table}}
\renewcommand{\thefigure}{A\arabic{figure}}

\begin{center}
	\Large \textbf{Online Appendix} 
\end{center}

\vspace{10pt}

\begin{abstract}
This appendix presents supplementary details not included in the main body of the paper.
\end{abstract}


\clearpage

\section{Estimation of the time-varying parameter VAR model}
\label{app:estimate}

Let $\mathbf{CIV}_{t}$ be an $N \times 1$ vector generated by a stable time-varying parameter (TVP) heteroskedastic VAR model with $p$ lags:
\begin{equation}\label{eq:VAR}
\mathbf{CIV}_{t,T}=\bPhi_{1}(t/T)\mathbf{CIV}_{t-1,T}+\ldots+\bPhi_{p}(t/T)\mathbf{CIV}_{t-p,T} + \bepsilon_{t,T},
\end{equation}
where 
$\bepsilon_{t,T}=\bSigma^{-1/2}(t/T)\bbeta_{t,T}, \bbeta_{t,T}\sim NID(0,\boldsymbol{I}_M)$ 
and 
$\bPhi(t/T)=(\bPhi_{1}(t/T),\ldots,\bPhi_{p}(t/T))^{\top}$ 
are the time-varying autoregressive coefficients.
Note that all roots of the polynomial $\chi(z)=\text{det}\left(\textbf{I}_{N}-\sum^{L}_{p=1}z^{p}\mathbf{B}_{p,t}\right)$ lie outside the unit circle, and $\bSigma^{-1}_{t}$ is a positive definite time-varying covariance matrix. Stacking the time-varying intercepts and autoregressive matrices in the vector $\phi_{t,T}$ with $\overline{\mathbf{CIV}}^{\top}_{t} = \left(\text{\textbf{I}}_{N} \otimes x_{t}\right),\: x_{t}=\left(1,x^{\top}_{t-1},\dots,x^{\top}_{t-p}\right)$ and denoting the Kronecker product by $\otimes,$ the model can be written as:
\begin{eqnarray}
\mathbf{CIV}_{t,T} = \overline{\mathbf{CIV}}^{\top}_{t,T}\phi_{t,T} + \bSigma^{-\frac{1}{2}}_{t/T}\bbeta_{t,T}.
\end{eqnarray}
We obtain the time-varying parameters of the model by employing the Quasi-Bayesian Local-Likelihood (QBLL) approach of \cite*{petrova2019quasi}. The estimation of Equation (\ref{eq:VAR}) requires re-weighting the likelihood function. The weighting function gives higher proportions to observations surrounding the time period whose parameter values are of interest. The local likelihood function at time period $k$ is given by:
\begin{gather}
\text{L}_{k}\left(\mathbf{CIV}|\theta_{k},\bSigma_{k},\overline{\mathbf{CIV}} \right) \propto\\ 
\nonumber
|\bSigma_{k}|^{\text{trace}(\mathbf{D}_{k})/2}\exp\left\{-\frac{1}{2}(\mathbf{CIV}-\overline{\mathbf{CIV}}^{\top}\phi_{k})^{\top}\left(\bSigma_{k}\otimes\mathbf{D}_{k}\right)(\mathbf{CIV}-\overline{\mathbf{CIV}}^{\top}\phi_{k})\right\},
\end{gather}
where $\mathbf{D}_{k}$ is a diagonal matrix whose elements hold the weights:
\begin{eqnarray}
\mathbf{D}_{k} &=& \text{diag}(\varrho_{k1},\dots,\varrho_{kT}), \\
\varrho_{kt} &=& \zeta_{Tk}w_{kt}/\sum^{T}_{t=1}w_{kt}, \\
\label{eq:weight}
w_{kt} &=& (1/\sqrt{2\pi})\exp((-1/2)((k-t)/H)^{2}),\quad\text{for}\: k,t\in\{1,\dots,T\}, \\
\zeta_{Tk} &=& \left(\left(\sum^{T}_{t=1}w_{kt}\right)^{2}\right)^{-1}, 
\end{eqnarray}
where $\varrho_{kt}$ is a normalised kernel function. $w_{kt}$ uses a Normal kernel weighting function. $\zeta_{Tk}$ gives the rate of convergence and behaves like the bandwidth parameter $H$ in Equation (\ref{eq:weight}). The kernel function puts a greater weight on the observations surrounding the parameter estimates at time $k$ relative to more distant observations.

We use a Normal-Wishart prior distribution for $\phi_{k}|\:\bSigma_{k}$ for $k\in\{1,\dots,T\}$:
\begin{eqnarray}
\phi_{k}|\bSigma_{k} \backsim \mathcal{N}\left(\phi_{0k},(\bSigma_{k} \otimes \mathbf{\Xi}_{0k})^{-1}\right), \\
\bSigma_{k} \backsim \mathcal{W}\left(\alpha_{0k},\mathbf{\Gamma}_{0k}\right), 
\end{eqnarray}
where $\phi_{0k}$ is a vector of prior means, $\mathbf{\Xi}_{0k}$ is a positive definite matrix, $\alpha_{0k}$ is a scale parameter of the Wishart distribution ($\mathcal{W}$), and $\mathbf{\Gamma}_{0k}$ is a positive definite matrix. The prior and weighted likelihood function implies a Normal-Wishart quasi posterior distribution for $\phi_{k}|\:\bSigma_{k}$ for $k=\{1,\dots,T\}$. Formally, let $\mathbf{A} = (\overline{x}^{\top}_{1},\dots,\overline{x}^{\top}_{T})^{\top}$ and $\mathbf{Y}=(x_{1},\dots,x_{T})^{\top}$, then:
\begin{eqnarray}
\phi_{k}|\bSigma_{k},\mathbf{A},\mathbf{Y} &\backsim & \mathcal{N}\left(\widetilde{\theta}_{k},\left(\bSigma_{k}\otimes\mathbf{\widetilde{\Xi}}_{k}\right)^{-1}\right), \\
\bSigma_{k} &\backsim & \mathcal{W}\left(\widetilde{\alpha}_{k},\mathbf{\widetilde{\Gamma}}^{-1}_{k} \right), 
\end{eqnarray}
with quasi-posterior parameters:
\begin{eqnarray}
\widetilde{\phi}_{k} &=& \left(\mathbf{I}_{N}\otimes \mathbf{\widetilde{\Xi}}^{-1}_{k}\right)\left[\left(\mathbf{I}_{N}\otimes \mathbf{A}^{\top}\mathbf{D}_{k}\mathbf{A}\right)\hat{\phi}_{k}+ \left(\mathbf{I}_{N}\otimes \mathbf{\Xi}_{0k}\right)\phi_{0k} \right], \\
\mathbf{\widetilde{\Xi}}_{k} &=& \mathbf{\widetilde{\Xi}}_{0k} + \mathbf{A}^{\top}\mathbf{D}_{k}\mathbf{A}, \\
\widetilde{\alpha}_{k} &=& \alpha_{0k}+\sum^{T}_{t=1}\varrho_{kt}, \\
\mathbf{\widetilde{\Gamma}}_{k} &=& \mathbf{\Gamma}_{0k} + \mathbf{Y}'\mathbf{D}_{k}\mathbf{Y} + \mathbf{\Phi}_{0k}\mathbf{\Gamma}_{0k}\mathbf{\Phi}^{\top}_{0k} - \mathbf{\widetilde{\Phi}}_{k}\mathbf{\widetilde{\Gamma}}_{k}\mathbf{\widetilde{\Phi}}^{\top}_{k},
\end{eqnarray}
where $\widehat{\phi}_{k} = \left(\mathbf{I}_{N}\otimes \mathbf{A}^{\top}\mathbf{D}_{k}\mathbf{A}\right)^{-1}\left(\mathbf{I}_{N} \otimes \mathbf{A}^{\top}\mathbf{D}_{k}\right)y$ is the local likelihood estimator for $\phi_{k}$. The matrices $\mathbf{\Phi}_{0k},\:\mathbf{\widetilde{\Phi}}_{k}$ are conformable matrices from the vector of prior means, $\phi_{0k}$, and a draw from the quasi posterior distribution, $\widetilde{\phi}_{k}$, respectively.

The motivation for employing these methods is threefold. First, we are able to estimate large systems that conventional Bayesian estimation methods do not permit. This is typical because the state-space representation of an $N$-dimensional TVP VAR ($p$) requires an additional $N(3/2 + N(p+1/2))$ state equations for every additional variable. Conventional Markov Chain Monte Carlo (MCMC) methods fail to estimate larger models, which in general confine one to (usually) fewer than 6 variables in the system. Second, the standard approach is fully parametric and requires a law of motion. This can distort inference if the true law of motion is misspecified. Third, the methods used here permit direct estimation of the VAR's time-varying covariance matrix, which has an inverse-Wishart density and is symmetric positive definite at every point in time. 

In estimating the model, we use $p$=2 and a Minnesota Normal-Wishart prior with a shrinkage value $\varphi=0.05$ and center the coefficient on the first lag of each variable to 0.1 in each respective equation. The prior for the Wishart parameters are set following \cite*{kadiyala1997numerical}. For each point in time, we run 500 simulations of the model to generate the (quasi) posterior distribution of parameter estimates. Note we experiment with various lag lengths, $p=\{2,3,4,5\}$; shrinkage values, $\varphi=\{0.01, 0.25, 0.5\}$; and values to center the coefficient on the first lag of each variable, $\{0, 0.05, 0.2, 0.5\}$. Network measures from these experiments are qualitatively similar. Notably, adding lags to the VAR  and increasing the persistence in the prior value of the first lagged dependent variable in each equation increases computation time.

Finally, the variance decompositions of forecast errors from the VMA($\infty$) representation require the truncation of the infinite horizon with a $H$ horizon approximation. As $H\rightarrow \infty$ the error disappears \citep*{lutkepohl2005new}. We note here that $H$ serves as an approximating factor and has no interpretation in the time domain. We obtain horizon-specific measures using Fourier transforms and set our truncation horizon $H$=100. The results are qualitatively similar for $H\in \{50,100,200\}$.

\section{Modeling volatility connectedness using mutually exciting jump processes}
\label{appendix: model derivations}

\subsection{The international cash flows with mutually exciting jumps}
\label{appendix: jumps}

Denoting log aggregate consumption in a country $i$ by $y_{i,t} = \ln\left(Y_{i,t}\right),$  we define the dynamics for endowments as:
\begin{align*}
\mathrm{~d} y_{i, t} &= \mu_i \mathrm{~d} t+\sum_{j=1}^n K_{i,j} \mathrm{~d} N_{j, t} \\
\mathrm{~d} \ell_{i, t} &= \kappa_i\left(\overline{\ell}_i-\ell_{i, t}\right) \mathrm{~d} t+\sum_{j=1}^n \beta_{i, j} \mathrm{~d} N_{j, t} 
\end{align*}
We apply the approach proposed in \cite*{ES2008} to solve for the equilibrium. The vector $X_i=\left(y_i, \ell_1, \ldots, \ell_n\right)^{\prime}$ follows the affine jump process
$$
\mathrm{~d} X_{i,t}=\mu\left(X_{i,t}\right) \mathrm{~d} t+\xi_{i,t} \mathrm{~d} N_t,
$$
\begin{itemize}
\item 
$\mu\left(X_{i,t}\right)=\mathcal{M}_i+\mathcal{K}_i X_{i,t}$
$$
\text { with } \mathcal{M}_i=\left(\begin{array}{l}
\mu_i \\
\kappa_1 \bar{\ell}_1 \\
\vdots \\
\kappa_n \bar{\ell}_n
\end{array}\right) \text { and } \mathcal{K}_i=\left(\begin{array}{llll}
0 & 0 & \ldots & 0 \\
0 & -\kappa_1 & \ldots & 0 \\
\vdots & \vdots & \ddots & \vdots \\
0 & 0 & \ldots & -\kappa_n \\
\end{array}\right) \text {, }
$$
\item $\ell_t=l_0+l_1 X_{i,t}$
$$
\text { with } l_0=\left(\begin{array}{l}
0 \\
\vdots \\
0
\end{array}\right) \text { and } l_1=\left(\begin{array}{cccc}
0 & 1 & \ldots & 0\\
\vdots & \vdots & \ddots & \vdots \\
0 & 0 & \ldots & 1 
\end{array}\right),
$$
\item $\xi_{i,t}=\left(\begin{array}{lll}\xi_{i, 1, t}, & \ldots, & \xi_{i, n, t}\end{array}\right)=\left(\begin{array}{lll}K_{i,1} & \ldots & K_{i,n} \\ \beta_{1,1} & \ldots & \beta_{1, n} \\ \vdots & \ddots & \vdots \\ \beta_{n, 1} & \ldots & \beta_{n, n} \end{array}\right)$.
\end{itemize}
The jump transformation $\varrho(u)=\mathbb{E}\left[\left(e^{u^{'} \xi_{i, 1, t}}, \ldots, e^{u^{'} \xi_{i, n, t}}\right)\right]^{\prime}$ is equal to $\left(e^{u^{'} \xi_{i, 1, t}}, \ldots\right.$, $\left.e^{u^{'} \xi_{i, n, t}}\right)^{\prime}$ because the jump sizes are all constant. We define the selection vectors $\delta_{y}$ and $\delta_{\ell_{i, t}}(i=1, \ldots$, $n)$ implicitly via $\mathrm{~d} y_{i,t}=\delta_{y_i}^{\prime} \mathrm{~d} X_t$ and $\mathrm{~d} \ell_{i, t}=\delta_{\ell, i}^{\prime} \mathrm{~d} X_{i, t}$.

\subsection{Deriving equilibrium conditions}
The continuous-time version of the Euler equation can be written as:
\begin{equation}
\label{eq: ee appendix}
0=\frac{1}{\mathrm{~d} t} \mathbb{E}_t\left[\frac{d\left(e^{\ln M_{i,t}+\ln R_{i,t}}\right)}{e^{\ln M_{i,t}+\ln R_{i,t}}}\right]
\end{equation}
where $R_i$ denotes the return on the claim to aggregate consumption. The logarithm of the pricing kernel follows the dynamics:
$$
\mathrm{~d} \ln M_{i,t}=-\delta \theta \mathrm{~d} t-(1-\theta) \mathrm{~d} \ln R_{i,t}-\frac{\theta}{\psi} \mathrm{~d} y_{i,t}
$$
We apply the usual affine conjecture for the log wealth-consumption ratio
\begin{align}
v_{i,t} & =A_i+B_i^{\prime} X_{i, t} \\
\nonumber 
& =A_i+\left(0, B_{i, 1}, \ldots, B_{i, n}\right) X_{i, t} \\
\nonumber 
& =A_i+\left(B_{i, 1}, \ldots, B_{i, n}\right) \ell_{t}
\end{align}
and use the Campbell-Shiller approximation for the return on the consumption claim
$$
\mathrm{~d} \ln R_{i, t}=k_{i, v, 0} \mathrm{~d} t+k_{i, v, 1} \mathrm{~d} v_{i, t}-\left(1-k_{i, v, 1}\right) v_{i, t} \mathrm{~d} t+\mathrm{~d} y_{i, t}
$$
Combining the Campbell-Shiller approximation, the affine guess for $v_{i, t}$, and the dynamics of the log pricing kernel, we obtain:
\begin{align*}
\mathrm{~d}\left( \ln M_{i,t} + \ln R_{i,t}\right) & = \left\{-\delta \theta+\theta k_{i, v, 0}-\theta\left(1-k_{i, v, 1}\right)\left(A_i+B_i^{\prime} X_{i, t}\right)+\chi_{y_i}^{\prime}\left(\mathcal{M}_i+\mathcal{K}_i X_{i,t}\right)\right\} \mathrm{~d} t \\
& + \chi_{y_i}^{\prime} \xi_{i,t} \mathrm{~d} N_t \\
\chi_{y_i} & =\theta\left[\left(1-\frac{1}{\psi}\right) \delta_{y_i}+k_{i, v, 1} B\right] \\
& =\left(-\theta\left(\frac{1}{\psi}-1\right), \theta k_{i, v, 1} B_1, \ldots, \theta k_{i, v, 1} B_n\right)^{\prime}
\end{align*}

Let the stochastic process $Y_t$ be defined as
 $\mathrm{~d} Y_t = \mu_t \mathrm{~d}t + g_t \mathrm{~d}N_t.$
Ito's formula implies that
 $\mathrm{~d}f(Y_t) = \mu_t f^\prime (Y_t) \mathrm{~d}t + \left[f(Y_t + g_t) - f(Y_t)\right] \mathrm{~d}N_t$. Using the function $f(x) = e^x$ and applying Ito's formula to the process   $Y_t = \ln M_{i,t} + \ln R_{i,t},$ we get:
\begin{align}
\label{eq: ee with CS approximation}
\frac{\mathrm{~d} \left(e^{\ln M_{i, t}+\ln R_{i, t}}\right)}{e^{\ln M_{i,t}+\ln R_{i,t}}} & = \left\{-\delta \theta+\theta k_{i, v, 0}-\theta\left(1-k_{i, v, 1}\right)\left(A_i+B_i^{\prime} X_{i, t}\right)+\chi_{y_i}^{\prime}\left(\mathcal{M}_i+\mathcal{K}_i X_{i,t}\right)\right\} \mathrm{~d} t \\
& +\left\{e^{\chi_{y_i}^{\prime} \xi_{i,t}}-1\right\} \mathrm{~d} N_t,
\nonumber
\end{align}
where 1 is a vector of ones with length $n$. We plug expression (\ref{eq: ee with CS approximation}) into the Euler equation (\ref{eq: ee appendix}) to get a system of equations for $A_i$ and $B_i:$
\begin{align}
\label{eq: Ai Bi 1}
0 & = \theta\left[-\delta+k_{i, v, 0}-\left(1-k_{i, v, 1}\right) A_i\right]+\mathcal{M}_i^{\prime} \chi_{y_i}+l_0^{\prime}\left[\varrho\left(\chi_{y_i}\right)-1\right], \\
\label{eq: Ai Bi 2} 
0 & = \mathcal{K}_i^{\prime} \chi_{y_i}-\theta\left(1-k_{i, v, 1}\right) B_i+l_1^{\prime}\left[\varrho\left(\chi_{y_i}\right)-1\right].
\end{align}
We have two additional equations for the log-linearization constants $k_{i, \nu, 0}$ and $k_{i, \nu, 1}:$
\begin{align}
\label{eq: ki 1}
0 & = -k_{i, v, 0}-\ln k_{i, v, 1}+\left(1-k_{i, v, 1}\right)\left[A_i+B_i^{\prime} \mu_X\right],
\\
\label{eq: ki 2}
0 & = A_i+B_i^{\prime} \mu_X-\ln \left(k_{i, v, 1}\right)+\ln \left(1-k_{i, v, 1}\right),
\end{align}
where $\mu_X$ is a vector with the $i$th component $\mathbb{E}\left[X_i\right]$ if that expectation is finite and 0 otherwise. Due to the presence of the mutually exciting jump terms, the long-run means $\overline{\overline{\ell_i}}$, that is, the unconditional expectations, are not equal to the respective mean reversion levels $\overline{\ell_i}$, as it would be the case, for example, for a standard square-root process. According to \cite{ait2015modeling}, the $\overline{\overline{\ell_i}}$ are the solution to the following system of equations:
$$
\overline{\overline{\ell_i}}=\frac{\kappa_i \bar{\ell}_i+\sum_{i \neq i} \beta_{i, j} \overline{\overline{\ell_i}}}{\kappa_i-\beta_{i, i}} \quad(i=1, \ldots, n).
$$
We assume $\kappa_i > \beta_{i,i}$ for $i = 1, \ldots , n$ to ensure that all the $\overline{\overline{\ell_i}}$ are positive.


\subsection{Deriving pricing kernel}
Combining the Campbell-Shiller approximation, the affine guess for $v_{i,t},$ and the dynamics of the log pricing kernel, we obtain:
\begin{align}
\nonumber
\mathrm{~d} \ln M_{i,t} & = \Big\{-\delta \theta - (1 - \theta) k_{i, v, 0} + (1 - \theta)\left(1-k_{i, v, 1}\right) \underbrace{v_{i,t}}_{A_i + B_i^{\prime} X_{i,t}}\Big\} \mathrm{~d} t \\
\nonumber
& - (1 - \theta) k_{i,v,1}  \underbrace{\mathrm{~d} v_{i,t}}_{= B_i^{\prime} \mathrm{~d} X_{i, t}} -  \underbrace{\Big[(1 - \theta) + \frac{\theta}{\psi}\Big]}_{= ~ \gamma}  \underbrace{\mathrm{~d} y_{i,t}}_{= ~ \delta_{y_i}^{\prime} \mathrm{~d} X_{i,t}} \\
\nonumber
& = \Big\{-\delta \theta - (1 - \theta) k_{i, v, 0} + (1 - \theta)\left(1-k_{i, v, 1}\right)\left(A_i+B_i^{\prime} X_{i, t}\right)\Big\} \mathrm{~d} t \\
\nonumber
& - \underbrace{\Big(\gamma \delta_{y_i} + (1 - \theta) k_{i,v,1} B_i\Big)^{\prime}}_{= \lambda_i^{\prime}} \underbrace{\mathrm{~d} X_{i, t}}_{=\mu\left(X_{i,t}\right) \mathrm{~d} t+\xi_{i,t} \mathrm{~d} N_t = (\mathcal{M}_i+\mathcal{K}_i X_{i,t})\mathrm{~d} t+\xi_{i,t} \mathrm{~d} N_t} \\
\nonumber
& = \Big\{-\delta \theta - (1 - \theta) k_{i, v, 0} + (1 - \theta)\left(1-k_{i, v, 1}\right)\left(A_i+B_i^{\prime} X_{i, t}\right) - \lambda_i^{\prime} (\mathcal{M}_i+\mathcal{K}_i X_{i,t})\mathrm{~d} t \Big\} \mathrm{~d} t \\
\label{eq: d ln M}
& - \lambda_i^{\prime} \xi_{i,t} \mathrm{~d} N_t \\
\nonumber 
\text{where} & \quad \lambda_i = \gamma \delta_{y_i}+(1-\theta) k_{i, v, 1} B_i = \left(\gamma,(1-\theta) k_{i, \nu, 1} B_{i,1}, \ldots,(1-\theta) k_{i,\nu, 1} B_{i,n}\right)^{\prime}
\end{align}
Using the function $f(x) = e^x$ and applying Ito's formula to the process $Y_t = \ln M_{i,t},$ we get:
\begin{align*}
\frac{\mathrm{~d} \left(e^{\ln M_{i, t}}\right)}{e^{\ln M_{i,t}}} & = \Big\{-\delta \theta - (1 - \theta) \underbrace{k_{i, v, 0}}_{= - \ln k_{i, v, 1}+\left(k_{i, v, 1}-1\right) [A_i + B_i^{\prime} \mu_X]} + (1 - \theta)\left(1-k_{i, v, 1}\right)\left(A_i+B_i^{\prime} X_{i, t}\right) \\ - \lambda_i^{\prime} (\mathcal{M}_i & + \mathcal{K}_i X_{i,t})\mathrm{~d} t \Big\} \mathrm{~d} t + \underbrace{\left\{e^{-\lambda_i^{\prime} \xi_{i,t}}-1\right\}}_{= -[1-\varrho(-\lambda_i)]^{\prime}} \mathrm{~d} N_t + \underbrace{0}_{
\begin{array}{ll}
= -[1-\varrho(-\lambda_i)]^{\prime}\ell_t  \mathrm{~d} t + [1-\varrho(-\lambda_i)]^{\prime}\ell_t  \mathrm{~d} t \\
= -[1-\varrho(-\lambda_i)]^{\prime}\left(\ell_0 + \ell_1 X_{i,t}\right)  \mathrm{~d} t + [1-\varrho(-\lambda_i)]^{\prime}\ell_t  \mathrm{~d} t 
\end{array}} \\
& = - \underbrace{\Big\{\theta \delta+(\theta-1)\left[\ln k_{i, v, 1}+\left(k_{i, v, 1}-1\right) B_i^{\prime} \mu_X\right]+\mathcal{M}_i^{\prime} \lambda_i + l_0^{\prime}[1 - \varrho(-\lambda_i)]\Big\}}_{\Phi_{i,0}} \mathrm{~d} t \\
& - \underbrace{\Big\{(1-\theta)\left(k_{i, v, 1}-1\right) B_i+\mathcal{K}_i^{\prime} \lambda_i + l_1^{\prime}[1 - \varrho(-\lambda_i)]\Big\}^{\prime}}_{\Phi_{i,1}^{\prime}} X_{i,t}\mathrm{~d} t \\
& -[1-\varrho(-\lambda_i)]^{\prime} \left(\mathrm{~d} N_t - \ell_t \mathrm{~d} t\right).
\end{align*}
Thus, the pricing kernel has dynamics:
$$
\frac{\mathrm{~d} M_{i,t}}{M_{i,t}}=-r_{i,t} \mathrm{~d} t-[1-\varrho(-\lambda_i)]^{\prime}\left(\mathrm{~d} N_t-\ell_t \mathrm{~d} t\right)
$$
where the risk-free rate is given as
\begin{align*}
r_{i,t} & = \Phi_{i,0}+\Phi_{i,1}^{\prime} X_{i,t} \\
\text{with} \quad \Phi_{i,0} & =\theta \delta+(\theta-1)\left[\ln k_{i, v, 1}+\left(k_{i, v, 1}-1\right) B_i^{\prime} \mu_X\right]+\mathcal{M}_i^{\prime} \lambda_i + l_0^{\prime}[1 - \varrho(-\lambda_i)] \\
\Phi_{i,1} & =(1-\theta)\left(k_{i, v, 1}-1\right) B_i+\mathcal{K}_i^{\prime} \lambda_i + l_1^{\prime}[1 - \varrho(-\lambda_i)]\\
\lambda_i & = \gamma \delta_{y_i}+(1-\theta) k_{i, v, 1} B_i \\
& = \left(\gamma,(1-\theta) k_{i, \nu, 1} B_{i,1}, \ldots,(1-\theta) k_{i, \nu, 1} B_{i,n}\right)^{\prime}.
\end{align*}
The market prices of jump risk are given as:
\begin{equation}
\label{eq: mpjr}
\mathrm{MPJR}_i = \left(\begin{array}{c}
\mathrm{MPJR}_{i,1} \\
\vdots \\
\mathrm{MPJR}_{i,n}
\end{array}\right) =[1-\varrho(-\lambda_i)] 
\end{equation}
\begin{equation*}
=\left(\begin{array}{c}
1-\exp \left(-\gamma K_{i,1}+k_{i, v, 1}(\theta-1)\left[B_{i, 1} \beta_{1,1}+\cdots+B_{i, n} \beta_{n, 1}\right]\right) \\
\vdots \\
1-\exp \left(-\gamma K_{i,n}+k_{i, v, 1}(\theta-1)\left[B_{i, 1} \beta_{1, n}+\cdots+B_{i, n} \beta_{n, n}\right]\right)
\end{array}\right).
\end{equation*}

\subsection{Approximating pricing kernel}
\label{appendix: dynamic model}

\subsubsection{First approximation step}
\label{appendix: first approximation step}

Assuming $\kappa_1=\cdots=\kappa_n=\kappa$ and $K_{i,1}=\cdots=K_{i,n}=K_i,$ we can rewrite Equation (\ref{eq: Ai Bi 2}) as the following system of equations:
$$
\begin{gathered}
0=B_{i, 1} \theta\left[k_{i, v, 1}(1-\kappa)-1\right]+\exp \left\{K_i(1-\gamma)+\theta k_{i, v, 1}\left(B_{i, 1} \beta_{1,1}+\cdots+B_{i, n} \beta_{n, 1}\right)\right\}-1 \\
\vdots \\
0=B_{i, n} \theta\left[k_{i, v, 1}(1-\kappa)-1\right]+\exp \left\{K_i(1-\gamma)+\theta k_{i, v, 1}\left(B_{i, 1} \beta_{i, 1, n}+\cdots+B_{i, n} \beta_{n, n}\right)\right\}-1
\end{gathered}
$$
and translate this into matrix notation:
$$
1=\theta\left[k_{i, v, 1}(1-\kappa)-1\right] B_i+\exp \{K_i(1-\gamma)\} \exp \left\{\theta k_{i, \nu, 1} \beta^{\prime} B_i\right\},
$$
where now and in the following, the "exp" operator, applied to a vector, stands for element-wise application of the "exp" operator to the vector.

Next, we apply the approximation $\exp (x)=1+x+O\left(x^2\right)$ and solve for $B_i:$
\begin{equation}
\label{eq: bi approx}
B_i=\left(I_{n \times n}+\frac{\exp \{K_i(1-\gamma)\}}{1-\kappa-\frac{1}{k_{i, v, 1}}} \beta^{\prime}\right)^{-1} \frac{1}{\theta\left[k_{i, v, 1}(1-\kappa)-1\right]}[1-\exp \{K_i(1-\gamma)\}]+O\left(\beta^2\right),
\end{equation}
where $I_{n \times n}$ denotes an $n \times n$ identity matrix and $\frac{\exp \{K_i(1-\gamma)\}}{1-\kappa-\frac{1}{k_{i, v, 1}}}<0$ since $\frac{1}{k_{i, v, 1}}>1-\kappa$ (due to $\frac{1}{k_{i, v, 1}}=\frac{1+\overline{e^v_i}}{\overline{e^v_i}}>1>1-\kappa$ for $0<\kappa<1$ ).
To conclude the first approximation step, we define:
\begin{equation}
\label{eq: bi star}
B^*_i=\left(I_{n \times n}+\frac{\exp \{K_i(1-\gamma)\}}{1-\kappa-\frac{1}{k_{i, v, 1}}} \beta \prime\right)^{-1} \frac{1}{\theta\left[k_{i, v, 1}(1-\kappa)-1\right]}[1-\exp \{K_i(1-\gamma)\}].
\end{equation}

\subsubsection{Second approximation step}
\label{appendix: second approximation step}

Since the inverse term in Equation (\ref{eq: bi approx}) has the structure of a Leontief inverse, $(I-A)^{-1}=I+A^1+A^2+\cdots$, we rewrite Equation (\ref{eq: bi approx}) as:
\begin{equation*}
B_i=\left[I_{n \times n} - \frac{\exp \{K_i(1-\gamma)\}}{1-\kappa-\frac{1}{k_{i, v, 1}}} \beta^{\prime} + \left(\frac{\exp \{K_i(1-\gamma)\}}{1-\kappa-\frac{1}{k_{i, v, 1}}} \beta^{\prime}\right)^2 - ... \right] \times
\end{equation*}
\begin{equation}
\label{eq: bi approx second step}
\times \frac{1}{\theta\left[k_{i, v, 1}(1-\kappa)-1\right]}[1-\exp \{K_i(1-\gamma)\}]+O\left(\beta^2\right)
\end{equation}
\begin{equation*}
= \left[I_{n \times n} - \frac{\exp \{K_i(1-\gamma)\}}{1-\kappa-\frac{1}{k_{i, v, 1}}} \beta^{\prime}\right] \times \frac{1}{\theta\left[k_{i, v, 1}(1-\kappa)-1\right]}[1-\exp \{K_i(1-\gamma)\}]+O\left(\beta^2\right).
\end{equation*}
To conclude the second approximation step, we define:
\begin{equation}
\label{eq: bi star star}
B_i^{**}= \left[I_{n \times n} - \frac{\exp \{K_i(1-\gamma)\}}{1-\kappa-\frac{1}{k_{i, v, 1}}} \beta^{\prime}\right] \times \frac{1}{\theta\left[k_{i, v, 1}(1-\kappa)-1\right]}[1-\exp \{K_i(1-\gamma)\}].
\end{equation}
Plugging Equation (\ref{eq: bi approx second step}) into the market price of risk given by Equation (\ref{eq: mpjr}) and rewriting this in matrix notation yields:
\begin{equation*}
\begin{split}
\mathrm{MPJR}_i & = 1 - \exp\left[-\gamma K_i  - \frac{k_{i, v, 1} (\theta - 1)}{\theta [k_{i, v, 1} (1 - \kappa) - 1]} \times \Big[\beta^{\prime}[1-\exp \{K_i(1-\gamma)\}]\Big]+O\left(\beta^2\right) \right] \\
& = 1 - \exp\left[-\gamma K_i  + \frac{(\theta - 1) (1-\exp \{K_i(1-\gamma)\}) }{\theta \left[(1 - \kappa) - \frac{1}{k_{i, v, 1}}\right]} \beta^{\prime} 1 + O\left(\beta^2\right) \right] \\
& = 1 - \exp\left[\mathcal{A}_i  + \mathcal{B}_i \beta^{\prime} 1 + O\left(\beta^2\right) \right],
\end{split}
\end{equation*}
\begin{equation}
\mathcal{A}_i = -\gamma K_i \quad \& \quad \mathcal{B}_i =  \frac{(\theta - 1) (1-\exp \{K_i(1-\gamma)\}) }{\theta \left[(1 - \kappa) - \frac{1}{k_{i, v, 1}}\right]}.
\end{equation}

\subsubsection{Deriving real exchange rate}
\label{appendix: deriving real exchange rate}

For a country $i,$ we define the real exchange rate as the ratio of SDF in this country $(M_{i,t})$ and the US $(M_{1,t}):$
\begin{equation}
\label{eq: Q}
Q_{i,t} = \frac{M_{i,t}}{M_{1,t}}.    
\end{equation}
Using the log pricing kernel given by Equation (\ref{eq: d ln M}), we obtain:
\begin{align*}
\mathrm{~d} \left(\ln \frac{M_{i,t}}{M_{1,t}}\right) 
& = \Big\{-\delta \theta - (1 - \theta) k_{i, v, 0} + (1 - \theta)\left(1-k_{i, v, 1}\right)\left(A_i+B_i^{\prime} X_{i, t}\right) - \lambda_i^{\prime} (\mathcal{M}_i+\mathcal{K}_i X_{i,t})\mathrm{~d} t \Big\} \mathrm{~d} t \\
& - \Big\{-\delta \theta - (1 - \theta) k_{1, v, 0} + (1 - \theta)\left(1-k_{1, v, 1}\right)\left(A_1+B_1^{\prime} X_{1, t}\right) - \lambda_i^{\prime} (\mathcal{M}_1+\mathcal{K}_i X_{1,t})\mathrm{~d} t \Big\} \mathrm{~d} t \\
& - \left(\lambda_i^{\prime} \xi_{i,t} - \lambda_1^{\prime} \xi_{1,t}\right)  \mathrm{~d} N_t.
\end{align*}
Using the function $f(x) = e^x$ and applying Ito's formula to the process $Y_t = \ln \frac{M_{i,t}}{M_{1,t}},$ we get:
\begin{align*}
\frac{\mathrm{~d} \left(e^{\ln \frac{M_{i, t}}{M_{1, t}}}\right)}{e^{\ln \frac{M_{i, t}}{M_{1, t}}}} & = - (r_{i,t} - r_{1,t}) \mathrm{~d} t + [1-\varrho(-\lambda_i)]^{\prime} \ell_t \mathrm{~d} t - [1-\varrho(-\lambda_1)]^{\prime}\ell_t \mathrm{~d} t - [1-\varrho(-\lambda_i) \varrho(\lambda_1)]^{\prime} \mathrm{~d} N_t  \\
&  = - (r_{i,t} - r_{1,t}) \mathrm{~d} t + [1-\varrho(-\lambda_i)]^{\prime} \ell_t \mathrm{~d} t - [1-\varrho(-\lambda_1)]^{\prime}\ell_t \mathrm{~d} t - [1-\varrho(-\lambda_i) \varrho(\lambda_1)]^{\prime} \ell_t \mathrm{~d} t \\
& - [1-\varrho(-\lambda_i) \varrho(\lambda_1)]^{\prime} \left(\mathrm{~d} N_t - \ell_t \mathrm{~d} t\right) \\
&  = - (r_{i,t} - r_{1,t}) \mathrm{~d} t + \Big\{[1-\varrho(\lambda_1)] \cdot [\varrho(-\lambda_1) - \varrho(-\lambda_i)]\Big\}^{\prime}\ell_t \mathrm{~d} t \\
& - \Big\{\varrho(\lambda_1) \cdot [\varrho(-\lambda_1)-\varrho(-\lambda_i)]\Big\}^{\prime} \left(\mathrm{~d} N_t - \ell_t \mathrm{~d} t\right).
\end{align*}
where the operator $"\cdot"$ denotes an element-wise product. Thus, the real exchange rate has dynamics:
\begin{align*}
\frac{\mathrm{~d} Q_{i,t}}{Q_{i,t}} & = \mu_{i,t} \mathrm{~d} t + \theta_{i} \left(\mathrm{~d} N_t - \ell_t \mathrm{~d} t\right) = \mu_{i,t} \mathrm{~d} t + \sum_{k=1}^n \theta_{i,k} \left(\mathrm{~d} N_{k,t} - \ell_{k,t} \mathrm{~d} t\right), \\
\mu_{i,t} &  = - (r_{i,t} - r_{1,t}) + \Big\{\mathrm{MPJR}_1 \cdot \theta_{i}\Big\}\ell_t = - (r_{i,t} - r_{1,t}) + \sum_{k=1}^n \mathrm{MPJR}_{1,k}  \theta_{i,k} \ell_{k,t}, \\
\theta_{i,k} &  =  - [1-\varrho(-\lambda_{i,k})\varrho(\lambda_{1,k})] = -\left[1 - \exp\left[(\mathcal{A}_i - \mathcal{A}_1)  + (\mathcal{B}_i - \mathcal{B}_1) \sum_{j=1}^n \beta_{j,k}\right]\right].
\end{align*}

\subsubsection{Propositions} 
This section formulates the key propositions used in the main text of the paper and provides their proofs.

\textbf{Proposition 1.} 
\begin{itemize}
\item[1.]
The stochastic discount factor in a country 
$i$
satisfies:
\begin{equation}
\frac{\mathrm{~d} M_{i,t}}{M_{i,t}}=-r_{i,t} \mathrm{~d} t-\mathrm{MPJR}_i \left(\mathrm{~d} N_t-\ell_t \mathrm{~d} t\right) = -r_{i,t} \mathrm{~d} t - \sum_{k=1}^n\mathrm{MPJR}_{i,k} \left(\mathrm{~d} N_{k,t}-\ell_{k,t} \mathrm{~d} t\right),
\end{equation}
where 
$r_{i,t}$
is the risk-free rate, 
$\mathrm{MPJR}_i = 1-\varrho(-\lambda_i)$
is the row vector of market prices of jump risk whose individual elements are given by:
\begin{equation}
\label{eq: rho appendix}
\mathrm{MPJR}_{i,k} = 1- \rho(-\lambda_{i,k}) = 1 - \exp\left[\mathcal{A}_i  + \mathcal{B}_i \sum_{j=1}^n \beta_{j,k}\right],
\end{equation}
\begin{equation}
\label{eq: A B appendix}
\mathcal{A}_i = -\gamma K_i \quad \& \quad \mathcal{B}_i =  \frac{(\theta - 1) (1-\exp \{K_i(1-\gamma)\}) }{\theta \left[(1 - \kappa) - \frac{1}{k_{i, v, 1}}\right]}.
\end{equation}
\item[2.]  $\mathcal{A}_i >0 ~ \wedge ~ \mathcal{B}_i > 0 ~ \wedge ~ \mathrm{MPJR}_{i,k} <0.$
\item[3.]  $|K_j|>|K_i| ~ \Rightarrow ~ \mathcal{A}_j > \mathcal{A}_i ~ \wedge ~ \mathcal{B}_j > \mathcal{B}_i ~ \wedge ~ |\mathrm{MPJR}_{j}| > |\mathrm{MPJR}_{i}|.$
\item[4.]  $ \sum_{j=1}^n \beta_{j,k} > \sum_{j=1}^n \beta_{j,l} ~ \Rightarrow ~ |\mathrm{MPJR}_{i,k}| > |\mathrm{MPJR}_{i,l}| ~ \forall i.$
\item[] Proof: 
\item[1.] Appendices \ref{appendix: jumps}-\ref{appendix: dynamic model} provide the pricing kernel derivations.
\item[2.]  Using $\gamma > 1, ~ \theta < 0, ~ 0 < \kappa < 1, ~ K_i < 0, ~ \frac{1}{k_{i, v, 1}}=\frac{1+\overline{e^{v_i}}}{\overline{e^{v_i}}}>1>1-\kappa,$ Equation (\ref{eq: A B appendix}) implies $\mathcal{A}_i >0$ and $\mathcal{B}_i > 0.$
\item[3.] Assume $|K_j|>|K_i|,$ then Equation (\ref{eq: A B appendix}) immediately implies $\mathcal{A}_j > \mathcal{A}_i.$ For the second inequality, note that the log-linearization constants $k_{i, \nu, 0}, k_{i, \nu, 1},$ and $\mathcal{B}_i$ are implicitly defined by Equations (\ref{eq: ki 1}), (\ref{eq: ki 2}) and (\ref{eq: A B appendix}). The latter two equations also imply a positive relationship between the parameters. Specifically, $k_{i, \nu, 1}$ increases with $\mathcal{B}_i.$ Consequently, Equation (\ref{eq: A B appendix}) implies that $k_{j, \nu, 1} > k_{i, \nu, 1}$ and $\mathcal{B}_j > \mathcal{B}_i$ must hold. Equation (\ref{eq: rho appendix}) implies the last inequality.
\item[4.] Equation (\ref{eq: rho appendix}) and the second statement of this proposition yield the result.
\end{itemize}

\textbf{Proposition 2.} 
\begin{itemize}
\item[1.]
The real exchange rate has dynamics:
\begin{align*}
\frac{\mathrm{~d} Q_{i,t}}{Q_{i,t}} & = \mu_{i,t} \mathrm{~d} t + \theta_{i} \left(\mathrm{~d} N_t - \ell_t \mathrm{~d} t\right) = \mu_{i,t} \mathrm{~d} t + \sum_{k=1}^n \theta_{i,k} \left(\mathrm{~d} N_{k,t} - \ell_{k,t} \mathrm{~d} t\right), \\
\mu_{i,t} &  = - (r_{i,t} - r_{1,t}) + \Big\{\mathrm{MPJR}_1 \cdot \theta_{i}\Big\}\ell_t = - (r_{i,t} - r_{1,t}) + \sum_{k=1}^n \mathrm{MPJR}_{1,k}  \theta_{i,k} \ell_{k,t}, \\
\theta_{i,k} &  =  - [1-\varrho(-\lambda_{i,k})\varrho(\lambda_{1,k})] = -\left[1 - \exp\left[(\mathcal{A}_i - \mathcal{A}_1)  + (\mathcal{B}_i - \mathcal{B}_1) \sum_{j=1}^n \beta_{j,k}\right]\right],
\end{align*}
where $r_{i,t}$ and $r_{1,t}$ are the risk-free rates in a country $i$ and US, and the operator $"\cdot"$ denotes an element-wise product.
\item[2.]  $|K_i|>|K_1| ~ \Rightarrow ~ \theta_{i,k} > 0 ~  \wedge ~ |K_1|>|K_i| ~ \Rightarrow ~ \theta_{i,k} < 0.$
\item[3.] $|K_j|>|K_i|>|K_1| ~ \vee ~ |K_1|>|K_i|>|K_j| ~ \Rightarrow ~ |\theta_{j,k}| > |\theta_{i,k}|.$
\item[4.] $\sum_{j=1}^n \beta_{j,k} > \sum_{j=1}^n \beta_{j,l} ~ \Rightarrow ~ |\theta_{i,k}| > |\theta_{i,k}| ~ \forall i.$
\item[] Proof: 
\item[1.] Appendix \ref{appendix: deriving real exchange rate} provides the real exchange rate derivations.
\item[2, 3, 4.] The first and third statements in Proposition 1 yield the results.
\end{itemize}

\subsection{Static model}
\label{appendix: static model}

The main text of the paper considers a static version of the model with unconstrained jump sizes. Specifically, we assume the dynamics for endowments as:
\begin{equation}
\label{eq: dy it main static}
\mathrm{~d} y_{i, t} = \mu \mathrm{~d} t+\sum_{j=1}^n K_{i,j} \mathrm{~d} N_{j, t},
\end{equation}
where 
$\mu$ 
is a constant drift, $N_{j,t}$ denotes jumps in a country $j$ whose impact on cash flows in a country $i$ is constant $K_{i,j}.$ For simplicity, we assume that all jump intensities are constant $\overline{\ell}.$ The coefficients $K_{i,j}$ can be positive or negative, but we assume $K_{i,j}<0$ for a clearer exposition. 

A static model does not require approximating the pricing kernel, which can be obtained in a closed form. Note that imposing the constraint $\beta_{i,j} = 0$ in Section \ref{appendix: first approximation step} immediately provides the solution for the vector $B_i.$ As a result, we obtain the analougous solution to Equation (\ref{eq: bi star star}):
\begin{equation}
B_{i,k}^{**}= \frac{1}{\theta\left[k_{i, v, 1}(1-\kappa)-1\right]}[1-\exp \{K_{i,k}(1-\gamma)\}].
\end{equation}
Similarly, the market prices of jump risk are given by:
\begin{equation*}
\mathrm{MPJR}_{i,k} = 1 - \exp\left[-\gamma K_{i,k}\right].
\end{equation*}
Using these identities, we obtain simplified versions of Propositions 1 and 2 in the main text.

\section{Additional exercises}
\label{section: additional exercises appendix}

This section reports additional results not included in the main text. Table \ref{tab: from-directional volatility connectedness portfolios} shows the results for from-directional volatility connectedness portfolios. Table \ref{tab: equity and hedge fund factors} presents time-series regression results of to-directional volatility connectedness portfolios on equity and hedge fund factors. Table \ref{tab: from-directional portfolios and benchmark strategies} time-series regression results of from-directional volatility connectedness portfolios on common currency factors. Tables \ref{tab: summary stats and alphas for two-month implied volatilities} and \ref{tab: diversification benefits for two-month implied volatilities} replicate the main results for connectedness measures based on two-month implied volatilities. Table \ref{table: allocation analysis from-directional volatility} presents the allocation analysis of from-directional volatility connectedness portfolios.

\clearpage
\begin{table}[t!]
\centering
\caption{From-directional volatility connectedness portfolios.} 
\begin{minipage}{\textwidth} 
\footnotesize
This table presents descriptive statistics for quintile $(\mathcal{P}_i: i = 1,\dots,5)$ and long-short $(\mathcal{P}_{5-1})$ portfolios sorted by from-directional volatility connectedness measures based on volatility shocks that exclude contemporaneous correlations. Panels A, B, and C report the results for shocks of any, short-, and long-term persistence. $\mathcal{P}_1 (\mathcal{P}_5)$ comprises currencies with the lowest (highest) levels of received volatility. $\mathcal{P}_{5-1}$ buys $\mathcal{P}_5$ and sells $\mathcal{P}_1.$ Mean, standard deviation, and Sharpe ratio are annualized, but the \cite{newey1987simple} t-statistic of mean, skewness, kurtosis, and the first-order autocorrelation are based on monthly returns. We also report the annualized mean of the exchange rate ($\text{fx} = \Delta s^k $) and interest rate ($\text{ir} = i^k - i$) components of excess returns and the average to- and from-directional volatility connectedness measures of portfolios. The sample spans March 2005 to December 2021, which corresponds to the out-of-sample estimation period.
\end{minipage}
\vspace{\medskipamount}
\footnotesize

\begin{tabular}{llD{.}{.}{4.2}D{.}{.}{4.2}D{.}{.}{4.2}D{.}{.}{4.2}D{.}{.}{4.2}D{.}{.}{4.2}}
\toprule
&       & \multicolumn{1}{r}{$\mathcal{P}_1$} & \multicolumn{1}{r}{$\mathcal{P}_2$} & \multicolumn{1}{r}{$\mathcal{P}_3$} & \multicolumn{1}{r}{$\mathcal{P}_4$} & \multicolumn{1}{r}{$\mathcal{P}_5$} & \multicolumn{1}{r}{$\mathcal{P}_{5-1}$} \\
\midrule
\multicolumn{8}{l}{\textbf{Panel A: $\mathcal{F}$-sorted portfolios}} \\
\midrule
mean (\%) &       & -0.39 & -1.89 & -1.25 & 0.27  & 1.63  & 2.02 \\
t-stat &       & -0.28 & -0.97 & -0.58 & 0.11  & 0.60  & 1.13 \\
\midrule
\multicolumn{1}{r}{fx (\%)} &       & -0.26 & -2.10 & -2.08 & -1.24 & -4.40 & -4.14 \\
\multicolumn{1}{r}{ir (\%)} &       & -0.13 & 0.21  & 0.84  & 1.51  & 6.03  & 6.16 \\
\multicolumn{1}{r}{$\mathcal{T}$} &       & 0.18  & 0.18  & 0.19  & 0.21  & 0.21  & 0.03 \\
\multicolumn{1}{r}{$\mathcal{F}$} &       & 0.17  & 0.23  & 0.27  & 0.31  & 0.42  & 0.25 \\
\midrule
Sharpe &       & -0.06 & -0.25 & -0.13 & 0.03  & 0.13  & 0.22 \\
std (\%) &       & 6.09  & 7.46  & 9.89  & 10.31 & 12.72 & 9.30 \\
skew  &       & -0.33 & -0.19 & -0.57 & -0.58 & -0.77 & -0.77 \\
kurt  &       & 6.23  & 3.84  & 5.41  & 5.27  & 4.26  & 5.24 \\
ac1   &       & 0.01  & 0.02  & 0.07  & 0.00  & 0.04  & -0.01 \\
\midrule
\multicolumn{8}{l}{\textbf{Panel B: $\mathcal{F}(S)$-sorted portfolios}} \\
\midrule
mean (\%) &       & -0.28 & -2.20 & -0.70 & -0.09 & 1.64  & 1.93 \\
t-stat &       & -0.20 & -1.13 & -0.33 & -0.04 & 0.60  & 1.06 \\
\midrule
\multicolumn{1}{r}{fx (\%)} &       & -0.12 & -2.41 & -1.46 & -1.61 & -4.48 & -4.36 \\
\multicolumn{1}{r}{ir (\%)} &       & -0.16 & 0.21  & 0.76  & 1.52  & 6.13  & 6.29 \\
\multicolumn{1}{r}{$\mathcal{T}(S)$} &       & 0.15  & 0.15  & 0.15  & 0.17  & 0.17  & 0.02 \\
\multicolumn{1}{r}{$\mathcal{F}(S)$} &       & 0.13  & 0.18  & 0.21  & 0.25  & 0.34  & 0.20 \\
\midrule
Sharpe &       & -0.05 & -0.29 & -0.07 & -0.01 & 0.13  & 0.21 \\
std (\%) &       & 6.05  & 7.58  & 9.61  & 10.46 & 12.73 & 9.33 \\
skew  &       & -0.34 & -0.34 & -0.32 & -0.55 & -0.80 & -0.78 \\
kurt  &       & 6.25  & 4.75  & 4.36  & 5.07  & 4.35  & 5.14 \\
ac1   &       & 0.02  & 0.01  & 0.07  & -0.01 & 0.04  & -0.01 \\
\midrule
\multicolumn{8}{l}{\textbf{Panel C: $\mathcal{F}(L)$-sorted portfolios}} \\
\midrule
mean (\%) &       & -0.43 & -2.25 & -1.03 & 0.38  & 1.70  & 2.13 \\
t-stat &       & -0.29 & -1.15 & -0.46 & 0.17  & 0.64  & 1.12 \\
\midrule
\multicolumn{1}{r}{fx (\%)} &       & -0.40 & -2.67 & -1.98 & -1.40 & -3.65 & -3.24 \\
\multicolumn{1}{r}{ir (\%)} &       & -0.03 & 0.41  & 0.95  & 1.78  & 5.35  & 5.38 \\
\multicolumn{1}{r}{$\mathcal{T}(L)$} &       & 0.02  & 0.02  & 0.02  & 0.02  & 0.03  & 0.01 \\
\multicolumn{1}{r}{$\mathcal{F}(L)$} &       & 0.02  & 0.03  & 0.03  & 0.04  & 0.06  & 0.04 \\
\midrule
Sharpe &       & -0.07 & -0.30 & -0.11 & 0.04  & 0.13  & 0.23 \\
std (\%) &       & 6.23  & 7.41  & 9.60  & 10.43 & 12.80 & 9.37 \\
skew  &       & -0.20 & -0.21 & -0.36 & -0.62 & -0.72 & -0.81 \\
kurt  &       & 5.09  & 3.91  & 4.51  & 5.59  & 4.21  & 5.15 \\
ac1   &       & -0.02 & 0.03  & 0.09  & -0.02 & 0.01  & -0.01 \\
\bottomrule
\end{tabular}%
\label{tab: from-directional volatility connectedness portfolios}%
\end{table}%

\clearpage
\begin{table}[t!]
\centering
\caption{To-directional volatility connectedness, equity, and hedge factors}
\begin{minipage}{\textwidth} 
This table presents time-series regression outputs of the excess returns of the to-directional volatility connectedness strategy on the equity and hedge fund factors, estimated one factor at a time. For the equity factors, we consider five Fama-French factors --- market (MKT), size (SMB), value (HML), profitability (RMW), investment (CMA) --- and momentum (MOM), which are constructed for the U.S., developed, and emerging markets. For the hedge fund factors, we consider Fung-Hsieh factors --- bond (BOND), currency (CURR), and commodity (COMM) trend-following factors, bond market (BMKT), and bond size spread (CSPREAD) factors. The values reported in the ``$\alpha$ (\%)'' column are expressed in percentage per annum. The t-statistics are based on \cite{newey1987simple} standard errors. The sample spans March 2005 to December 2021, which corresponds to the out-of-sample estimation period for volatility-connectedness-sorted portfolios.
\end{minipage}
\vspace{\medskipamount}
\footnotesize

\begin{tabular}{llD{.}{.}{4.2}D{.}{.}{4.2}D{.}{.}{4.2}D{.}{.}{4.2}D{.}{.}{4.2}}
\toprule
\multicolumn{1}{r}{ } &       & \multicolumn{1}{r}{$\alpha$ (\%)} & \multicolumn{1}{r}{t-stat} & \multicolumn{1}{r}{$\beta$} & \multicolumn{1}{r}{t-stat} & \multicolumn{1}{r}{$R^2$ (\%)} \\
\midrule
MKT   &       & 5.28  & 3.41  & -0.08 & -2.46 & 2.69 \\
SMB   &       & 4.44  & 2.74  & 0.03  & 0.39  & 0.10 \\
HML   &       & 4.38  & 2.63  & -0.03 & -0.58 & 0.20 \\
RMW   &       & 4.10  & 2.49  & 0.10  & 1.01  & 0.65 \\
CMA   &       & 4.46  & 2.72  & -0.02 & -0.20 & 0.02 \\
MOM   &       & 4.47  & 2.75  & -0.02 & -0.59 & 0.12 \\
\midrule
MKT (D) &       & 5.26  & 3.44  & -0.09 & -3.15 & 4.12 \\
SMB (D) &       & 4.46  & 2.74  & 0.01  & 0.08  & 0.00 \\
HML (D) &       & 4.30  & 2.58  & -0.09 & -1.13 & 0.78 \\
RMW (D) &       & 3.44  & 1.98  & 0.25  & 1.88  & 1.94 \\
CMA (D) &       & 4.46  & 2.73  & -0.01 & -0.07 & 0.00 \\
MOM (D) &       & 4.54  & 2.72  & -0.02 & -0.50 & 0.07 \\
\midrule
MKT (E) &       & 4.99  & 3.15  & -0.06 & -2.64 & 2.86 \\
SMB (E) &       & 4.42  & 2.68  & 0.08  & 0.94  & 0.40 \\
HML (E) &       & 4.87  & 2.71  & -0.09 & -1.21 & 0.61 \\
RMW (E) &       & 4.01  & 2.37  & 0.22  & 1.99  & 1.71 \\
CMA (E) &       & 4.51  & 2.77  & -0.02 & -0.19 & 0.02 \\
MOM (E) &       & 4.89  & 2.82  & -0.05 & -0.97 & 0.40 \\
\midrule
Bond  &       & 4.57  & 2.81  & 0.01  & 1.95  & 1.16 \\
Curr  &       & 4.58  & 2.76  & 0.01  & 1.36  & 0.63 \\
Comm  &       & 4.47  & 2.75  & 0.02  & 1.60  & 1.21 \\
Bmkt  &       & 4.28  & 2.18  & 0.00  & -0.07 & 0.00 \\
Cspread &       & 4.29  & 2.20  & 0.00  & 0.06  & 0.00 \\
\bottomrule
\end{tabular}%
\label{tab: equity and hedge fund factors}
\end{table}

\clearpage
\begin{table}[t!]
\centering
\caption{From-directional volatility connectedness and alternative currency strategies.}
\begin{minipage}{\textwidth} 
This table presents time-series regression outputs of the excess returns of the from-directional volatility connectedness strategy on alternative benchmarks, estimated one factor at a time. The values reported in the ``$\alpha$ (\%)'' column are expressed in percentage per annum. The t-statistics are based on \cite{newey1987simple} standard errors. The sample spans March 2005 to December 2021, which corresponds to the out-of-sample estimation period for volatility-connectedness-sorted portfolios.
\end{minipage}
\vspace{\medskipamount}
\footnotesize

\begin{tabular}{llD{.}{.}{4.2}D{.}{.}{4.2}D{.}{.}{4.2}D{.}{.}{4.2}D{.}{.}{4.2}}
\toprule
\multicolumn{1}{r}{ } &       & \multicolumn{1}{r}{$\alpha$ (\%)} & \multicolumn{1}{r}{t-stat} & \multicolumn{1}{r}{$\beta$} & \multicolumn{1}{r}{t-stat} & \multicolumn{1}{r}{$R^2$ (\%)} \\
\midrule
DOL   &       & 2.25  & 1.73  & 0.71  & 7.89  & 42.65 \\
CAR   &       & 0.90  & 0.85  & 0.65  & 8.32  & 60.60 \\
VAL   &       & 1.00  & 0.52  & 0.57  & 3.48  & 25.10 \\
MOM (ST) &       & 1.82  & 1.06  & -0.34 & -4.10 & 8.19 \\
MOM (LT) &       & 1.71  & 0.89  & -0.24 & -1.74 & 5.25 \\
\midrule
VOL   &       & 0.01  & 0.01  & 0.93  & 27.12 & 78.44 \\
IVOL  &       & 1.00  & 1.53  & 0.94  & 30.78 & 91.40 \\
VRP   &       & 2.00  & 1.13  & -0.05 & -0.34 & 0.26 \\
SRP   &       & 1.27  & 1.44  & 0.73  & 7.21  & 59.32 \\
RR    &       & 0.13  & 0.16  & 0.81  & 11.63 & 73.40 \\
NFA   &       & 1.01  & 0.89  & 0.81  & 6.28  & 44.81 \\
\midrule
CBC   &       & -0.19 & -0.12 & 0.99  & 9.31  & 39.97 \\
TTNC  &       & 1.50  & 1.22  & 1.13  & 8.08  & 32.12 \\
TImg  &       & 2.03  & 1.15  & 0.12  & 0.43  & 0.37 \\
GImb  &       & 2.46  & 1.45  & -0.25 & -0.83 & 1.53 \\
V-weighted TImb &       & 2.71  & 1.71  & -0.46 & -4.78 & 22.51 \\
HKM   &       & 2.16  & 1.18  & -0.10  & -0.33 & 0.19 \\
\bottomrule
\end{tabular}%
\label{tab: from-directional portfolios and benchmark strategies}%
\end{table}%

\clearpage
\begin{table}[t!]
\centering
\caption{To-directional volatility connectedness portfolios: two-month implied volatilities.}
\begin{minipage}{\textwidth} 
Panel A presents descriptive statistics for the to-directional volatility connectedness strategies, which are based on volatility transmission estimated from two-month implied volatilities. The panel presents the results for shocks of any, short-, and long-term persistence. Mean, standard deviation, and Sharpe ratio are annualized, but the \cite{newey1987simple} t-statistic of mean, skewness, kurtosis, and the first-order autocorrelation are based on monthly returns. We also report the annualized mean of the exchange rate ($\text{fx} = \Delta s^k $) and interest rate ($\text{ir} = i^k - i$) components of excess returns and the average to- and from-directional volatility connectedness measures of portfolios. Panel B presents the results of time-series regressions of the excess returns of the $\mathcal{T}$ strategy on alternative benchmarks, estimated one factor at a time. The values reported in the ``$\alpha$ (\%)'' column are expressed in percentage per annum. The t-statistics are based on \cite{newey1987simple} standard errors. The sample spans March 2005 to December 2021, which corresponds to the out-of-sample estimation period.
\end{minipage}
\vspace{\medskipamount}
\footnotesize

\begin{tabular}{llD{.}{.}{4.2}D{.}{.}{4.2}D{.}{.}{4.2}D{.}{.}{4.2}D{.}{.}{4.2}}
\toprule
\multicolumn{7}{l}{\textbf{Panel A: Summary statistics}} \\
\midrule
&       &       & \multicolumn{1}{c}{$\mathcal{T}(S)$} & \multicolumn{1}{c}{$\mathcal{T}(L)$} & \multicolumn{1}{c}{$\mathcal{T}$} &  \\
\midrule
mean (\%) &       &       & 3.31  & 3.94  & 3.78  &  \\
t-stat &       &       & 2.10  & 2.43  & 2.37  &  \\
\multicolumn{1}{r}{fx (\%) \quad\quad\quad}    &       &       & 2.80  & 3.40  & 3.26  &  \\
\multicolumn{1}{r}{ir (\%) \quad\quad\quad}    &       &       & 0.51  & 0.54  & 0.52  &  \\
\multicolumn{1}{r}{$\mathcal{T}$ \quad\quad\quad}  &       &       & -0.30 & -0.06 & -0.40 &  \\
\multicolumn{1}{r}{$\mathcal{F}$ \quad\quad\quad}  &       &       & -0.03 & -0.01 & -0.04 &  \\
Sharpe &       &       & 0.44  & 0.51  & 0.50  &  \\
std (\%) &       &       & 7.55  & 7.71  & 7.63  &  \\
skew  &       &       & 0.54  & 0.50  & 0.51  &  \\
kurt  &       &       & 3.93  & 3.60  & 3.69  &  \\
ac1   &       &       & 0.00  & 0.01  & 0.01  &  \\
\midrule
\multicolumn{7}{l}{\textbf{Panel B: Alternative currency strategies.}} \\
\midrule
\multicolumn{1}{c}{ } &       & \multicolumn{1}{c}{$\alpha$ (\%)} & \multicolumn{1}{c}{t-stat} & \multicolumn{1}{c}{$\beta$} & \multicolumn{1}{c}{t-stat} & \multicolumn{1}{c}{$R^2$ (\%)} \\
\midrule
DOL   &       & 3.72  & 2.35  & -0.18 & -1.92 & 4.06 \\
CAR   &       & 3.87  & 2.47  & -0.06 & -0.82 & 0.68 \\
VAL   &       & 3.71  & 2.42  & 0.03  & 0.36  & 0.13 \\
MOM (ST) &       & 3.87  & 2.40  & 0.16  & 1.82  & 2.71 \\
MOM (LT) &       & 3.62  & 2.29  & -0.12 & -1.35 & 1.87 \\
\midrule
VOL   &       & 4.12  & 2.63  & -0.16 & -1.60 & 3.37 \\
IVOL  &       & 3.90  & 2.47  & -0.12 & -1.26 & 2.11 \\
VRP   &       & 3.75  & 2.46  & -0.08 & -0.87 & 0.93 \\
SRP   &       & 3.82  & 2.43  & -0.04 & -0.59 & 0.31 \\
RR    &       & 4.04  & 2.63  & -0.11 & -1.33 & 2.13 \\
NFA   &       & 3.96  & 2.47  & -0.15 & -1.32 & 2.24 \\
\midrule
CBC   &       & 4.03  & 2.60  & -0.11 & -0.76 & 0.77 \\
TTNC  &       & 3.73  & 2.32  & 0.11  & 0.69  & 0.43 \\
TImg  &       & 3.78  & 2.35  & 0.07  & 0.42  & 0.17 \\
GImb  &       & 3.22  & 1.98  & 0.31  & 2.17  & 3.61 \\
V-weighted TImb &       & 3.44  & 2.25  & 0.22  & 2.76  & 7.56 \\
HKM   &       & 3.65  & 2.33  & 0.09  & 0.45  & 0.22 \\
\bottomrule
\end{tabular}%
\label{tab: summary stats and alphas for two-month implied volatilities}%
\end{table}%

\clearpage
\begin{table}[t!]
\centering
\caption{Diversification gains: two-month implied volatilities.}
\begin{minipage}{\textwidth} 
This table illustrates the impact of adding the $\mathcal{T}$ strategy, which is based on volatility transmission estimated from two-month implied volatilities, to common currency factors. Specifically, we construct a naive 50\%-50\% portfolio of the $\mathcal{T}$ strategy and another factor. Mean, standard deviation, and Sharpe ratio are annualized, but the \cite{newey1987simple} t-statistic of mean, skewness, kurtosis, and the first-order autocorrelation are based on monthly returns. The last column shows the percentage increase in the Sharpe ratio of a naive portfolio relative to the original currency strategy. The sample spans March 2005 to December 2021, which corresponds to the out-of-sample estimation period.
\end{minipage}
\vspace{\medskipamount}
\footnotesize

\begin{tabular}{llD{.}{.}{4.2}D{.}{.}{4.2}D{.}{.}{4.2}D{.}{.}{4.2}D{.}{.}{4.2}D{.}{.}{4.2}D{.}{.}{4.2}D{.}{.}{4.2}}
\toprule
 & &      \multicolumn{1}{r}{mean (\%)} & \multicolumn{1}{r}{t-stat} & \multicolumn{1}{r}{Sharpe} & \multicolumn{1}{r}{std (\%)} & \multicolumn{1}{r}{skew} & \multicolumn{1}{r}{kurt} & \multicolumn{1}{r}{ac(1)} & \multicolumn{1}{r}{$\%\Delta$ Sharpe} \\
\midrule
DOL   & &  1.72  & 1.46  & 0.34  & 5.11  & -0.07 & 4.44  & 0.04  & \multicolumn{1}{c}{\quad\quad n.a.} \\
CAR   & & 2.75  & 1.94  & 0.43  & 6.46  & 0.04  & 5.10  & 0.04  & 173.96 \\
VAL   & & 2.79  & 2.00  & 0.49  & 5.70  & 0.18  & 4.95  & 0.02  & 122.75 \\
MOM (ST) & & 1.60  & 1.36  & 0.27  & 5.86  & 0.82  & 6.31  & 0.01  &  \multicolumn{1}{c}{\quad\quad n.a.} \\
MOM (LT) & & 1.25  & 1.01  & 0.23  & 5.38  & -0.22 & 3.56  & 0.00  & \multicolumn{1}{c}{\quad\quad n.a.} \\
\midrule
VOL   & & 2.97  & 2.78  & 0.56  & 5.28  & -0.12 & 6.60  & -0.01 & 130.04 \\
IVOL  & & 2.43  & 2.09  & 0.43  & 5.61  & -0.25 & 5.48  & 0.00  & 278.07 \\
VRP   & & 1.74  & 1.42  & 0.32  & 5.49  & 0.77  & 7.52  & 0.03  & \multicolumn{1}{c}{\quad\quad n.a.} \\
SRP   & & 2.40  & 2.04  & 0.40  & 6.04  & 0.11  & 5.63  & 0.03  & 280.65 \\
RR    & & 3.05  & 2.77  & 0.53  & 5.75  & 0.05  & 6.13  & 0.01  & 123.19 \\
NFA   & & 2.51  & 2.24  & 0.50  & 4.99  & -0.51 & 6.68  & 0.10  & 210.09 \\
\midrule
CBC   & & 3.01  & 2.92  & 0.65  & 4.63  & 0.10  & 5.12  & 0.05  & 72.78 \\
TTNC  & & 2.12  & 2.21  & 0.46  & 4.59  & 0.17  & 5.20  & 0.03  & 367.06 \\
TImg  & & 1.82  & 2.09  & 0.40  & 4.59  & 0.09  & 3.20  & 0.05  & \multicolumn{1}{c}{\quad\quad n.a.} \\
GImb  & & 2.77  & 2.94  & 0.58  & 4.81  & 0.73  & 5.58  & 0.03  & 49.43 \\
V-weighted TImb & & 2.64  & 1.85  & 0.38  & 6.87  & 0.90  & 6.06  & -0.03 & 143.32 \\
HKM   & & 2.63  & 2.94  & 0.59  & 4.42  & 0.51  & 4.29  & 0.02  & 65.72 \\
\bottomrule
\end{tabular}%
\label{tab: diversification benefits for two-month implied volatilities}%
\end{table}%

\clearpage
\begin{table}[t!]
\centering
\caption{Allocation analysis: from-directional volatility connectedness.}
\begin{minipage}{\textwidth} 
This table presents the allocation frequencies of currencies across quintile $(\mathcal{P}_i: i = 1,\dots,5)$ portfolios sorted by from-directional volatility connectedness. $\mathcal{P}_1 (\mathcal{P}_5)$ comprises currencies with the lowest (highest) levels of received volatility. The columns report the fraction of months each currency belongs to the portfolios considered. The sample spans March 2005 to December 2021, which corresponds to the out-of-sample estimation period for volatility-connectedness-sorted portfolios.
\end{minipage}
\vspace{\medskipamount}
\footnotesize

\begin{tabular}{llD{.}{.}{2.2}D{.}{.}{2.2}D{.}{.}{2.2}D{.}{.}{2.2}D{.}{.}{2.2}D{.}{.}{4.0}}
\toprule
&& \multicolumn{1}{r}{$\mathcal{P}_1$} & \multicolumn{1}{r}{$\mathcal{P}_2$} & \multicolumn{1}{r}{$\mathcal{P}_3$} & \multicolumn{1}{r}{$\mathcal{P}_4$} & \multicolumn{1}{r}{$\mathcal{P}_5$}  & \multicolumn{1}{r}{Obs.} \\
\midrule
Australia &       & 0.00  & 0.11  & 0.42  & 0.45  & 0.02  & 202 \\
Brazil &       & 0.00  & 0.03  & 0.07  & 0.16  & 0.73  & 202 \\
Canada &       & 0.30  & 0.55  & 0.12  & 0.02  & 0.00  & 202 \\
Czech Republic &       & 0.12  & 0.26  & 0.28  & 0.26  & 0.08  & 202 \\
Denmark &       & 0.42  & 0.40  & 0.17  & 0.02  & 0.00  & 202 \\
Euro Area &       & 0.07  & 0.67  & 0.23  & 0.03  & 0.00  & 202 \\
Hungary &       & 0.01  & 0.10  & 0.18  & 0.26  & 0.45  & 202 \\
Japan &       & 0.19  & 0.33  & 0.29  & 0.13  & 0.06  & 202 \\
Mexico &       & 0.11  & 0.12  & 0.17  & 0.22  & 0.37  & 202 \\
New Zealand &       & 0.00  & 0.01  & 0.24  & 0.56  & 0.18  & 202 \\
Norway &       & 0.00  & 0.02  & 0.32  & 0.55  & 0.10  & 202 \\
Poland &       & 0.00  & 0.02  & 0.25  & 0.33  & 0.40  & 202 \\
Singapore &       & 0.99  & 0.01  & 0.00  & 0.00  & 0.00  & 202 \\
South Africa &       & 0.00  & 0.00  & 0.01  & 0.04  & 0.94  & 202 \\
South Korea &       & 0.24  & 0.26  & 0.23  & 0.19  & 0.07  & 202 \\
Sweden &       & 0.00  & 0.07  & 0.42  & 0.48  & 0.02  & 202 \\
Switzerland &       & 0.21  & 0.45  & 0.26  & 0.07  & 0.01  & 202 \\
Taiwan &       & 1.00  & 0.00  & 0.00  & 0.00  & 0.00  & 202 \\
Turkey &       & 0.04  & 0.13  & 0.17  & 0.12  & 0.53  & 202 \\
United Kingdom &       & 0.29  & 0.45  & 0.15  & 0.10  & 0.01  & 202 \\
\bottomrule
\end{tabular}%
\label{table: allocation analysis from-directional volatility}
\end{table}%







\end{document}